\documentclass[11pt]{amsart}
\usepackage{amsmath}
\usepackage{amssymb}
\usepackage{dsfont}
\usepackage{comment}
\usepackage{ifthen}
\newboolean{bl_anonymize_for_subm}
\setboolean{bl_anonymize_for_subm}{false}
\usepackage{color}
\usepackage{xcolor}
\usepackage{url}
\usepackage{amsthm}
\usepackage{bbm}
\usepackage[linesnumbered,ruled,vlined]{algorithm2e}
\usepackage{bm}
\usepackage{xy}
\usepackage{enumitem}
\usepackage[ruled,vlined]{algorithm2e}
\usepackage{mathrsfs}
\usepackage[gen]{eurosym}
\usepackage{graphicx}
\usepackage{subcaption}
\usepackage[
    top=1.25in,
    bottom=1.25in,
    left=1.25in,
    right=1.25in
]{geometry}
\usepackage{fancyhdr}
\usepackage{booktabs}
\usepackage{tikz}
\usetikzlibrary{decorations.markings}
\usetikzlibrary{decorations.pathmorphing}
\usetikzlibrary{arrows, snakes}
\usetikzlibrary{shapes.geometric, calc,decorations.pathreplacing}
\usetikzlibrary{backgrounds}
\usetikzlibrary{fit}
\usetikzlibrary{decorations.pathreplacing}
\input xy
\xyoption{all}

\tikzset{middlearrow/.style={
		decoration={markings,
			mark= at position 0.5 with {\arrow{#1}} ,
		},
		postaction={decorate}
	}
}

\newcommand\myprime{\mkern-3.5mu\raise0.6ex\hbox{$\scriptstyle\prime$}}

\theoremstyle{plain}
\newtheorem{theorem}{Theorem}[section]

\newtheorem{proposition}[theorem]{Proposition}

\newtheorem{definition}[theorem]{Definition}

\theoremstyle{remark}
\newtheorem{remark}[theorem]{Remark}

\numberwithin{equation}{section}

\allowdisplaybreaks[1]

\usepackage{hyperref}
\hypersetup{
  colorlinks=true, linkcolor=blue, citecolor=blue, urlcolor=blue,
  pdfkeywords = {Automated Market Maker, Liquidity Provision, Fees, Adverse Selection},
  pdftitle = {Optimal Fees for Liquidity Provision in Automated Market Makers},
  pdfsubject = {Optimal Fees for Liquidity Provision in Automated Market Makers},
  pdfpagemode = UseNone
}
\ifthenelse{\boolean{bl_anonymize_for_subm}}
{\hypersetup{pdfauthor={}}}
{\hypersetup{pdfauthor={Steven Campbell, Philippe Bergault, Jason Milionis, Marcel Nutz}}}
\usepackage[capitalize,noabbrev]{cleveref}
\usepackage[citestyle=numeric,bibstyle=numeric,maxcitenames=2,maxbibnames=999]{biblatex}

\makeatletter
\newcommand*\bigcdot{\mathpalette\bigcdot@{.8}}
\newcommand*\bigcdot@[2]{\mathbin{\vcenter{\hbox{\scalebox{#2}{$\m@th#1\bullet$}}}}}
\makeatother
\usepackage{setspace}

\title{Optimal Fees for Liquidity Provision in Automated Market Makers}

\ifthenelse{\boolean{bl_anonymize_for_subm}}
{}
{
  \usepackage{amsaddr} %

  \author{
    Steven Campbell$^{1}$,
    Philippe Bergault$^{2}$,
    Jason Milionis$^{3,4}$,
    Marcel~Nutz$^{1}$
  }

  \address{
    $^{1}$Dept.\ of Statistics, Columbia University \\
    $^{2}$CEREMADE, Université Paris Dauphine-PSL \\
    $^{3}$Dept.\ of Computer Science, Columbia University \\
    $^{4}$Category Labs %
  }

  \email{sc5314@columbia.edu, bergault@ceremade.dauphine.fr, jm@cs.columbia.edu}
  \email{\quad \quad \quad \quad \quad \quad \quad \quad  \ mnutz@columbia.edu}
}
\date{\today}

\begin{document}

\begin{abstract} 
Passive liquidity providers (LPs) in automated market makers (AMMs) face losses due to adverse selection (LVR), which static trading fees often fail to offset in practice. We study the key determinants of LP profitability in a dynamic reduced-form model where an AMM operates in parallel with a centralized exchange (CEX), traders route their orders optimally to the venue offering the better price, and arbitrageurs exploit price discrepancies. Using large-scale simulations, we analyze how LP profits vary with market conditions such as volatility and trading volume, and characterize the profit-maximizing AMM fee at fixed liquidity. We then endogenize liquidity through competitive LP entry and characterize the fee that maximizes equilibrium total value locked (TVL). We highlight the mechanisms driving these relationships through extensive comparative statics, and confirm the model's relevance through market data calibration. A key trade-off emerges: fees must be low enough to attract volume, yet high enough to earn sufficient revenues and mitigate arbitrage losses. We find that under normal market conditions, the profit-maximizing AMM fee is competitive with the trading cost on the CEX and remarkably stable, whereas in periods of very high volatility, a high fee protects passive LPs from severe losses. Similarly, under competitive entry, the TVL-maximizing fee typically lies slightly below the CEX trading cost, and equilibrium liquidity falls sharply with volatility. Our findings suggest that a threshold-type dynamic fee schedule is robust to market conditions and improves LP outcomes.
\end{abstract}

\maketitle

\section{Introduction}

Decentralized exchanges (DEXs) are an important class of trading venues in decentralized finance (DeFi), allowing users to trade assets or provide liquidity without intermediation. %
Most DEXs operate using automated market makers (AMMs) where traders interact with a liquidity pool rather than being matched directly as in a limit order book.\footnote{See \url{https://defillama.com/dexs} for up-to-date trading volume statistics. We lightly abuse terminology in using the terms DEX and AMM interchangeably.} Exchange rates are set by predetermined formulas, most often following the constant function market maker (CFMM) model which enforces a fixed relationship between asset quantities in the pool. A key attraction of CFMMs is \emph{passive} liquidity provision---liquidity providers (LPs) can supply liquidity without managing their positions continuously as in limit order book markets. In return for providing liquidity, LPs earn a pro-rata share of fees that are levied on trades in the AMM. However, the convenience of not managing positions comes at a cost: LPs are exposed to losses due to adverse selection (see \textcite{CapponiJia2025ExAdoption,jasonLVR}, among others). Indeed, rational market participants monitoring real-time prices on other venues where price discovery occurs can exploit stale prices offered by the AMM. Even after hedging away price risk, LPs incur a systematic loss known as loss-versus-rebalancing (LVR, see \textcite{jasonLVR}, also identified as ``convexity cost,'' see \textcite{Cartea24,cartea2023predictable}). While fees would ideally compensate for that, practice has shown that LPs often suffer persistent negative returns, especially in volatile markets (e.g., \cite{canidio2023arbitrageurs,crocswap_usage_2022,atis2023}). This poses a serious challenge to the long-term sustainability of passive liquidity provision (e.g., \textcite{hasbrouck2025concentrated}).
Therefore, a plurality of recent works study this issue and possible solutions from various angles, such as dynamic fees adjusting to volatility (\textcite{cao2023automated}), modifications to blockchain frequency (\textcite{jason_arb_profits}), or auction-based AMM designs redistributing arbitrage gains to LPs (\textcite{adams2024amm}), among many others. The practical relevance of such approaches is also evidenced by the numerous works from industry groups aiming to mitigate LVR (e.g., \cite{algebradexengine,catalabs2023,fenbushi2023,gyroscope2023,sorella}).
Despite this rapidly growing body of literature, a simulation-based, model-agnostic framework for analyzing LP outcomes is absent.

The key goal of the present work is to understand the PnL of passive liquidity provision \emph{as a function of market conditions} such as trading fees, price volatility, trading volume, and liquidity. In particular, we want to determine the AMM fee that maximizes expected LP PnL given market conditions, and whether that PnL is positive. Moreover, we want to understand how much the PnL deteriorates when the AMM fee deviates from this optimum, perhaps because market conditions evolve over time. As markets mature, passive liquidity provision may equilibrate according to its profitability over longer horizons. Building on the previous analysis, we thus perform an additional investigation with competitive LP entry,  which endogenizes liquidity, and study the fee that maximizes equilibrium total value locked (TVL). One motivation for the present work is the proposal of \emph{dynamic fees} (that vary as a function of market conditions) as a way to reduce LVR. In the context of Uniswap~v4 hook design, Uniswap's guidance highlights volatility and trade volume dependence as a best practice in fee design~\cite{uniswapDynamicFees}.

In contrast to much of the existing related literature (see \cref{sec:literature} for a literature review), our analysis begins with a reduced-form model where market conditions are \emph{input} parameters. Using large-scale simulations, we can then explore the hypersurface of all parameter combinations. A dynamic market where conditions evolve jointly over time can be seen as a curve traveling on that hypersurface, so that our study can inform LP profitability in a dynamic situation: a simulation with constant parameters over a short time horizon serves as a building block of a dynamic model obtained by concatenating several simulations with different parameters. While the evolution would certainly include market and asset-specific dynamics, the design of the building block is agnostic to those. %

\subsection{Model}
In our baseline model, the num\'eraire asset $X$ and the risky asset $Y$ are traded on two venues, a centralized exchange (CEX) and a decentralized exchange operating by an AMM, more precisely a constant product market maker (first analyzed by \textcite{angeris2019analysis}) akin to Uniswap~v2.\footnote{While Uniswap v3 with concentrated liquidity carries higher trading volume at the time of writing, and v4 with hooks is the most recent Uniswap protocol, we prefer a parsimonious model with few parameters to study the mechanics of CEX-DEX interaction. In such a model, producing realistic price ratios is a nontrivial sanity check (see \cref{sec:empirics}), whereas with a very large parameter set as in v3 (even infinite in v4), it would be difficult to draw any conclusions.} The price of $Y$ (in units of $X$) on the CEX evolves exogenously and the assets can be traded at a fixed proportional cost~$\eta^0$ which stands in for all trading costs on the CEX---nominal fees levied by the exchange, bid-ask spread, etc. The liquidity for the AMM is supplied by a representative LP at the initial time. After that, no shares are minted or burnt, so that pool reserves on the AMM evolve only due to swaps by traders. Swaps on the AMM incur a proportional fee $\eta^1$ which is paid to the LP. There are two types of traders. \emph{Arbitrageurs} act whenever prices on the two venues allow for a profitable roundtrip after accounting for fees. Their trades thus align the marginal price on the AMM with the CEX price, up to fees. Traders of the second type are called \emph{fundamental traders} and differ only by having an exogenous trading demand. To satisfy their demand, they route their order flow optimally to the two venues, meaning that each marginal trade is routed to the venue offering a better marginal price after fees.\footnote{Equivalently, fundamental traders send all their orders to a smart router (aggregator) monitoring both venues.} As a result, some of the fundamental flow may go to the AMM and contribute to aligning the price with the CEX, even when there is no profitable roundtrip. We evaluate the expected PnL of the representative LP, which comprises the fees accrued plus the change in marked-to-CEX value of the pool reserves, conditional on the initial pool depth.

In a second step, we add competitive entry for LPs, endogenizing initial pool depth. Given the AMM fee and other market parameters, capital enters or exits according to whether the expected hedged LP return is above or below an outside benchmark. Equilibrium pool depth, when it exists, equates the two. Once that depth is determined, the trading stage proceeds as above.

Before moving on to our results, let us pause to highlight how our setup differs from the existing literature studying the interaction of an AMM with a CEX (see also \cref{sec:literature}). %
Most previous works assume that the CEX is frictionless---while trading on the AMM is subject to a fee, trading on the CEX does not incur any fee or slippage. Then, trading on the AMM should be attractive only to one side of the informed market; for instance, buyers if the price on the AMM, inclusive of fee, is lower than the CEX price. The resulting trades would be the ones with the highest adverse selection cost to LPs, and it is intuitive that liquidity provision would never be beneficial in this situation. Previous works include noise traders who trade on the AMM \emph{regardless} of prices on the CEX. This adds trades that are on average profitable to LPs as soon as the AMM fee is positive, so that a sufficiently large volume of noise trades renders liquidity provision profitable. By contrast, our baseline model does not have noise traders. While our fundamental traders have an exogenous trading motive, they are informed about CEX prices and route their flow optimally between the two venues. 

In the absence of noise traders, a key reason that liquidity provision can be profitable is that trading on the CEX \emph{is not free either}. Indeed, both venues offer a combination of price and slippage. Even when the AMM marginal price is inferior to the CEX price, the AMM can be competitive for fundamental flow if its fee is lower. Specifically, a marginal fundamental buyer prefers the AMM if AMM marginal price plus AMM fee is lower than CEX price plus CEX fee. For an LP who marks assets to CEX prices and earns the AMM fee, this means that the trade is potentially profitable.\footnote{If the AMM price inclusive of fee exceeds the CEX price without fee, the difference will be the LP's marginal net profit (which is therefore bounded from above by the CEX fee). If not, the trade results in a loss for the~LP.}

In the absence of free money from noise traders, the AMM needs to compete with the CEX in terms of trading costs. This idea will be underscored by our quantitative results on fee choice below, which will show that the AMM can make up for its stale pricing by offering a competitive fee, and carve out a regime where installed liquidity earns positive expected PnL without noise traders. This is a conservative assessment: We agree with the literature that in reality there are traders who will avoid the CEX even when its price is more attractive, for instance, to avoid counterparty risk or Know Your Customer regulations. Adding noise traders to our model as a third type of trader will only improve LP outcomes, as is discussed in \cref{sec:extensions} among other extensions of our baseline model.

\subsection{Contributions and Results} Before discussing the insights derived from our model, let us briefly highlight the implementation itself. %
To meet the computational demands of large-scale experimentation, we provide an efficient, parallelized code base ready to deploy on high-performance computing clusters. The implementation is modular and extensible, so that interested researchers can run their own experiments. The code is publicly available.\footnote{\url{https://github.com/JasonSome/cpmm-trading/tree/master}}

Thanks to our reduced-form modeling approach, we can produce comparative statics with respect to any parameter, and that is key to our goal of understanding liquidity provision. Assuming that the CEX price follows a driftless geometric Brownian motion,\footnote{While we focus here on geometric Brownian motion, the simulation accepts any time series as input for the CEX price, be it a sample from a stochastic process or real-world price data as discussed in \cref{sec:empirics}.} the price dynamics are summarized by a single volatility parameter $\sigma$. In line with previous findings, our simulation shows that all else equal, expected LP PnL is decreasing in volatility. On the other hand, it is increasing in fundamental demand, which was not considered before.\footnote{PnL is also increasing in noise trader volume if such traders are included. This agrees with previous findings and is also straightforward analytically as fee income from noise trader volume amounts to a direct transfer to LPs.} See \cite{capponi2026viability,Cartea24} for related work that connects trading activity to liquidity provision through fee income and noise demand that is sensitive to pool depth. We highlight the mechanics leading to those effects by extensively analyzing (a) the distribution of the ratio between AMM and CEX prices, which captures the adverse selection cost of trades, and (b) the volume attracted by the AMM, which multiplies the gross PnL of those trades and additionally the LP's fee earnings. Greater CEX price volatility is reflected in wider dispersion of the price ratio distribution, thus increasing adverse selection and reducing PnL. Increasing fundamental demand generally means that the AMM captures more volume, increasing fee revenue. More importantly, this additional fundamental flow tends to trade at prices that imply smaller adverse selection costs or even profits (depending on the relationship between AMM and CEX fees): When CEX prices sufficiently deviate from AMM prices, the AMM attracts one-sided fundamental flow (only buyers or only sellers) which acts as a \emph{reverting drift} on the price ratio process even before arbitrage opportunities arise.\footnote{If prices deviate even more, the ratio process is \emph{reflected} at the threshold where the arbitrage opportunity arises.} This effect is clearly seen in the narrowing price ratio distribution, and exemplifies how volatility and fundamental demand tend to act in opposition in shaping the ratio distribution. Such an effect has not been observed in previous works.

While our reduced-form model allows us to discuss comparative statics individually for each parameter, it is important to remember that market characteristics are likely correlated in practice. For instance, our simulation also shows that if a volatility increase is accompanied by a surge in demand, the net effect on LP PnL can be approximately flat. However, all else fixed, a sufficiently large volatility will always render liquidity provision unprofitable at the given pool depth: in the limit $\sigma\to\infty$, both the arbitrage volume and the per-trade adverse-selection loss tend to infinity. A detailed discussion of LP profitability and comparative statics can be found in \cref{sec:profitability}.

Next, we highlight the complex role of the AMM fee, likely the most interesting parameter in light of the current discussions about AMM design. %
Higher fees increase per-trade revenues for the LP and defend against adverse selection. On the other hand, higher fees reduce the volume captured by the AMM, reducing fee income but also increasing price stickiness as the reverting force of fundamental flow is reduced and arbitrage requires greater price disparities. This economic narrative is central to our analysis: the fee is not chosen in a vacuum; instead, fee‐setting happens in direct competition with the CEX, and every basis‐point change carries downstream consequences for trade volume, price action, revenue generation, and arbitrage activity. %

For a broad range of volatility and fundamental demand parameters (which we interpret as normal market conditions), we find that given fixed initial liquidity, there is a unique optimal AMM fee maximizing expected LP PnL. This fee \emph{undercuts} the representative CEX trading cost $\eta^0$, in line with the aforementioned idea that the AMM needs to be competitive with the CEX in terms of slippage. Indeed, slippage on the AMM is given by AMM fee plus price impact whereas slippage on the CEX is given by the CEX fee alone, hence undercutting by a modest amount is an intuitive choice for a competitive AMM fee. Nevertheless, it seems that we are the first to make this direct connection between the PnL-maximizing AMM fee and trading costs on the CEX.

At this point, it is important to remember that our CEX fee represents all trading costs on the CEX. While assuming a single representative number is useful for comparative statics and the empirical experiments below, we also note that in reality, different traders incur different costs when trading on a CEX, for instance because nominal fees differ by volume tiers\footnote{For instance, \url{https://www.binance.com/en/fee/schedule} lists a maker/taker fee of 10 basis points (bps) for users with a 30-day trading volume under \$1M, decreasing to 4/5.2 bps for a user with a \$100M volume (as of June 30, 2025).} or because the slippage of a sophisticated execution algorithm is smaller than for a simple market order. On the other hand, the AMM fee is the same for all traders. Thus, in practice, the message of undercutting may be interpreted as offering a better fee for \emph{some} traders, perhaps retail customers that form a relatively important fraction in cryptocurrency markets. See also the discussion in \cref{sec:extensions_fees}. In \cref{sec:empirics}, the representative CEX fee is determined by \emph{calibrating} to the price ratio distribution rather than using posted nominal fees.

Returning to the discussion of simulation results, the PnL-maximizing AMM fee is increasing in volatility and decreasing in demand throughout the regime of normal markets, reflecting the changing tradeoff between volume capture and defending against adverse selection losses.\footnote{In a regime of small fundamental demand, the optimal fee is decreasing-then-increasing. An explanation can be found in \cref{sec:optimal.fees}.} Both the sensitivity and the level of the fee are significantly lower than in previous studies. While our PnL-maximizing fee is below the CEX fee, previous models implied optimal AMM fees at very elevated levels, for instance around 125--250 basis points (bps) for the volatility range presented in the recent work of \textcite[Figure~5]{he2024optimal}. While the methodology in previous works is quite different, it is intuitive that inclusion of fundamental traders generally lowers the optimal AMM fee. Our findings align with the recent popularity of 5 bps pools (see also \cref{sec:conclusions}).

Our simulation allows us to study in detail the PnL regret, i.e., how much PnL is diminished when the fee is not optimal. We find that in the regime of normal markets, and more generally as long as volatility does not exceed the threshold where liquidity provision becomes unprofitable, the regret is quite moderate when the AMM fee is kept at a well-chosen value below the CEX fee. This has important practical implications for AMM design, as discussed in \cref{sec:conclusions}.

When the volatility exceeds that threshold, the optimal fee ceases to exist: while the optimum is quite stable up to the threshold (hovering slightly below the CEX fee), the optimal fee is infinite above the threshold as the LP prefers to halt trading on the AMM.
The precise threshold depends on the market parameters---if an increase in volatility is accompanied by a sufficiently large increase in fundamental trading volume, the optimal fee remains fairly stable.

We next use the same return function to determine pool depth in the partial equilibrium setting with competitive LP entry. Scale invariance analytically implies that equilibrium liquidity scales proportionally with fundamental demand. For typical parameter values, a unique stable equilibrium exists and the optimum can be reliably identified numerically. Our simulations show that equilibrium TVL falls sharply with volatility.  The fee that maximizes equilibrium TVL lies close to, and typically just below, the CEX trading cost. Because competitive entry pins LP returns to the outside option, equilibrium TVL measures the amount of capital the pool can support rather than LP profitability. It is not a complete welfare criterion: fees near the TVL maximum may reduce fundamental participation and increase trader execution costs. Our results on AMM fees are reported in \cref{sec:optimal.fees}, and some practical conclusions for a threshold-type dynamic fee schedule are discussed in~\cref{sec:conclusions}.
 
Our empirical analysis in \cref{sec:empirics} demonstrates that our stylized framework, despite its parsimony, captures relevant features of real-world markets. Because our simulation accommodates arbitrary price trajectories, we can feed observed CEX time series directly into the model for calibration and testing. Using high-frequency data for ETH/USDC from Binance and Uniswap, we estimate the model’s three core parameters: the CEX fee, price volatility, and fundamental demand. With only these inputs, the model reproduces a price ratio distribution that closely aligns with empirical observations. The calibrated CEX fee converges near posted taker rates, and estimated volatility falls within plausible intraday ranges. Importantly, when testing LP performance on real price data, we find that lowering the AMM fee below the \emph{effective} CEX level improves LP profitability---a conclusion that reinforces our simulation results and underscores the model’s practical relevance. Applying competitive entry in the calibrated environment similarly produces substantially greater equilibrium reserves and a TVL-maximizing fee below the effective CEX trading cost.

To confirm that our conclusions are robust, \cref{sec:extensions} considers several extensions of the baseline model. These extensions include adding noise traders that always prefer the AMM to the CEX, additional fee levels, changing trading frequency, and CEX prices with drift.

\subsection{Related Literature}\label{sec:literature}

A first group of papers studies how adverse selection and uninformed trading determine aggregate AMM liquidity. \textcite{aoyagi2022liquidity} considers a two-period CPMM model with risk-neutral LPs, an informed trader, an exogenous noise trader, and no fees. Under this model, informed trading generates adverse-selection losses, whereas random noise trading generates a positive expected LP return. The paper derives a unique stable competitive pool size, which decreases with adverse selection and volatility and increases with the magnitude of noise trading, and also studies strategic competition among finitely many LPs. In contrast to our setting, there is no AMM fee and no competition with a CEX.

\textcite{CapponiJia2025ExAdoption} emphasize a different friction arising from blockchain infrastructure. Because arbitrage losses are shared across the pool while withdrawal costs are borne individually, LPs have insufficient incentives to withdraw before adverse selection losses occur, producing a tragedy of the commons. The authors show that arbitrage rents flow primarily to validators and that LPs respond defensively to heightened adverse-selection risk by adopting more convex pricing curves. Their analysis focuses on strategic withdrawal and curve design, whereas we take the CPMM invariant as given and study how fees and market conditions affect liquidity provision.

\textcite{LeharParlour.23} characterize equilibrium liquidity and complement their theoretical analysis with comprehensive transaction data. In their model, uninformed trading generates fee revenue, while fundamental price innovations induce arbitrage and hence losses for LPs. Equilibrium pool depth decreases with volatility and increases with uninformed trading volume. Empirically, the authors observe stable pools and absence of long-lived arbitrage opportunities. Our equilibrium comparative statics share the prediction that volatility reduces pool depth and trading demand supports greater liquidity. The order-flow mechanism differs, however: in their model, liquidity traders choose their trade size on the AMM as a function of pool depth, but do not route orders between two venues. In our model, fundamental traders observe contemporaneous CEX prices and route each marginal trade to the venue offering the better all-in execution price. Consequently, fundamental trades in our model can generate gains or losses for LPs and can move the AMM price toward the CEX price before arbitrage becomes profitable.

The feedback among fees, liquidity, price impact, and venue choice is central to \textcite{hasbrouck2022need}. In their one-period model, investors allocate capital between DEX inventory and an outside investment, while liquidity demanders choose between the DEX and CEX based on all-in execution costs. A higher DEX fee can increase the return to supplying inventory, thus attract additional liquidity and reduce price impact sufficiently that the marginal cost of DEX trading falls and DEX volume increases. The same indirect channel is present in our equilibrium formulation: the fee affects fundamental routing directly and also through the pool depth. A key difference is that in \textcite{hasbrouck2022need}, the DEX and CEX marginal prices remain aligned, as any DEX price movement caused by a liquidity trade is immediately reversed by an opportunistic trade. Thus, the effect of endogenous DEX inventory on subsequent venue choice operates through price impact. In our model, by contrast, the DEX price is itself a state variable. Greater depth reduces price impact, but a higher fee also makes the AMM uncompetitive in more price states and weakens the price-aligning effect of fundamental flow. In our simulations, the net effect is that a higher AMM fee generally reduces the AMM's share of fundamental demand, even when the induced entry response supports a deeper pool.

Closest to our model in terms of order flow routing is the work of \textcite{AoyagiIto.24}, which examines the amount of liquidity in a one-period equilibrium model where LPs factor in opportunity cost. The main similarity with our modeling is that in addition to noise traders, there are informed traders choosing between the AMM and CEX depending on prices, bid-ask spread on CEX, and fee on AMM. In their model, informed traders always cause a loss to LPs (and noise traders always cause a gain, see Lemmas~5 and~6 in \cite{AoyagiIto.24}), whereas our optimally routed fundamental trades may be profitable or unprofitable depending on relative prices and fees. The equilibrium amount of liquidity provision is found to be hump-shaped in volatility. Moreover, there is a positive spillover effect between liquidity on AMM and CEX, contrary to the centralization phenomenon usually observed between venues of the same type.

Among papers combining competitive liquidity provision with fee design, \textcite{MalinovaPark.23} provide a benchmark in the context of equity markets. They ask how trading costs and liquidity provision would change if empirical equity volumes were intermediated by AMMs. Their model features competitive liquidity provision and arbitrageurs aligning the AMM price with an exogenous efficient price. Liquidity providers earn fee income from noise trader flow while suffering adverse selection losses due to stale prices. A one-period equilibrium is constructed by assuming that liquidity providers achieve the same expected return as a buy-and-hold benchmark. The authors then determine the optimal fee for minimizing the transaction cost of a representative trade. Calibrating the model to U.S.\ equity data and assuming that the observed equity trading volume is intermediated by AMMs with asset-specific fees, they find that optimal AMM fees are on the order of a few basis points and that AMM trading costs could be substantially below equity market spreads. The optimal fee increases with the volatility of the (open-to-close) return and with the square root of order size relative to trading volume. Their paper is close to ours in emphasizing asset-specific fees. Both papers study AMMs with passive liquidity provision and find that volatility and trading volume are key determinants of the appropriate fee. The two analyses differ in environment and objective. \textcite{MalinovaPark.23} study AMMs as a stand-alone trading mechanism in a one-period model: order flow is exogenous except for arbitrage trades, the LP benchmark is buy-and-hold, and the fee is chosen to minimize equilibrium transaction costs for liquidity demanders. We instead study a dynamic setting where an AMM competes with a centralized exchange and fundamental traders route orders to the venue offering the better price. While their fee trade-off primarily operates through the effect on equilibrium pool depth and trading costs, ours primarily operates through volume capture and protection against LVR.

\textcite{he2024optimal} study AMM design in an infinite-horizon dynamic portfolio model. A representative risk-averse LP allocates wealth among AMM liquidity, the two underlying assets on a CEX, and a risk-free asset, while also choosing consumption. Liquidity traders optimize under heterogeneous beliefs, and arbitrageurs respond to discrepancies between AMM and CEX exchange rates. Using dynamic programming, the authors solve the LP's portfolio problem and choose both the unit trading fee and the CGM pricing function to maximize the LP's utility. They find that the optimal fee increases with volatility and that the pricing function is chosen to make the pool's asset allocation efficient for the LP. The qualitative volatility dependence is consistent with our results. Their liquidity decision is the dynamic portfolio choice of a single representative risk-averse LP, whereas our equilibrium depth is determined by competitive entry of risk-neutral LPs. Moreover, they jointly optimize the fee and pricing function, while we hold the CPMM invariant fixed and focus on the interaction between fees and routing.

\textcite{drissi2026equilibrium} develop a partial-equilibrium model in which a single risk-averse LP chooses initial DEX depth, dynamically offsets inventory risk through CEX trading, and anticipates price-sensitive liquidity demand and arbitrage. The authors establish equilibrium existence for general trading functions and derive explicit results for constant-product markets. Equilibrium liquidity decreases with volatility, risk aversion, and CEX replication costs and increases with the profitability of liquidity demand, whereas private information can have a nonmonotone effect. Their treatment of the CEX as the primary venue for price discovery is close to ours. On the other hand, their equilibrium depth is the optimal capacity choice of a monopolistic LP subject to risk preferences and a budget constraint, whereas ours is determined by competitive entry. More importantly, we focus on the AMM fee as the principal design variable.

In a different direction, \textcite{hasbrouck2025concentrated} endogenize the distribution of liquidity across price intervals in a concentrated-liquidity AMM. They show that an interval-specific liquidity position is related to a covered call whose embedded option is sold at intrinsic rather than market value. Equilibrium interval liquidity is proportional to expected fee revenue and inversely related to the associated option time premium. Expected fee revenue, in turn, depends on the expected time the price spends in that interval.

Relative to this literature, our key contribution is a reduced-form approach that is directly applicable by practitioners. We propose a dynamic model of venue competition and analyze the mapping from market conditions to LP returns that it induces. This mapping yields two complementary fee-design questions: which fee maximizes the return on installed LP capital, and which fee supports the greatest TVL under competitive LP entry. As shown in \cref{sec:empirics}, real-world time series can be used for asset prices and latent parameters can be calibrated to observables such as asset price ratios. Then, our model produces realistic patterns of order flow and LP PnL at existing market conditions.

As mentioned, dynamic AMM fees (that vary with market conditions) have been proposed to reduce LVR and are one motivation for the present study. There are existing protocols that leverage dynamic fees in an empirical manner to provide liquidity providers with better arbitrage protection (e.g., \textcite{lfj,e_dynamic_2024}), including in the context of Uniswap~v4 hook design, but they are mostly using ad hoc solutions based on real-time empirics. For instance, \textcite{lfj} implements the variable fee as a quadratic function of instantaneous volatility. Attempts to understand the benefits of dynamic fees have appeared in the work of \textcite{cao2023automated} who conduct an optimization over volatility-dependent fees in a static single-period model \emph{without} CEX competition. More similar to the present work, \textcite{volosnikov2024impact} study profitability depending on fixed or sigmoid volatility-dependent fees by simulation, but again without taking competition by a CEX into account. \textcite{fritsch2021optimalfeesCFMMs} studies fees in a static Nash equilibrium between competing AMMs, also without a CEX. The systematic study of fees as a function of market conditions in a competitive setting is a crucial gap in the literature that is addressed by our work. 

A different stream of literature studies potential AMM designs with a price oracle, essentially meaning that the CEX price is a state variable of the AMM. \textcite{bergault_automated_2024} use a mean-variance framework inspired by portfolio theory to investigate the use of non-adversarial quotes for AMMs and perform simulations under various market conditions; see also \cite{mvstables,priceaware} for extensions. %
\textcite{aqsha2025equilibriumrewardliquidityproviders} use a principal-agent framework to study the optimal mechanism to redistribute a CFMM's fixed fee to LPs in order to maximize order flow. In a different direction, \textcite{baggiani2025optimal} study optimal dynamic fees in a CFMM using stochastic control. We emphasize that the present setting is fundamentally different as the adverse selection issue is directly related to the latency in price discovery which would be removed by the oracle.\footnote{For example, when (as in \cite{baggiani2025optimal}) there are two dynamic fees (for buying and selling) and both can depend on the CEX price, then one possible fee schedule is to make AMM prices (inclusive of fees) equal to CEX prices, eliminating adverse selection.}

There is by now a substantial literature on AMMs, in addition to the classical literature on market microstructure. For a more extensive review, see for instance \textcite[Section~1.1]{he2024optimal}. 

\subsection{Organization} The remainder of this paper is structured as follows. \Cref{sec:ecosystem} describes the market model consisting of CEX, DEX, traders, and LP, together with the sequence of their interactions. \Cref{sec:profitability} investigates how market characteristics affect LP returns and highlights the underlying mechanisms by analyzing the price ratio and volume distributions. \Cref{sec:equilibrium} introduces competitive liquidity provision and analyzes equilibrium pool depth and market outcomes. \Cref{sec:optimal.fees} studies fee choice with both fixed and equilibrium liquidity. \Cref{sec:empirics} contains the empirical analysis applying our model to Binance and Uniswap data. \Cref{sec:extensions} confirms that our main findings are robust to several extensions of the model. \Cref{sec:conclusions} concludes. \Cref{se:simulation-algorithm} gives pseudocode for the simulation algorithm. \Cref{se:summary.stats} details various summary statistics for the simulations with different parameters underlying the price ratio histograms shown in \cref{fig:histograms} of \cref{sec:profitability}. 
\Cref{sec:model.properties} establishes the competitive
arbitrage and scale invariance properties of the model, and analyzes the limiting regime of large demand and small volatility discussed in \cref{sec:optimal.fees}.
\Cref{app:equilibrium.extensions} expands on the equilibrium discussion in \cref{sec:extensions}, while \cref{app:modeling-assumptions} supplies supplementary empirical evidence in support of the modeling assumptions.

\section{The Model}\label{sec:ecosystem}

We propose a stylized model of an economy with two assets and two trading venues. The first asset, denoted $X$, serves as a reference currency or \emph{numéraire}. The second asset, denoted $Y$, plays the role of the risky asset. In our examples, we take $X$ to be the stablecoin USDC and $Y$ to be ETH. The two assets are traded against one another on a centralized exchange (CEX) and a decentralized exchange (DEX). Trading evolves over a time interval $[0,T]$ and is simulated as $N$ discrete periods of equal length $\delta t = T/N$. There will be three types of agents: arbitrageurs, fundamental traders, and liquidity providers. (\Cref{sec:extensions} additionally considers noise traders.) Below, we first detail each of these ingredients, then explain how they interact in the simulation.

\subsubsection*{CEX} The price $S_t^0$ of the risky asset $Y$ at time~$t$ (denominated in units of $X$) on the CEX follows exogenous dynamics. Most of our exposition uses a simulated price, namely, a geometric Brownian motion with zero drift: $S_t^0$ has continuous dynamics
\begin{equation}\label{eqn:price.evolution}
    \log(S_t^0/S_0^0) = -\frac{1}{2} \sigma^2t + \sigma W_t, \ \ \ t\geq0,
\end{equation}
where $W=(W_t)_{t\geq0}$ is a standard Brownian motion, and this process is sampled at the discrete trading periods. We choose this model for simplicity and parsimony. That said, our implementation can accept an arbitrary time series as input for $S_t^0$, and \cref{sec:empirics} presents an empirical analysis where $S_t^0$ follows a historical time series from Binance instead of a simulated price. \Cref{sec:extensions} considers simulated processes with drift.

As we are modeling a situation where the CEX is a highly liquid trading venue (relative to the DEX), trading on the CEX incurs a proportional fee $\eta^0 > 0$, but no price impact. We emphasize that this fee stands in for \emph{all trading costs} on the CEX (such as bid-ask spread for market orders, adverse selection costs for limit orders, etc.), thus may be larger than the nominal fee posted on exchanges like Binance. The role of $\eta^0$ will be highlighted in greater detail below, and generalizations of this modeling are discussed in \cref{sec:extensions}. 

The mechanics of the CEX are then straightforward: traders can swap any quantity $y>0$ of $Y$ against $yS_t^0$ of $X$ while paying the fee $\eta^0yS_t^0$ (in units of $X$) to third parties.

\subsubsection*{DEX} The DEX is implemented as a constant product market maker (CPMM), akin to Uniswap v2.\footnote{Our simulation code is modular and can accommodate arbitrary AMM designs, for instance G3Ms.} In our baseline model, the initial reserves $(X_0,Y_0)$ of the two assets are seeded by a single representative liquidity provider (LP) who passively holds inventory of both assets and collects all trading fees that traders incur on the AMM.

Prices on the AMM are governed by the constant product invariant $X_t Y_t = k$, where $(X_t, Y_t)$ are the reserves of the two assets at time $t$. The marginal price of $Y$ is $S_t^1 = X_t / Y_t$, and the unit cost for trading a quantity $\Delta$ of $Y$ (positive for buy, negative for sell) is
\begin{equation}\label{eq:executionPrice}
P_t[\Delta] = \frac{X_t}{Y_t - \Delta}.
\end{equation}
After such a trade, the reserves update via
\begin{align}\label{eq:reservesUpdate}
X_{t+\delta t} = X_t + P_t[\Delta] \cdot \Delta, \quad Y_{t+\delta t} = Y_t - \Delta,
\end{align}
and the new marginal price becomes
\begin{align}\label{eq:newMarginalPrice}
S_{t+\delta t}^1 = \frac{X_{t+\delta t}}{Y_{t+\delta t}} = \left( \frac{Y_t}{Y_t - \Delta} \right)^2 S_t^1.
\end{align}
A proportional fee $\eta^1 > 0$ is charged on every DEX trade and paid directly to the LP (i.e., the amount $\eta^1 P_t[\Delta] |\Delta|$ is paid in currency $X$).\footnote{In practice, two methods have been adopted for the collection of fees. Either fees are paid into a separate account, as in our model, or fees are deposited into the AMM pool. We have implemented both methods and found very similar results, thus use the method which leads to simpler formulas and has been favored in recent practice.} Since we assume that LPs do not mint or burn shares over time, the evolution of the reserves $(X_t, Y_t)$ is entirely dictated by trader activity via~\eqref{eq:reservesUpdate}.

\subsubsection*{Traders} In addition to the LP, two types of traders interact with the venues: fundamental traders and arbitrageurs. Fundamental traders have an exogenous trading motive whereas arbitrageurs only act when they can attain a round-trip profit from buying on one venue and immediately selling on the other. 

Indeed, if the DEX offers a better sell price after fees than the CEX buy price after fees (or vice versa), arbitrageurs act to exploit the mispricing. For instance, if 
\[S_t^0(1+\eta^0)<S_t^1(1-\eta^1)\]
it would be profitable to buy on the CEX and sell on the DEX. The size of the arbitrage trade is determined endogenously such as to maximize the profit, or equivalently, to align post-trade marginal prices (inclusive of fees). Explicitly, the arbitrageur's trade on the DEX is
\begin{align}\label{eq:arbTradeSize}
\Delta_t^{A} = Y_t\left(1 - \sqrt{\frac{S_t^1(1-\eta^1)}{S_t^0(1+\eta^0)}} \right) \wedge 0 + Y_t\left(1 - \sqrt{\frac{S_t^1(1+\eta^1)}{S_t^0(1-\eta^0)}} \right) \vee 0.
\end{align}
On the other hand, fundamental traders have exogenous demand. They consist of fundamental buyers and sellers who must execute quantities $\Delta_t^B > 0$ and $\Delta_t^S < 0$, respectively. While those quantities can be specified arbitrarily for each trading period of the simulation, we shall keep them constant over time for parsimony. Buyers and sellers act sequentially in randomized order: either buyers act first, or  sellers. Their order flow is routed optimally between the venues, as though processed by a perfectly efficient aggregator (see  \cref{sec:extensions} for a generalization). This means that traders send their marginal volume to the DEX if and only if the marginal price inclusive of fee is better than the price on the CEX inclusive of fee. Volume will be routed to the DEX until either the order has been fully executed or the price on the DEX becomes aligned with the CEX (inclusive of fees). At that point, the remaining volume will be routed to the CEX as there is no price impact on the CEX. In formulas, buyers trade
\begin{align}\label{eq:buyTradeSize}
\Delta_t^{B,DEX} = \left(\Delta_t^B \wedge Y_t \left(1 - \sqrt{\frac{S_t^1(1+\eta^1)}{S_t^0(1+\eta^0)}} \right)\right) \vee 0
\end{align}
on the DEX and sellers trade
\begin{align}\label{eq:sellTradeSize}
\Delta_t^{S,DEX} = \left(\Delta_t^S \vee Y_t \left(1 - \sqrt{\frac{S_t^1(1-\eta^1)}{S_t^0(1-\eta^0)}} \right)\right) \wedge 0.
\end{align}

\begin{figure}[!htb]
        \centering
        \includegraphics[width=.99\linewidth]{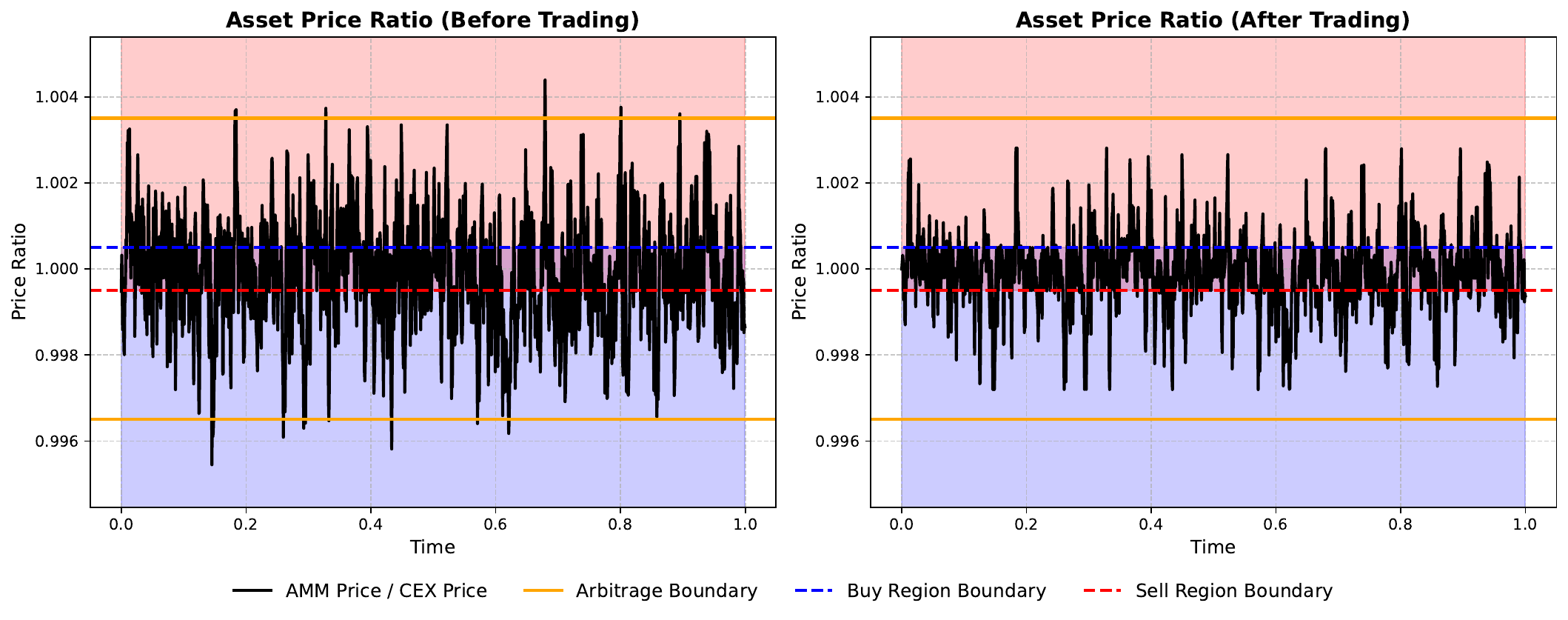}
        \caption{Sample trajectory of the asset price ratio $R_t = S_t^1/S_t^0$ before and after trading on the DEX. Outside of the orange lines arbitrageurs act and reflect the price ratio. Above the red line the AMM captures fundamental sellers and the ratio is pushed down. Below the blue line the AMM captures fundamental buyers and the ratio is pushed up. The center band represents the region where the AMM captures both buy and sell flow. The figure shows the cumulative effect of arbitrageurs and fundamental traders.}
        \label{fig:ratio.process}
\end{figure}

We observe in~\cref{eq:arbTradeSize,eq:buyTradeSize,eq:sellTradeSize} that the decisions of all traders can be neatly expressed in terms of the \emph{price ratio} process $R_t = S_t^1/S_t^0$. Comparing the above expressions, we see that buyers send marginal volume to the DEX if $R_t<(1+\eta^0)/(1+\eta^1)$, while seller volume arrives if $R_t> (1-\eta^0)/(1-\eta^1)$. Similarly, arbitrageurs trade if $R_t< (1-\eta^0)/(1+\eta^1)$ or $R_t>(1+\eta^0)/(1-\eta^1)$. We call those thresholds the \emph{buy region} boundary, \emph{sell region} boundary, and \emph{arbitrage region} boundaries, respectively. These regions will play a crucial role in our analysis of LP profitability; they are visualized alongside an illustrative evolution of $R_t$ in Figure \ref{fig:ratio.process}.

\subsubsection*{LP PnL} 
\begin{figure}[!htb]
        \centering
        \includegraphics[width=0.75\linewidth]{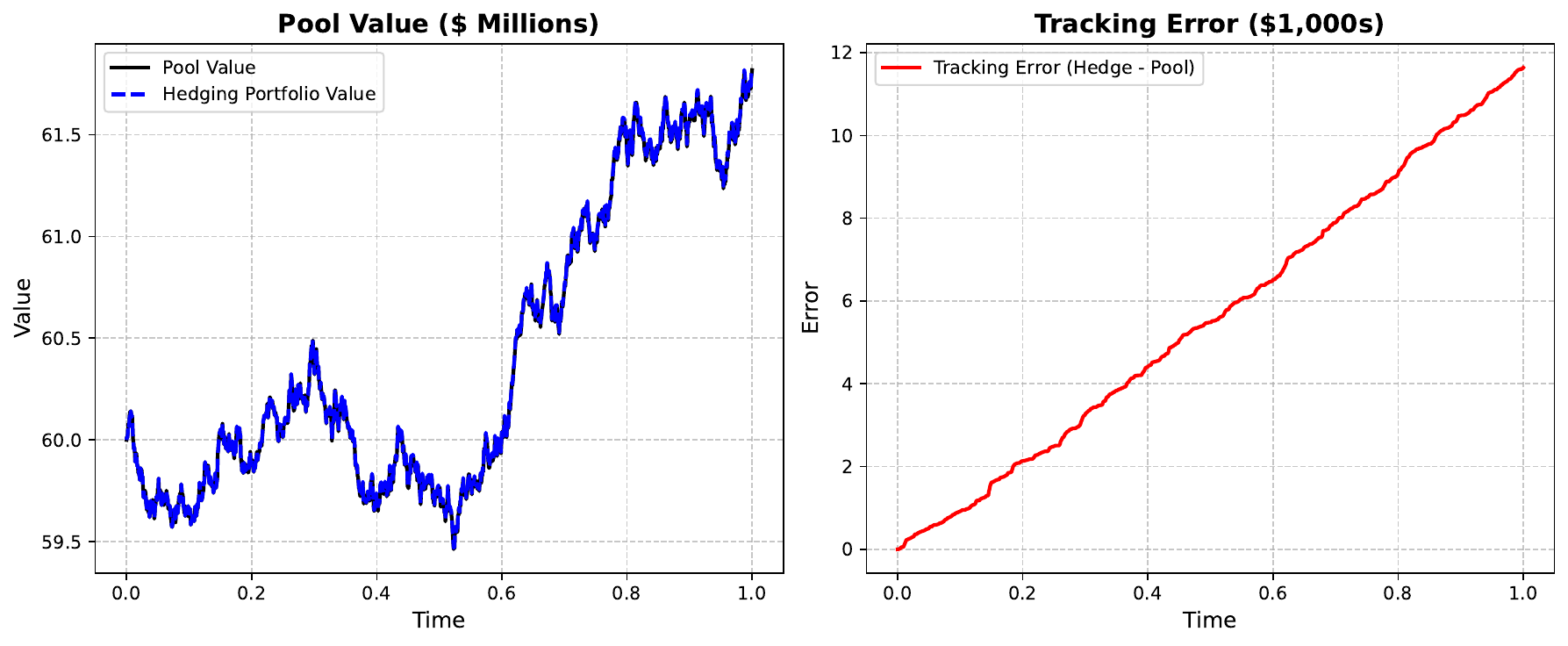}
        \includegraphics[width=0.75\linewidth]{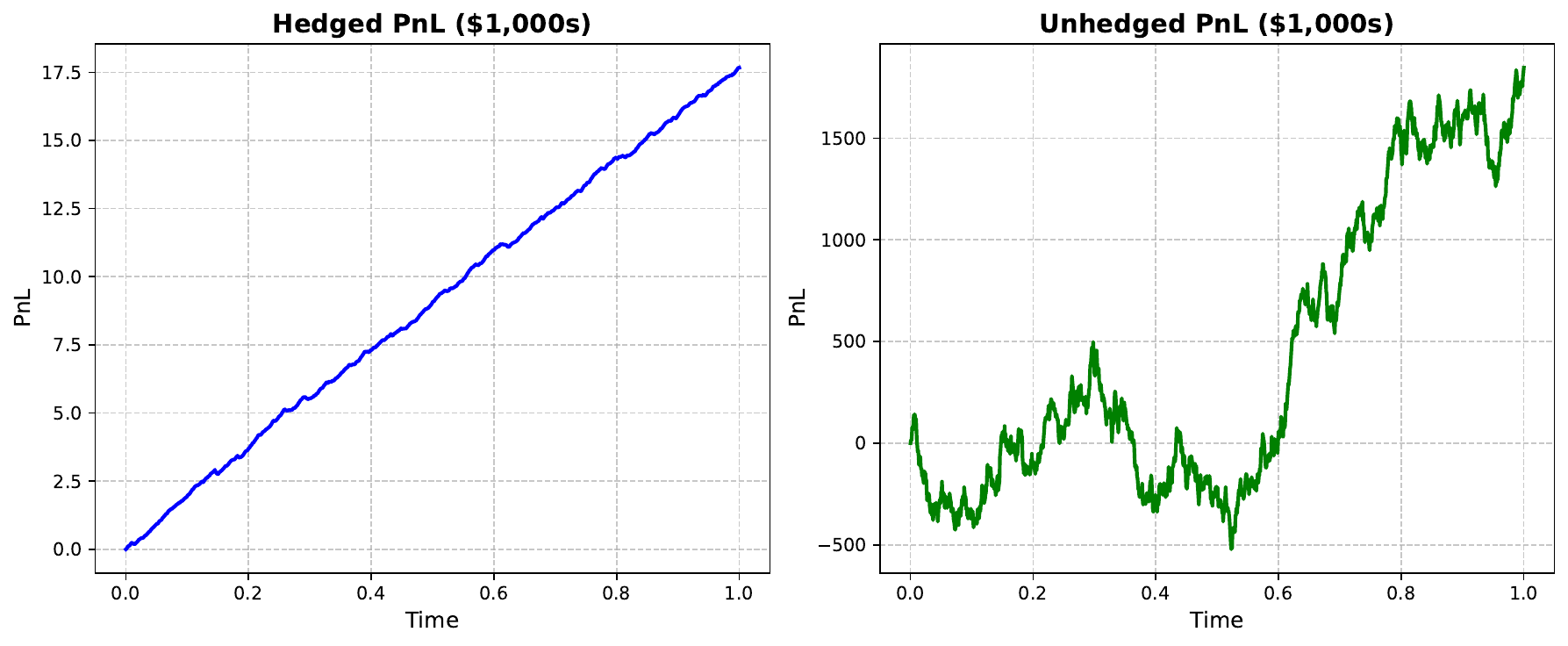}
        \caption{(Upper Left) Sample trajectory of the Pool Value $X_t + Y_t S^0_t$ and hedging portfolio value $H_t$. (Upper Right) Corresponding evolution of the tracking error over time. (Bottom Left) Sample trajectory of Hedged PnL, to be compared with the Unhedged PnL (Bottom Right). }
        \label{fig:hedged.vs.unhedged}
\end{figure}

We evaluate LP performance through both unhedged and hedged profit-and-loss (PnL). Since we consider the CEX to be the primary venue of price formation, we mark holdings in $Y$ to the CEX price (rather than the DEX price). The \emph{unhedged (LP) PnL} is defined as the change in the marked-to-market value of the pool reserves plus the total of fees collected,
\[
\text{Unhedged PnL} = X_T + Y_T \cdot S_T^0 - (X_0 + Y_0 \cdot S_0^0) + \text{Fees Accrued}.
\]
When the price $S_t^0$ is volatile, the unhedged PnL tends to be very noisy and primarily driven by the exposure to the price fluctuations of the $Y$ holdings. 
Practitioners are more interested in the \emph{hedged PnL} which isolates the effects of trading and fees by removing the price risk. Specifically, hedging takes place on the CEX; it proceeds by initializing a position of $Y_0$ units of the risky asset and rebalancing each period to match the current $Y_t$ reserve level. We let $H=(H_t)_{t\geq0}$ denote the value of the hedging portfolio which is initialized at $H_0 = X_0 + Y_0 S^0_0$ and updated in a self-financing fashion according to
\[H_{t+\delta t} = H_t + Y_{t} \cdot (S^0_{t+\delta t} - S^0_t).\]
The cost of adjusting the hedge is computed using CEX prices and is frictionless in our baseline simulation.\footnote{For large LPs this simplifying assumption may be justified by internal netting. As an extension, our implementation also allows one to specify a separate, non-zero fee for hedge trades; this is discussed in \cref{sec:extensions}.} The hedged LP PnL is therefore
\[
\text{Hedged PnL} = \text{Unhedged PnL} - \sum_{i=1}^N Y_{(i-1)\delta t} \cdot \left(S_{i\delta t}^0 - S_{(i-1)\delta t}^0\right).
\]
We see that when $S^0$ is a martingale, the value of the hedging portfolio is given in terms of a martingale transform which is again a martingale. %
Throughout the paper, the \emph{expected} PnL will be an important metric. Because of the martingale nature of the hedge, the expected value of the hedged and unhedged PnL are \emph{identical}, hence either could be used. On a trajectory basis, the hedged PnL is significantly less noisy---as seen in the order of magnitude difference between the representative trajectories displayed in the lower panel of Figure \ref{fig:hedged.vs.unhedged}.

It is useful to group the pool value and hedge value into a single term, the \emph{tracking error},
\[
\text{Tracking Error} = H_t - X_t - Y_t \cdot S_t^0 ,
\]
so that we may write
\[
\text{Hedged PnL} =  \text{Fees Accrued} - \text{Tracking Error}.
\]
The tracking error is exactly the loss-versus-rebalancing (LVR) of \cite{jasonLVR} except that we mark asset values to the CEX price instead of the DEX. It can be thought of as reflecting LP losses that are attributable to the adverse selection on the DEX brought on by stale prices. A visualization of the tracking error is given in the upper right panel of Figure~\ref{fig:hedged.vs.unhedged}.

\subsubsection*{Simulation Steps}

We adopt a turn-based simulation in which each period of length $\delta t$ is subdivided into four further steps:

\begin{enumerate}
    \item[1.] The CEX price is incremented. %
    \item[2.] Arbitrageurs compare the new CEX price with the DEX price and, if applicable, execute a roundtrip trade of size~\eqref{eq:arbTradeSize}. This trade changes the marginal DEX price according to~\eqref{eq:newMarginalPrice}, removing the arbitrage opportunity. 
    \item[3.] An independent coin-flip decides whether fundamental buyers or sellers move first. In the former case, buyers trade quantity~\eqref{eq:buyTradeSize} on the DEX whose marginal price adjusts according to~\eqref{eq:newMarginalPrice}. Next, sellers trade quantity~\eqref{eq:sellTradeSize} on the DEX, again changing its price. In the latter case, the order is reversed.
    \item[4.] The LP updates their hedging portfolio on the CEX. 
\end{enumerate}

The simulation loops over those steps; Appendix~\ref{se:simulation-algorithm} provides pseudocode summarizing the implementation. Some models in the literature use noise traders whose actions can lead to arbitrage opportunities, hence those models include a second intervention of arbitrageurs exploiting those. In the present baseline model, the actions of fundamental traders in Step~3 never create arbitrage opportunities, hence it would be futile to include another arbitrageur trade.

\subsection{Scale Invariance}
\label{sec:scale.invariance}
The model has a useful scale invariance that we will leverage in our analysis. Namely, we will see that scaling demand and initial pool reserves by the \emph{same} factor leaves prices and LP returns unchanged.

We first introduce notation for the total value locked in the pool. At time $t$, define
\begin{equation}\label{eq:TVL}
    \mathrm{TVL}_t:=X_t+S_t^1Y_t.
\end{equation}
Since $S_t^1=X_t/Y_t$ for the CPMM, $\mathrm{TVL}_t=2X_t$. We choose the convention that the holdings in $Y$ are marked at the \emph{internal price}. The total value locked at time $0$ will form the reference with respect to which we compute returns. In the analysis that follows, we initialize the system with $S_0^1=S_0^0$. Thus, the internally and externally marked values of the pool coincide at time zero, and returns normalized by either convention are identical.

Fix initial reserves $(X_0,Y_0)$, a CEX price path $S^0$, and a path of desired fundamental orders $(\Delta_t^B,\Delta_t^S)_t$. Consider a second instance of the model in which the initial reserves and fundamental orders are multiplied by the same factor $L>0$:
\begin{equation}\label{eqn:scaling}
    (X_0,Y_0)
    \longmapsto
    (LX_0,LY_0),
    \qquad
    (\Delta_t^B,\Delta_t^S)
    \longmapsto
    (L\Delta_t^B,L\Delta_t^S).
\end{equation}
We also assume that the fee parameters and realization of the buyer-seller ordering are held fixed so that the common scale factor is the only change between the two model instances. We write $X_t(L)$ and $Y_t(L)$ for the reserves in the scaled model (and similarly for other pool quantities). The scaling behavior is summarized in the following proposition whose proof can be found in Appendix \ref{app:scale.invariance}.

\begin{proposition}
\label{prop:scale.invariance}
For every $L>0$, if the reserves and demand are scaled as in
\eqref{eqn:scaling}, then, at every time $t$,
\[
\begin{aligned}
(X_t(L),Y_t(L))
&=L(X_t(1),Y_t(1)),\\
\bigl(\Delta_t^{B,\mathrm{DEX}}(L),
      \Delta_t^{S,\mathrm{DEX}}(L),
      \Delta_t^A(L)\bigr)
&=
L\bigl(\Delta_t^{B,\mathrm{DEX}}(1),
       \Delta_t^{S,\mathrm{DEX}}(1),
       \Delta_t^A(1)\bigr).
\end{aligned}
\]
Moreover, the DEX price is scale invariant (i.e., $S^1(L)=S^1(1)$) and both hedged and unhedged LP returns do not depend on $L$:
\[
\frac{\mathrm{Unhedged \ PnL}(L)}{\mathrm{TVL}_0(L)}=\frac{\mathrm{Unhedged \ PnL}(1)}{\mathrm{TVL}_0(1)}, \qquad \frac{\mathrm{Hedged \ PnL}(L)}{\mathrm{TVL}_0(L)}=\frac{\mathrm{Hedged \ PnL}(1)}{\mathrm{TVL}_0(1)}.
\]
\end{proposition}

As a consequence of these properties, normalized model outcomes depend on fundamental demand only relative to pool liquidity. More precisely, they depend on the normalized demand path $\left(
        \Delta_t^B/Y_0,
        \Delta_t^S/Y_0
    \right)_t$.
In the baseline specification of the next section, the shape of demand over time is fixed and only its magnitude varies. We may therefore summarize relative demand by
\[
    q
    :=
    \frac{\Delta^B+|\Delta^S|}{Y_0},
\]
where $\Delta^B$ and $\Delta^S$ denote total buy and sell demand over the simulation horizon. Thus, with all other parameters held fixed, quantities such as LP returns depend on demand and liquidity only through the demand-to-liquidity ratio $q$.  This observation will be useful in \cref{sec:equilibrium}, where aggregate pool
liquidity is determined endogenously. In particular, scale invariance allows us
to formulate each analysis in terms of the economically natural variable,
focusing on demand when studying trading activity and on liquidity when
studying market depth or equilibrium, since the variable held fixed introduces
no additional independent scale effect.

\section{Determinants of LP Profitability}\label{sec:profitability}
 
In this section, we examine how LP profitability is influenced by fundamental demand, market volatility, and fee levels, while holding the initial liquidity fixed. To that end, we investigate the comparative statics of the simulation model described in the preceding section. %

Our baseline considers a one-day trading horizon with a one-minute time resolution, corresponding to $T = 1$ and $N = 1440$. The AMM is initialized with \$30{,}000{,}000 in numéraire reserves ($X$) and 10{,}000 units of the risky asset ($Y$), implying an initial price of $S_0^1 = \$3{,}000$ and initial TVL of
\$60{,}000{,}000. Prices on the CEX are set to the same initial value, $S_0^0 = \$3{,}000$, and evolve according to \eqref{eqn:price.evolution} with volatility $\sigma = 0.04$. The default CEX fee is $\eta^0 = 20$ basis points (bps). Fundamental buy and sell demand arrives at a rate of 5{,}000 units of $Y$ per day, uniformly distributed across time (i.e., $\Delta_t^B=-\Delta_t^S=5{,}000 \cdot \delta t$ per minute), which corresponds to a total unsigned daily demand of 10,000 units or 100\% of the pool's initial inventory. This configuration serves as a reference point for our analysis. A detailed discussion of the plausibility of these benchmark values, including an empirical calibration to real market data, is provided in \cref{sec:empirics}. 
Unless otherwise stated, reported PnL is normalized by
initial TVL and expressed in basis points. When fundamental demand is varied
in this section, initial pool liquidity is held fixed at its baseline level.

\begin{figure}[htb]
  \centering
  \includegraphics[width=0.75\textwidth]{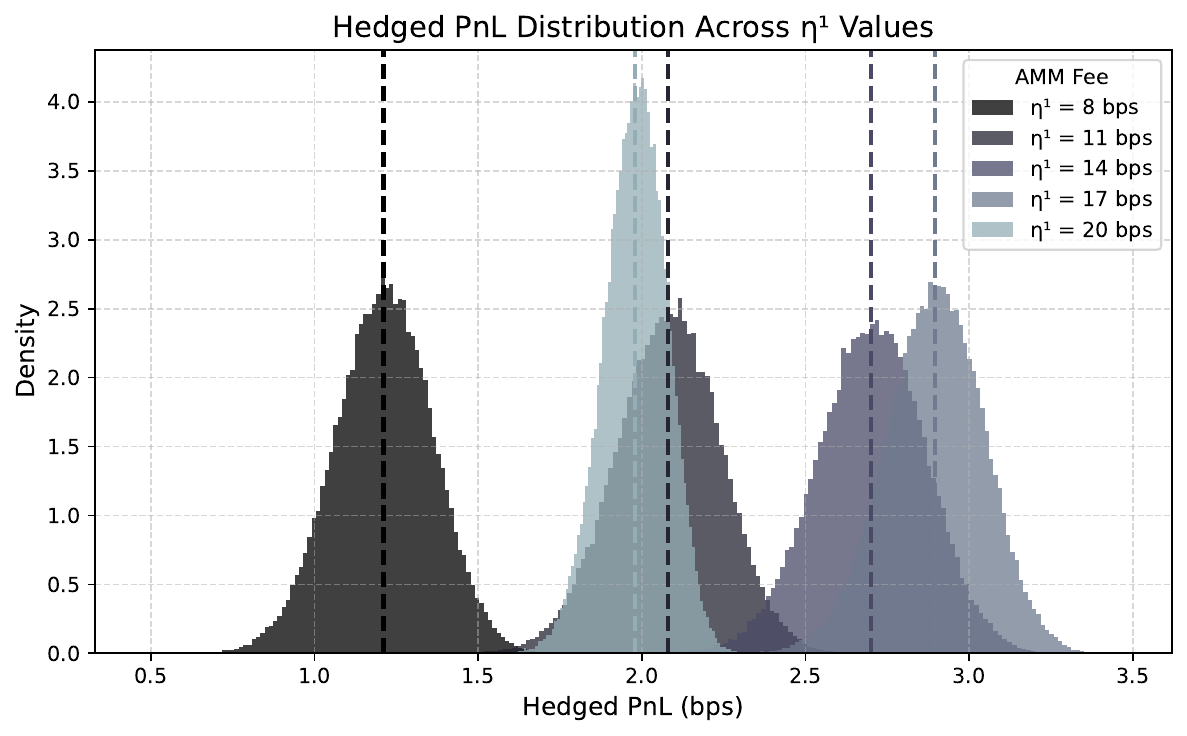}
  \caption{Histograms of terminal hedged PnL (bps) across 100{,}000 simulations, colored by AMM fee level. Dashed lines mark the distribution means.}
  \label{fig:pnl_hist_eta1}
\end{figure}

We begin by motivating our use of \emph{expected} PnL as a performance metric for LPs. \Cref{fig:pnl_hist_eta1} shows the distribution of terminal hedged PnL across different AMM fee levels. While the PnL variance changes only modestly with the fee, the expected PnL is much more sensitive. Moreover, some fee levels consistently outperform others, and PnL is \emph{not} monotonic in the fee level, highlighting its central role in determining LP returns.

\begin{figure}[htb]
  \centering
  \begin{minipage}{0.49\textwidth}
    \centering
    \includegraphics[width=\linewidth]{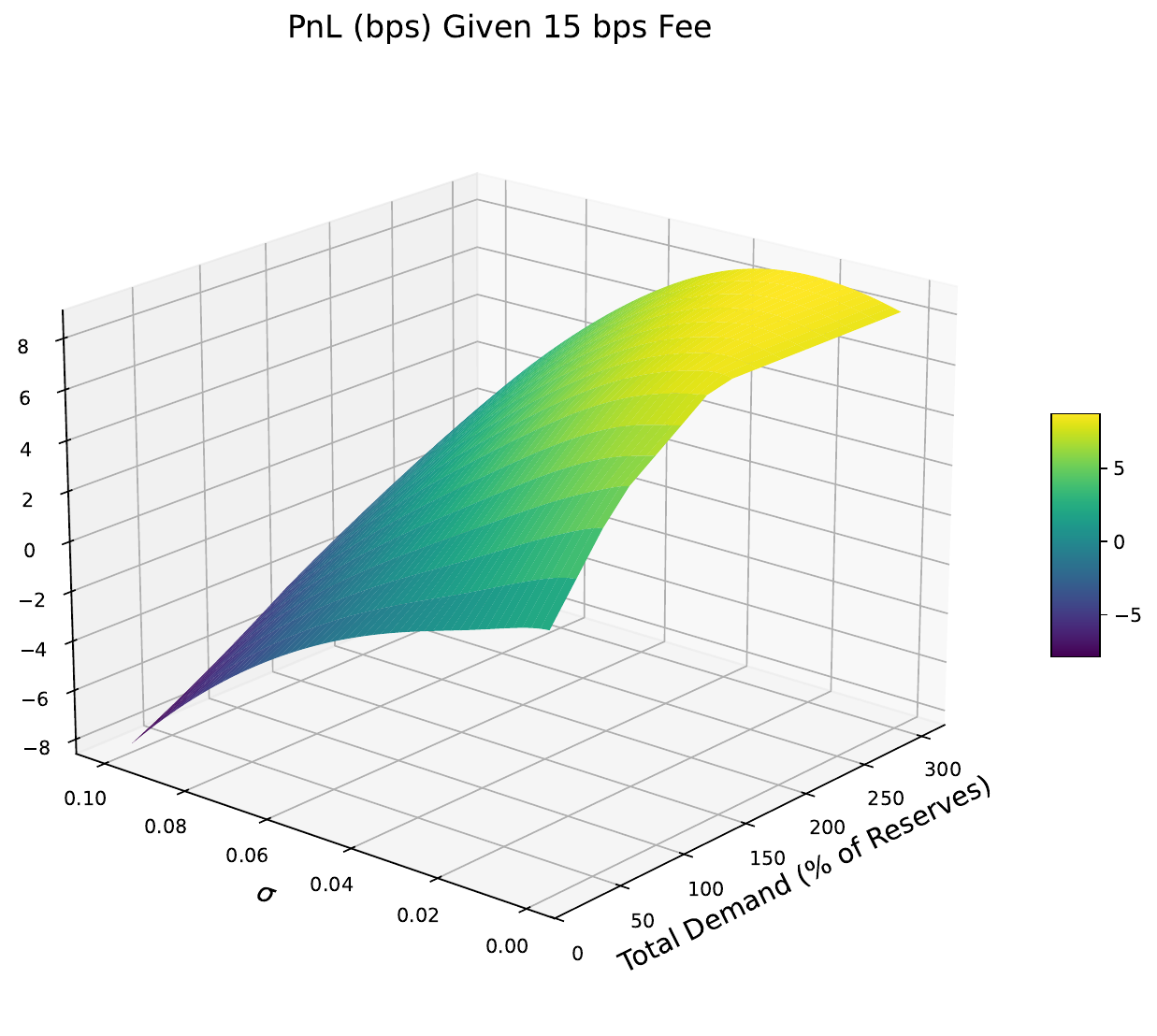}%
  \end{minipage}%
  \hfill
  \begin{minipage}{0.49\textwidth}
    \centering
    \includegraphics[width=\linewidth]{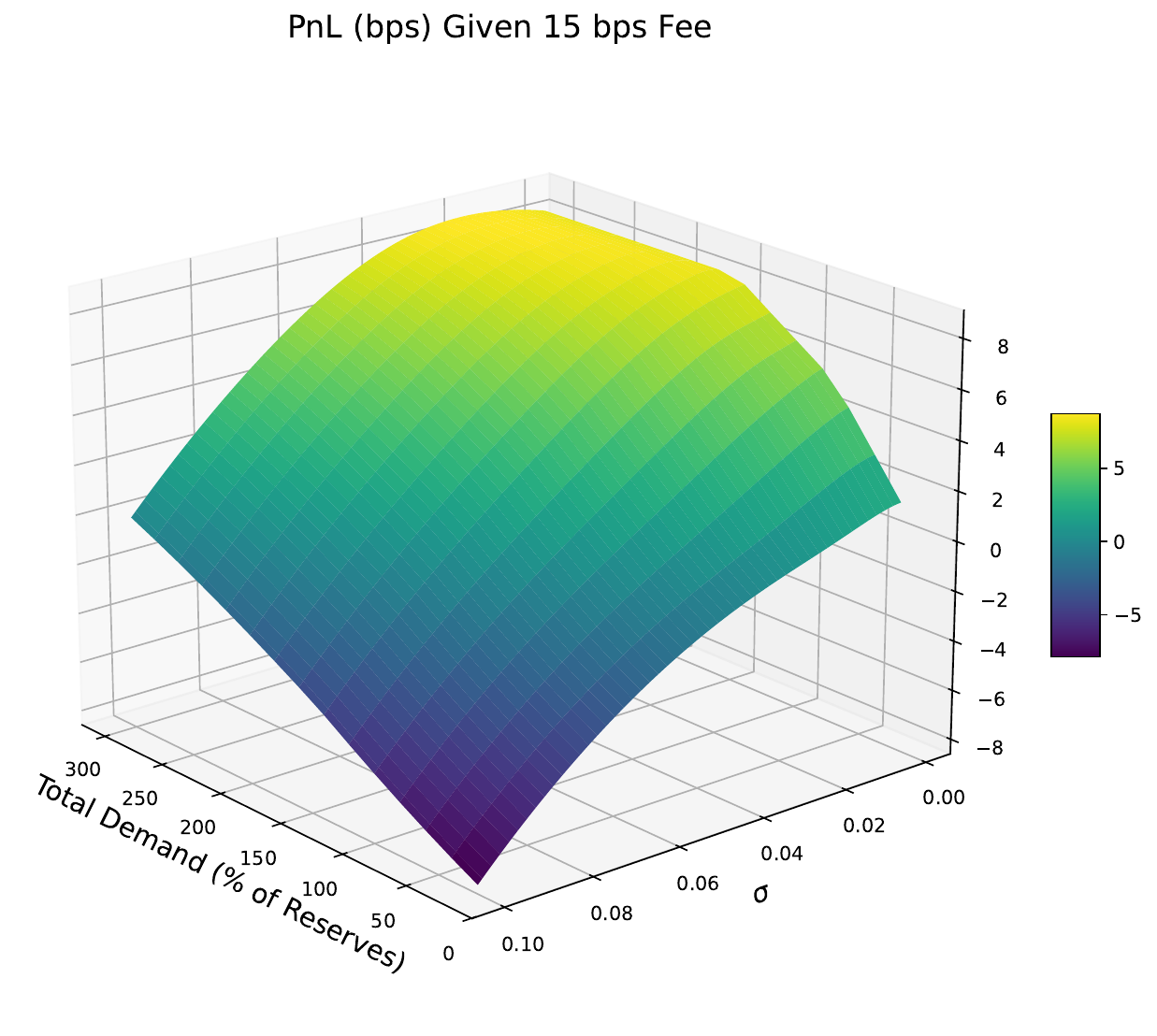}
  \end{minipage}
  \caption{Expected PnL (bps) as a function of fundamental demand and volatility under the baseline AMM fee of 15 bps. The left and right figures show two different orientations of the same surface.}
  \label{fig:pnl_vs_demand_vol}
\end{figure}

Before delving into the details, we observe two basic features of LP profitability. \Cref{fig:pnl_vs_demand_vol} plots average PnL across varying levels of fundamental demand and volatility, holding the AMM fee fixed at $\eta^1 = 15$ bps. 

Two broad patterns emerge: expected PnL declines with volatility and increases with the size of fundamental demand. The detailed analysis in this section will show how greater price dispersion increases the likelihood of stale execution and adverse selection, lowering PnL. On the other hand, when more flow interacts with the AMM, the LP earns more fee revenue. More importantly, we will see that this additional fundamental flow tends to trade at prices that are less stale, an effect not seen in previous models with only (uninformed) noise traders.

\subsection{The Price Ratio}\label{subsec:priceRatio}

The ratio between AMM and CEX price is a key quantity to understand why trades generate gains or losses for the LP. Following the concept of toxic vs.\ benign flow in classical finance, it is tempting to think of arbitrageur trades as inherently ``bad'' for the LP and fundamental trades as ``good.'' However, this dichotomy is misleading. The next part of our analysis builds on a key insight: whether a trade benefits or harms the LP does not depend on the identity or intent of the trader, but rather on the \emph{price ratio} at which the trade occurs.

\begin{table}[bht]
\centering
\caption{Trading and Profitability Thresholds}
\label{tab:thresholds}
\begin{tabular}{@{}lcc@{}}
\toprule
 & \textbf{Trade Condition} & \textbf{Profit Condition} \\ 
\midrule
\textbf{Fundamental Buyers} & 
$R_t \le \dfrac{1 + \eta^0}{1 + \eta^1}$ & 
$R_t \ge \dfrac{1}{1 + \eta^1}$ \\ 

\textbf{Fundamental Sellers} & 
$R_t \ge \dfrac{1 - \eta^0}{1 - \eta^1}$ & 
$R_t \le \dfrac{1}{1 - \eta^1}$ \\ 

\textbf{Arbitrage Buyers} & $R_t\leq \dfrac{1-\eta^0}{1+\eta^1}$ & N/A\\

\textbf{Arbitrage Sellers} & $R_t\geq \dfrac{1+\eta^0}{1-\eta^1}$ &  N/A\\
\bottomrule
\end{tabular}
\end{table}

\begin{figure}[!htb]
    \centering
    \begin{minipage}{.95\textwidth}
        \centering
        \includegraphics[width=.75\linewidth]{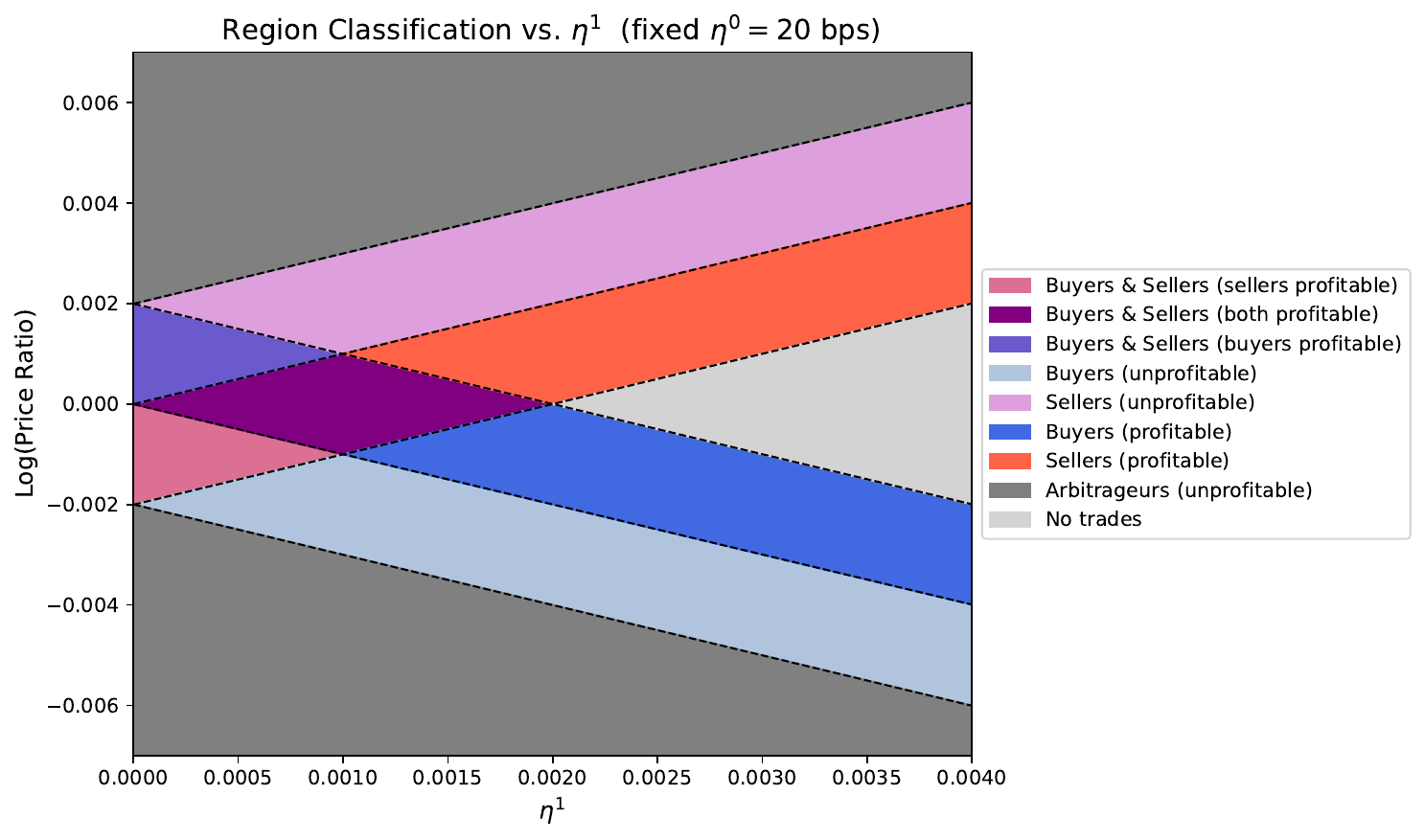}
    \end{minipage}
    \caption{ Illustration of how the critical regions from Table~\ref{tab:thresholds} change with the AMM fee $\eta^1$, shown in terms of the log price ratio.}
    \label{fig:region_classification}
\end{figure}

To build intuition, we characterize the conditions under which a marginal (infinitesimally small) trade on the DEX yields a profit for the LP, accounting for hedging activity on the CEX. Suppose a trader executes a small buy order of size $ \Delta > 0 $ units of the risky asset $Y$. The LP gives up $\Delta$ units of $Y$, receives approximately $ \Delta \cdot S_t^1 $ units of the numéraire asset $X$ at the current AMM price, and collects an additional fee of $ \eta^1 \cdot \Delta \cdot S_t^1 $. To maintain a neutral inventory, the LP rebalances by purchasing $\Delta$ units of $Y$ on the CEX at price $S_t^0$, incurring a cost of $ \Delta \cdot S_t^0 $. The marginal change in the LP's hedged PnL (expressed in units of $X$) is therefore
\[
\text{Change in Hedged PnL} \approx \Delta \cdot \left[ (1 + \eta^1) S_t^1 - S_t^0 \right] 
= \Delta \cdot S_t^0 \left[ (1 + \eta^1) R_t - 1 \right].
\]
Similarly, for a small sell trade $ \Delta < 0 $, the LP receives $|\Delta|$ units of $Y$, pays $ |\Delta| \cdot S_t^1 $ in $X$, and hedges by selling $|\Delta|$ units of $Y$ on the CEX at price $S_t^0$, yielding a revenue of $ |\Delta| \cdot S_t^0 $. The resulting impact on the hedged PnL is
\[
\text{Change in Hedged PnL} \approx \Delta \cdot \left[ (1 - \eta^1) S_t^1 - S_t^0 \right] 
= \Delta \cdot S_t^0 \left[ (1 - \eta^1) R_t - 1 \right].
\]
Evidently, for a buy trade ($\Delta > 0$), the LP earns a profit on a marginal trade if and only if $R_t > 1/(1 + \eta^1)$. For a sell trade ($\Delta < 0$), the LP earns a profit on a marginal trade if and only if $R_t < 1/(1 - \eta^1)$. We combine these thresholds with those of the preceding section in Table \ref{tab:thresholds}, and illustrate their sensitivity to the choice of the AMM fee $\eta^1$ in Figure \ref{fig:region_classification}.

We see that the LP benefits from trades that occur when the AMM price does not deviate too far from the CEX reference. In particular, \emph{both} buy and sell trades are profitable for the LP when the price ratio lies within the interval
\[
\left[ \frac{1}{1 + \eta^1}, \, \frac{1}{1 - \eta^1} \right].
\]
We refer to this as the \emph{profit region}: any marginal trade occurring within this band yields a net gain for the hedged LP, regardless of trade direction. Outside this region, one side of the market results in a loss for the LP, even if the flow originates from fundamental traders.

A related concept is the \emph{buy-sell region}, defined (when $ \eta^1 \leq \eta^0 $) by
\[
\left[ \frac{1 - \eta^0}{1 - \eta^1}, \, \frac{1 + \eta^0}{1 +\eta^1} \right].
\]
This is the range of price ratios in which both fundamental buyers and sellers prefer trading on the AMM over the CEX, allowing the LP to capture two-sided flow. When $\eta^1 > \eta^0$, this region becomes empty and the DEX is never simultaneously optimal for both sides of the market.

Crucially, the profit and buy-sell regions respond differently to the AMM fee $ \eta^1 $, and there is no consistent inclusion relationship between them. As $\eta^1$ increases, the profit region expands but the buy-sell region shrinks. Figure~\ref{fig:region_classification} illustrates this interaction. When $\eta^1$ is small, it is possible for the buy and sell thresholds to lie outside the profit region, meaning the LP captures two-sided flow even when some trades are marginally unprofitable. By contrast, as $\eta^1 \uparrow \eta^0$, the profit region will fully contain the buy-sell boundaries.

When the price ratio leaves the broader \emph{no-arbitrage region}
\[
\left[ \frac{1 - \eta^0}{1 + \eta^1}, \, \frac{1 + \eta^0}{1 - \eta^1} \right],
\]
arbitrageurs intervene to realign prices across venues. The no-arbitrage region always contains the profit region, since $ \eta^0 \geq 0 $, and arbitrage trades---by construction---are \emph{always unprofitable} for the LP. Figure~\ref{fig:region_classification} demonstrates that as $\eta^1$ increases, the arbitrage thresholds are pushed outwards. While this reduces the frequency of loss-making arbitrage trades, from the LP's perspective this should be balanced by the effects on the other regions.

Our model can be compared to that of Glosten and Milgrom~\cite{glosten1985bid}. There, market makers defensively widen spreads because they cannot distinguish benign flow from informed flow that predicts future price movements. Here, fees play an analogous role, as the AMM fee cannot depend on the current CEX price or the trader's motive.

\subsection{Distribution of the Price Ratio}\label{sec:price.ratio.dist}

The preceding classification of regions provides a static view of trade incentives and profitability but offers no insight into how often the AMM/CEX price ratio falls within each region. Next, we study the distribution of the price ratio, which determines the occupation times of the profit, buy-sell, and arbitrage bands, and hence offers first insights into the LP PnL.

\subsubsection{Sensitivity to Fees}
\begin{figure}[!htb]
    \centering
    \begin{minipage}{.99\textwidth}
        \centering
        \includegraphics[width=.99\linewidth]{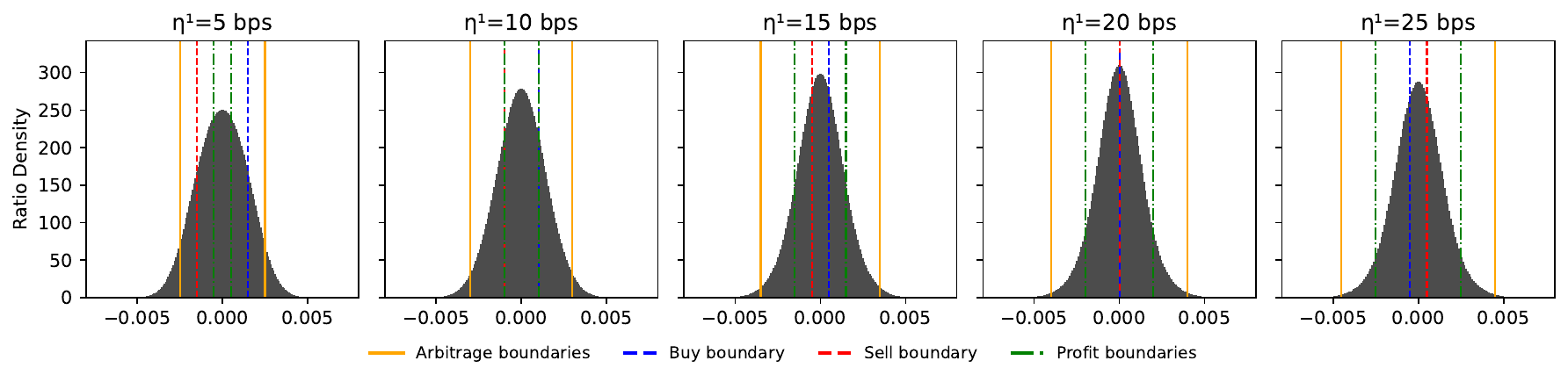}
        \includegraphics[width=.99\linewidth]{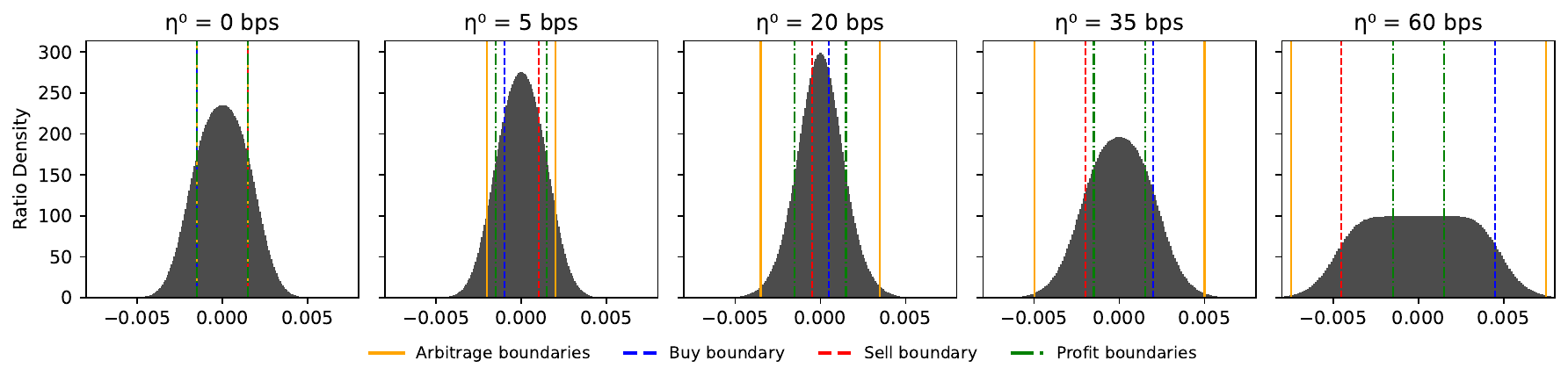}
        \includegraphics[width=.99\linewidth]{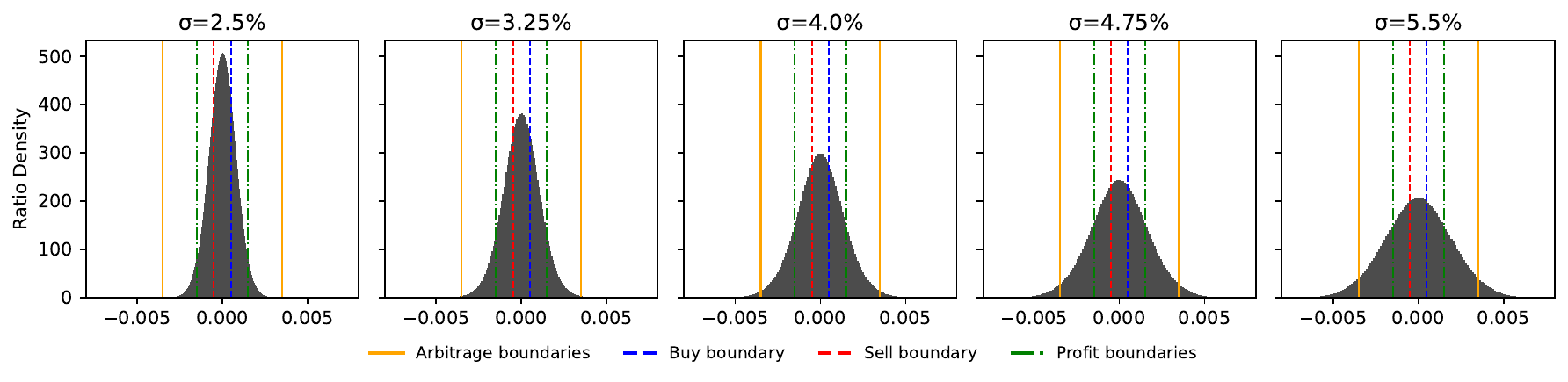}
        \includegraphics[width=.99\linewidth]{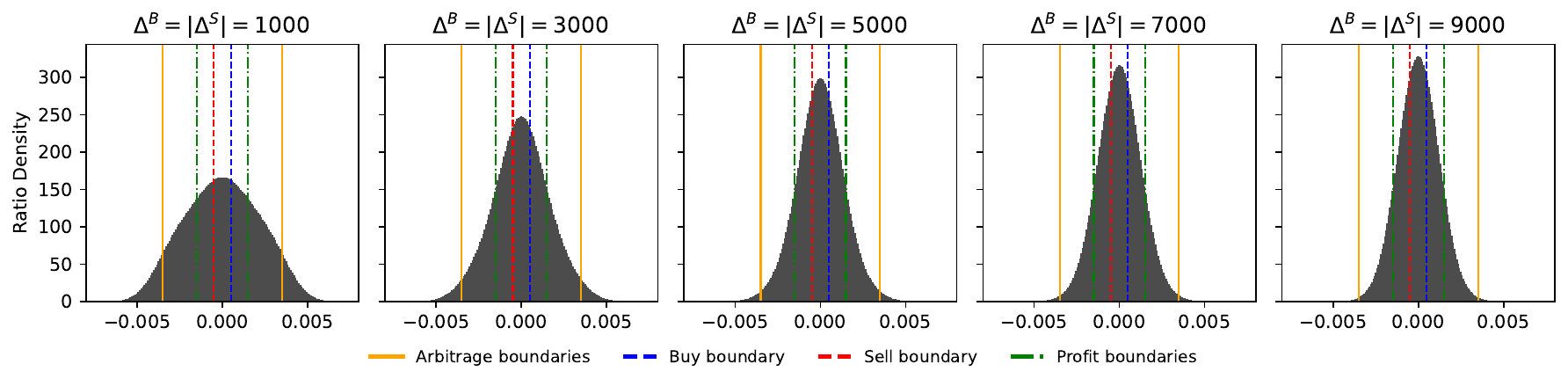}
    \end{minipage}
    \caption{Distributions of the log price ratio $\log(R_t) = \log(S^1_t / S^0_t)$ sampled immediately before trade execution.  The center panel of each row corresponds to the common benchmark setting: $\eta^1 = 15$ bps, $\eta^0 = 20$ bps, $\sigma = 4\%$, and $\Delta^B = |\Delta^S| = 5{,}000$. Each row illustrates how the shape of this distribution varies with (top) the AMM fee $\eta^1$, (second) the CEX fee $\eta^0$, (third) market volatility $\sigma$, and (bottom) fundamental demand. Distributions are aggregated over 100{,}000 simulation paths. All other parameters follow the baseline specification in Section~\ref{sec:profitability}.  Statistics for each configuration are reported in Appendix \ref{se:summary.stats}.}
    \label{fig:histograms}
\end{figure}

The first row of \cref{fig:histograms} shows histograms of the log price ratio, $\log(R_t) = \log(S^1_t / S^0_t)$, sampled immediately \emph{before} trade execution (i.e., between Steps~1 and~2 as listed at the end of \cref{sec:ecosystem}), across various AMM fee levels ranging from 5 to 25 basis points. The second row of \cref{fig:histograms} illustrates the closely related influence of the CEX fee on the price ratio distribution. These distributions are computed using data from 100{,}000 simulation runs, aggregated over all time steps. When one parameter is varied, the others are set to the baseline values introduced at the beginning of this section. Simulation statistics like expected profit and volume captured are reported in \cref{se:summary.stats}.

At low AMM fee levels relative to the CEX (e.g., $\eta^1 = 5$ bps and $\eta^0=20$ bps, or $\eta^1=15$ bps and $\eta^0=60$ bps), the AMM offers competitive execution and captures a large share of trading volume. The buy-sell region is wide, and the resulting bidirectional fundamental flow flattens the central portion of the distribution. Indeed, when buy and sell trades compensate one another, the log ratio process locally resembles a Brownian motion without drift. The flat part of the distribution (see the second row and last column of Figure \ref{fig:histograms} for a pronounced example) can be compared to the stationary distribution of Brownian motion reflected in a strip, which is a uniform distribution. Overall, this regime produces a broad and diffuse shape. While a large fraction of the transactions originates from fundamental traders, many of the trades occur at price ratios outside the LP’s profit region, leading to poor PnL despite high turnover.

As the AMM fee increases, the widening of the profit region and the decline in the AMM’s competitiveness restrict participation to trades that remain viable despite the higher cost. This leads to a contraction of the price ratio distribution around its center due to the price impact of asymmetric fundamental flow: when only one side of the market interacts with the AMM, trade flow pushes prices directionally. At high (relative) fees (e.g., $\eta^1 \geq \eta^0 = 20 $ bps), only unilateral flow is captured and this effect becomes most pronounced. In fact, when $\eta^1>\eta^0$ the order of the buy and sell boundaries switches so that they enclose a region where the AMM receives \emph{no} order flow. At the same time, a widening of the no-arbitrage region leads to slightly longer tails of the distribution at large values of $\eta^1$. The overall effect of increasing $\eta^1$ is thus a transition from high-turnover, low-margin activity with relatively diffuse pricing to high-margin, selective trading centered near the CEX price.

While both the AMM and CEX fees expand the no-arbitrage region by pushing the arbitrage thresholds outward, an important distinction arises in their impact on the buy-sell region. Specifically, increasing the AMM fee shrinks this region, while increasing the CEX fee expands it by making AMM execution comparatively more favorable for fundamental traders. The net effect is a broader, flatter ratio distribution as $\eta^0$ increases. Notably, although the CEX fee modifies the attractiveness of AMM execution, it leaves the boundaries of the LP’s profit region unaffected.

\subsubsection{Sensitivity to Volatility}

The third row of \Cref{fig:histograms} depicts how the log price ratio distribution changes with market volatility $\sigma$ when the AMM fee $\eta^1$ is fixed at $15$ bps. At low volatility (e.g., $\sigma = 2.5\%$), the distribution is sharply concentrated near zero. In this regime, the reference price evolves gradually, and fundamental order flow acts as a stabilizing force: when the price ratio escapes the buy-sell region, the impact from one-sided trading is often enough to push the ratio back in and keep prices closely aligned with the CEX, without any action from arbitrageurs. As a result, the AMM captures a large share of fundamental flow, with much of it arriving in the profit region for this fee level (see \cref{se:summary.stats} for the numerical values).

As volatility increases, the distribution of $\log(R_t)$ becomes increasingly dispersed. At high values of $\sigma$, fundamental demand is unable to adjust the AMM price by enough to track the large swings in the reference price. This leads to more frequent and severe price misalignments. A large number of trades occur outside of the profit region and at the highest value illustrated, $\sigma=5.5\%$, a nontrivial fraction of the trades occur outside of the arbitrage boundaries, implying that they are arbitrageur trades. This widening of the price ratio distribution at high volatility levels contributes to the deterioration in LP profitability observed above.

\subsubsection{Sensitivity to Demand}
The last row of Figure \ref{fig:histograms} illustrates how the distribution of the log price ratio evolves with increasing levels of fundamental demand when $\eta^1=15$ bps. At low levels of demand, the ratio distribution has a broad and bulbous shape, and a large fraction of the trading volume arrives while the price ratio is outside of the profit region.  As demand increases, we observe a notable increase in the concentration of the distribution about zero.

This behavior can be attributed to the stronger directional pressure exerted by fundamental order flow. When demand is weak, price adjustments are sluggish and the AMM quotes often lag behind the CEX value, which exposes the pool to adverse selection. As demand increases, the consistent push from one-sided flow outside of the buy-sell region creates a stabilizing drift that keeps AMM prices more tightly aligned with the CEX. The result is twofold: not only does higher demand increase traded volume and hence the fee revenue, but it also enhances the quality of execution from the LP's perspective. (Of course, there is a limit to this effect: even if fundamental demand were infinite, the finite liquidity in the pool ensures that price impact would grow until traders are incentivized to transact on the CEX.)

\subsection{Trading Volume and Market Demand}

\begin{figure}[htb]
    \centering
    \begin{minipage}{.48\textwidth}
        \includegraphics[width=\linewidth]{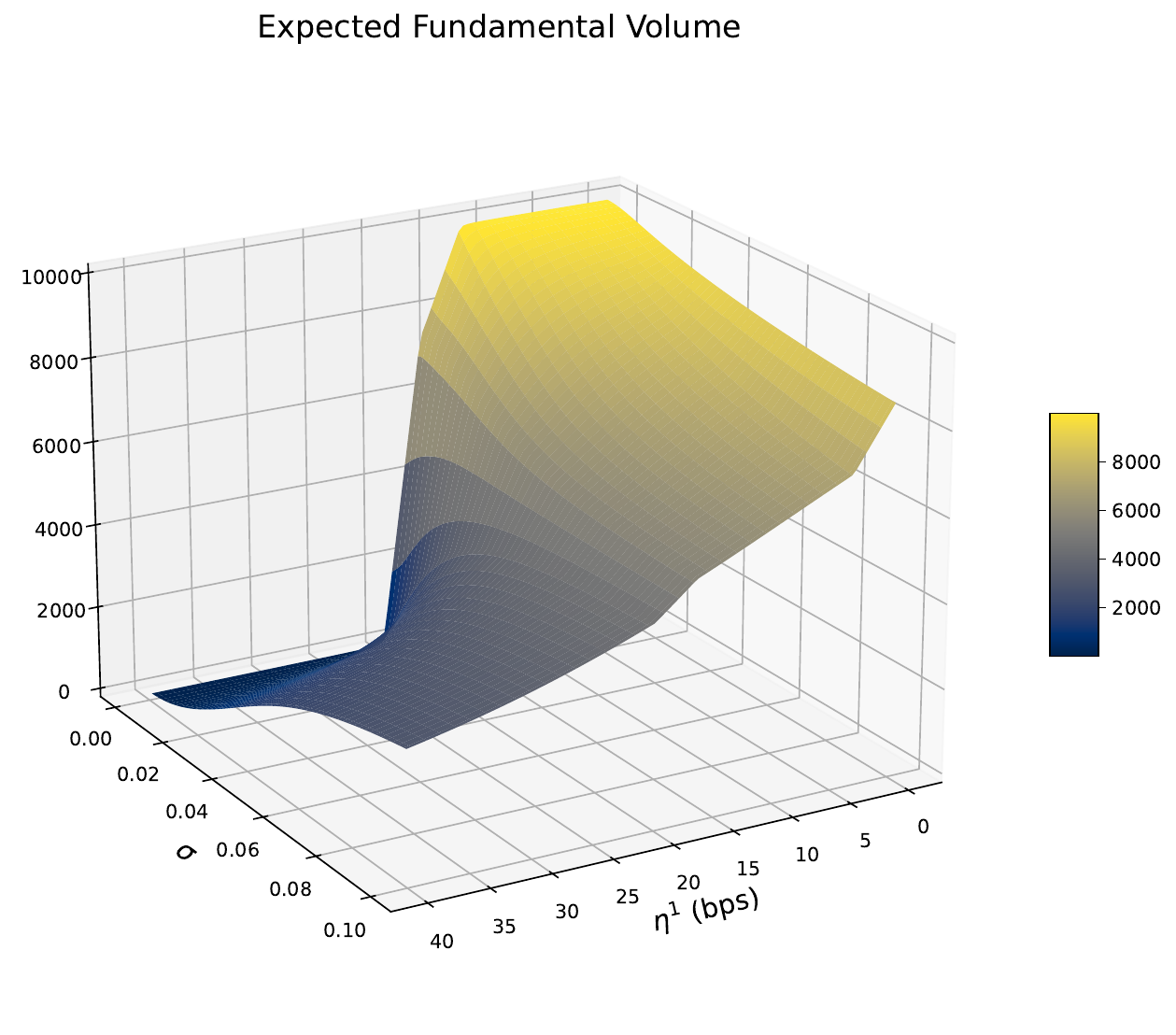}
    \end{minipage}\hfill
    \begin{minipage}{.48\textwidth}
        \includegraphics[width=\linewidth]{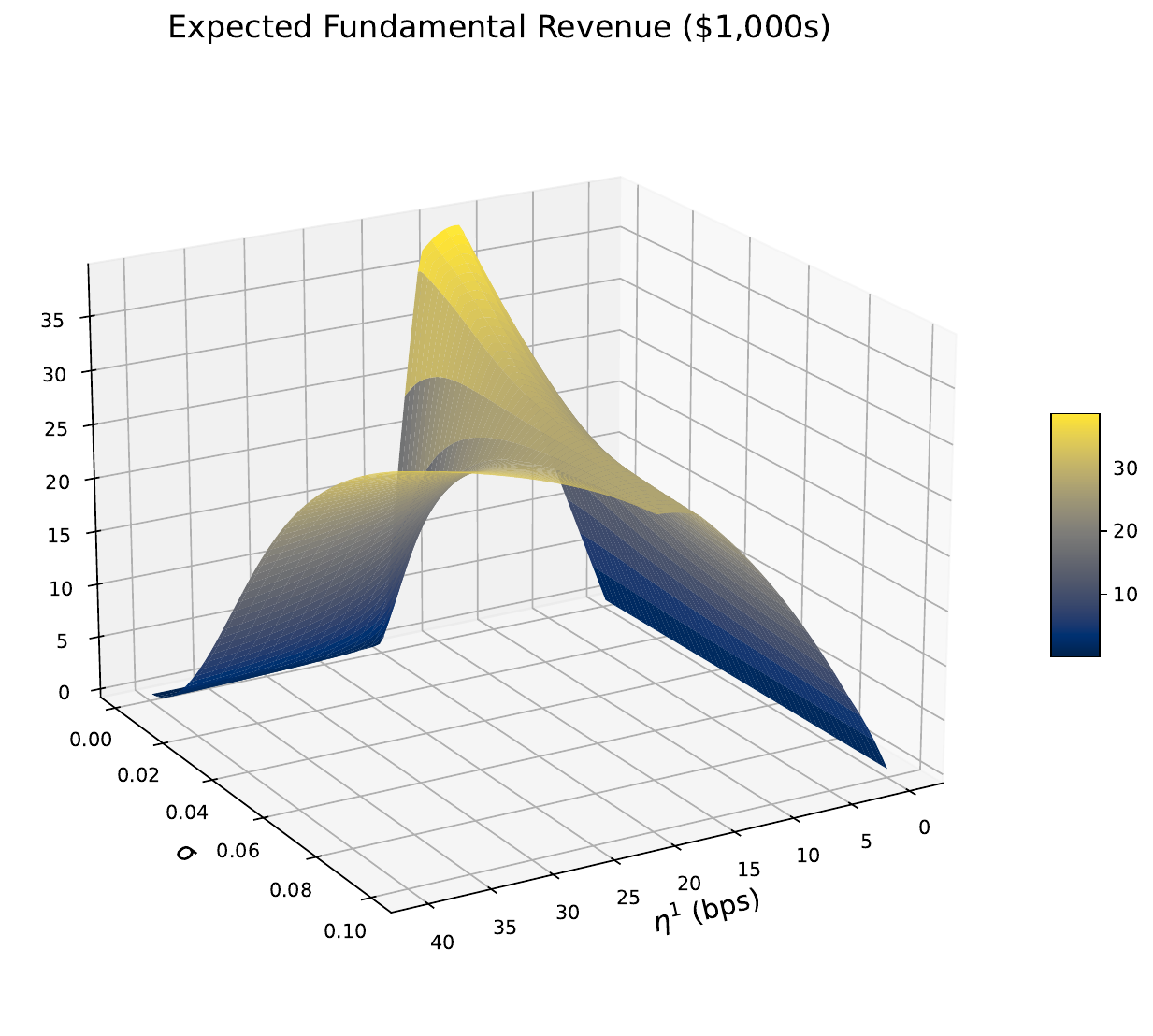}
    \end{minipage}
    \caption{(Left) Expected fundamental trader volume received by the AMM as a function of $\eta^1$ and $\sigma$. (Right) Corresponding fee revenue. Lower volatility favors high capture rates, while revenue reflects the interaction of volume and fee level.}
    \label{fig:expected_volume_revenue}
\end{figure}

\begin{figure}[htb]
    \centering
    \begin{minipage}{.48\textwidth}
        \includegraphics[width=\linewidth]{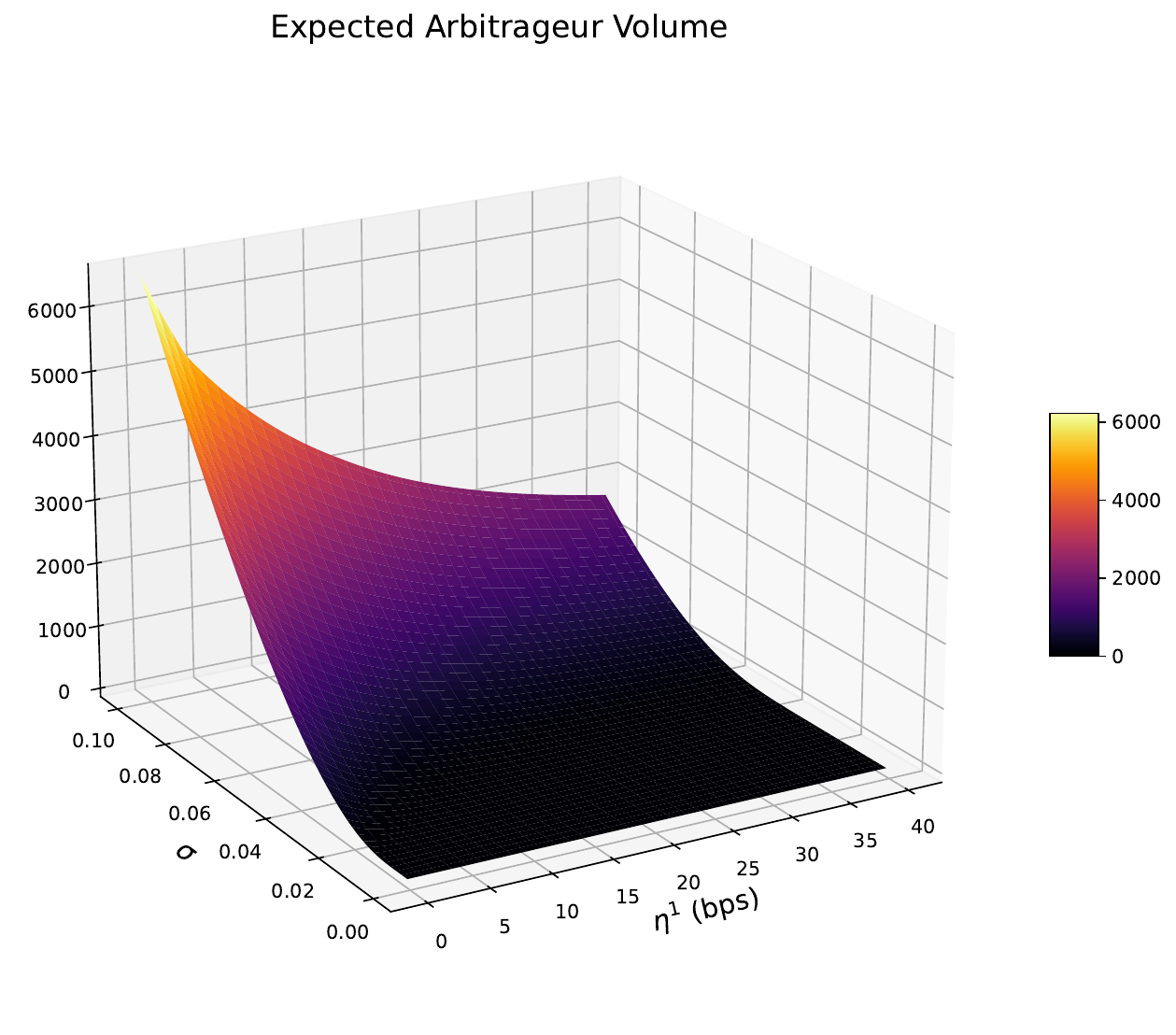}
    \end{minipage}\hfill
    \begin{minipage}{.48\textwidth}
        \includegraphics[width=\linewidth]{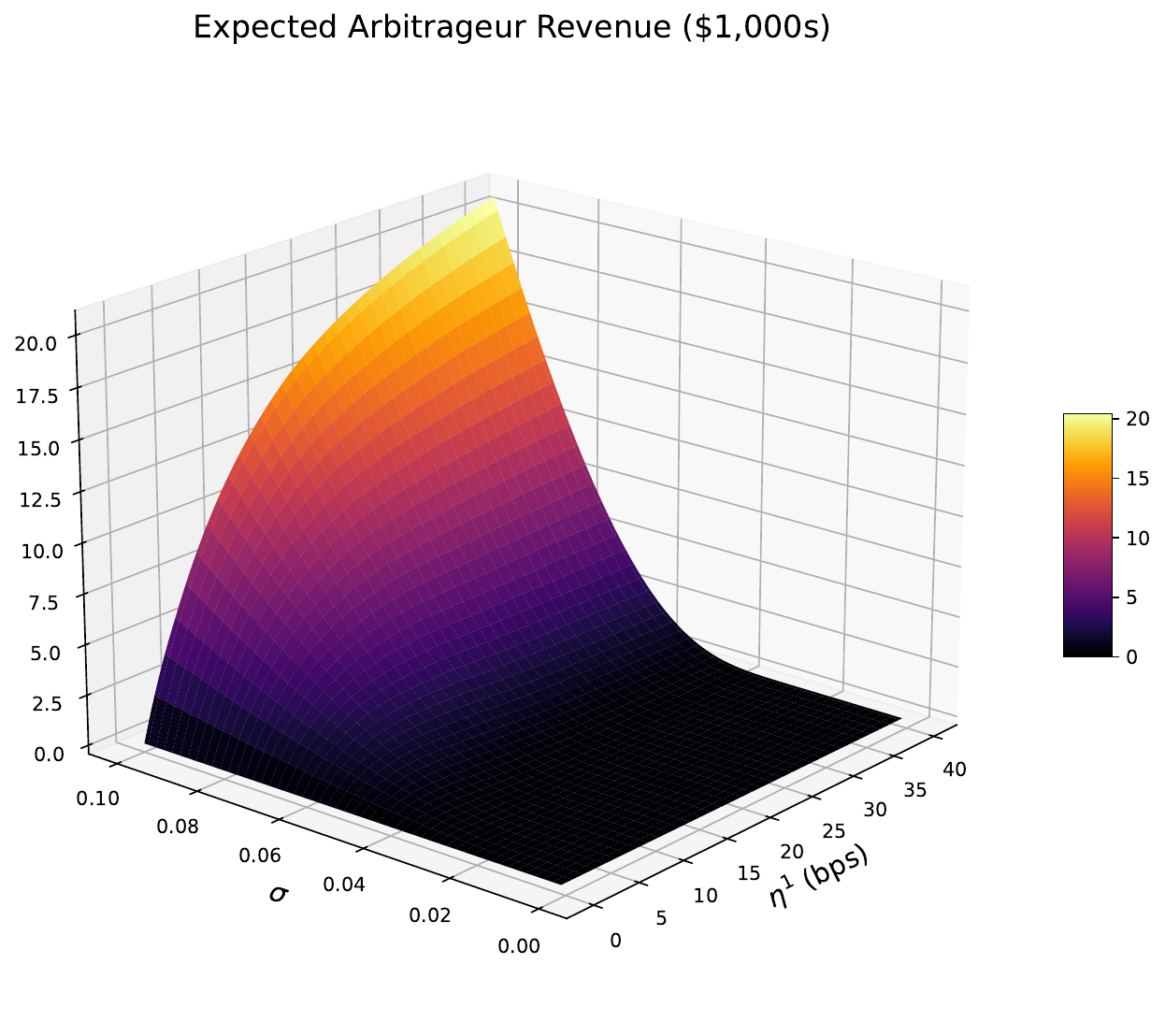}
    \end{minipage}
    \caption{(Left) Average arbitrageur volume received by the AMM across combinations of AMM fee $\eta^1$ and volatility $\sigma$. Higher volatility increases arbitrage activity, while higher fees reduce it. (Right) Associated fee revenue earned from these trades.}
    \label{fig:arb_volume_revenue}
\end{figure}

Understanding where the price ratio spends time is only part of the picture---LP profitability hinges on how much volume is transacted in each region and the fees earned per trade. Fundamental order flow is directed to the AMM when it offers better execution than the CEX for buy or sell orders. The competitiveness of the AMM's pricing, and hence the volume it captures, is shaped by the AMM fee~$\eta^1$ and market volatility $\sigma$. Because fundamental traders are the \emph{only} source of profitable volume (see \cref{subsec:priceRatio}), it is critical to understand this relationship.

\Cref{fig:expected_volume_revenue} shows the expected volume of fundamental trades executed by the AMM and the associated fee revenue over the simulation horizon. Since prices remain relatively stable over the course of a day, fee revenue is roughly proportional to the product of captured volume and the fee level $\eta^1$. The most striking insight of \cref{fig:expected_volume_revenue} is that, except at very high volatility, volume and revenue strongly depend on the relationship between AMM and CEX fee. At low volatility, even modestly undercutting the CEX fee enables the AMM to capture almost all of the fundamental flow; a more detailed discussion of this undercutting will follow in \cref{sec:optimal.fees} in the context of optimizing the AMM fee. While this effect is reduced as volatility increases, it remains important in a range of volatilities. As volatility increases, AMM prices become stale more quickly, reducing their competitiveness. Maintaining volume under these conditions requires lowering fees, but this comes at the cost of reduced revenue per trade. The result is a fundamental economic trade-off: greater competitiveness erodes margins, and this tension intensifies in more volatile markets.

On the other hand, the relationship between arbitrage activity, AMM fee, and volatility is relatively straightforward. Higher volatility increases price discrepancies between the AMM and the CEX and hence the arbitrageur volume, as shown in the left panel of \cref{fig:arb_volume_revenue}. By contrast, higher AMM fees reduce arbitrage incentives and thus lower arbitrage flow. The right panel reports the associated fee revenue from these trades. Although this revenue can be considerable in high-volatility and high-fee settings, it is not sufficient to offset the losses from adverse selection (as shown analytically in \cref{subsec:priceRatio}).

\section{Competitive Liquidity Provision}
\label{sec:equilibrium}

The analysis so far conditions on the initial reserves supplied to the AMM. This is a natural simplifying assumption over the short trading horizon of our model. 
Over a longer horizon, however, the amount of capital supplied to the pool should respond to the return earned by LPs. For instance, a change in the AMM fee affects LP profitability, which creates incentives for entry or exit. The resulting change in pool depth alters price impact and hence the amount of fundamental order flow routed to the AMM. 

We model competitive liquidity provision through an entry decision that precedes the trading period: given AMM fee and market parameters, liquidity providers assess their expected hedged PnL and choose whether to supply capital to the pool. Once aggregate liquidity has been determined, the trading stage proceeds exactly as in \cref{sec:ecosystem}, and no provision or withdrawal decisions occur on $[0,T]$. In this way, the equilibrium considered below determines the initial pool depth for the market dynamics studied in the preceding sections.

\subsection{Competitive Entry}
\label{sec:competitive.entry}
Fix reference reserves $(X_0,Y_0)$ and let $L>0$ denote the scale
of liquidity supplied to the pool, holding fundamental demand fixed.  That is, we seed the pool with the initial reserve values $(X_0(L), Y_0(L))=(LX_0,LY_0)$. Let $\textsf{Hedged PnL}_T(L;\eta^1)$ denote the terminal hedged LP PnL when the liquidity scale is $L$ and the AMM fee is $\eta^1$. We define the corresponding expected LP return by
\[
\mathcal R(L;\eta^1)
:=
\frac{\mathbb E\left[\textsf{Hedged PnL}_T(L;\eta^1)\right]}
{\mathrm{TVL}_0(L)}.
\]
The remaining market parameters, including fundamental demand, volatility, and the CEX trading cost, are suppressed from the notation.

Liquidity providers are assumed to have access to an outside investment yielding an expected return $\bar r$ over the same horizon. Under perfect competition, capital enters the pool when the expected LP return exceeds $\bar r$ and exits when it falls below $\bar r$. This motivates the following definition.

\begin{definition}[Competitive Equilibrium]
\label{def.equilibrium}
For a given AMM fee $\eta^1$, a positive liquidity scale $L^\star(\eta^1)>0$ is a competitive equilibrium if $\mathcal R(L^\star(\eta^1);\eta^1)=\bar r$.
\end{definition}

This construction is a partial equilibrium. It endogenizes aggregate LP capital while retaining total fundamental demand, the CEX price process, and CEX trading costs as market primitives. The quantity of fundamental flow executed by the AMM remains endogenous. For any given aggregate liquidity, traders route their orders according to the execution prices on the two venues. Consequently, a change in $\eta^1$ affects AMM volume directly through the fee charged to traders and indirectly through the equilibrium response of pool depth. For illustration purposes we fix $\bar r=0$ as a reasonable proxy for the short run benchmark return on the time scale of our simulation (1 day).

Note that an equilibrium need not exist. If $\mathcal{R}(L;\eta^1)<\bar r$ for every $L>0$, competitive LPs do not supply capital to the pool. Conversely, if $\mathcal{R}(L;\eta^1)>\bar r$ for every $L>0$, the model does not determine a finite equilibrium pool size. Likewise, multiple equilibria may also exist in general. We say that a positive equilibrium $L^\star$ is locally stable if there exists $\varepsilon>0$ such that $\mathcal R(L;\eta^1)>\bar r$ for $L\in(L^\star-\varepsilon,L^\star)$ and $\mathcal R(L;\eta^1)<\bar r$ for $L\in(L^\star,L^\star+\varepsilon)$.
Thus a small reduction in liquidity induces entry, while a small increase induces exit. We distinguish these cases in the numerical analysis that follows. For the vast majority of parameter specifications, we find a unique stable equilibrium over the search space for $L$.

\begin{figure}[htb]
    \centering
    \begin{minipage}{.54\textwidth}
        \includegraphics[width=\linewidth]{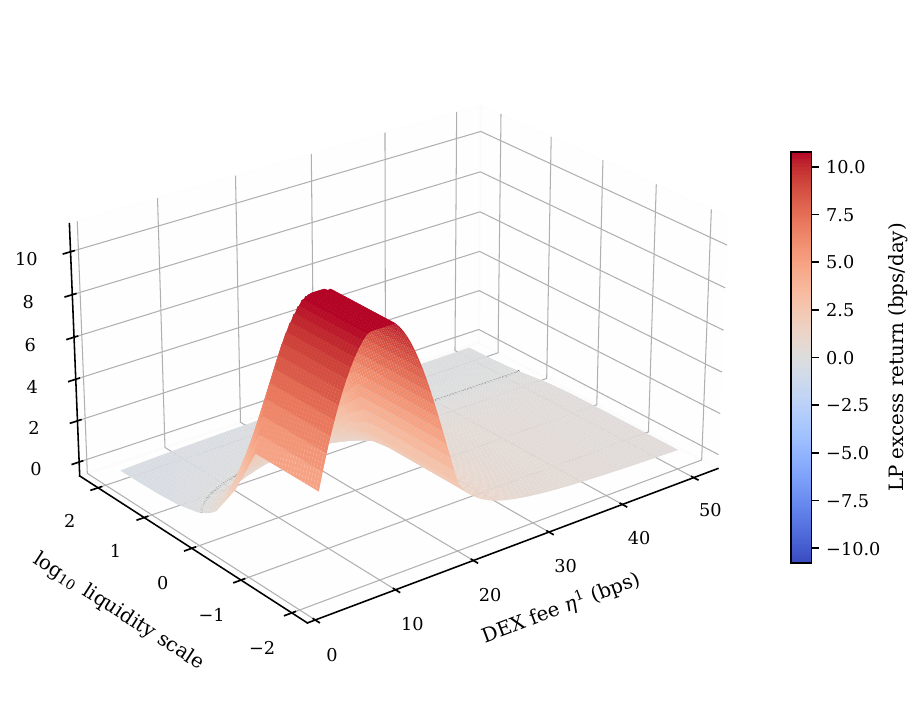}
    \end{minipage}
    \begin{minipage}{.44\textwidth}
        \includegraphics[width=\linewidth]{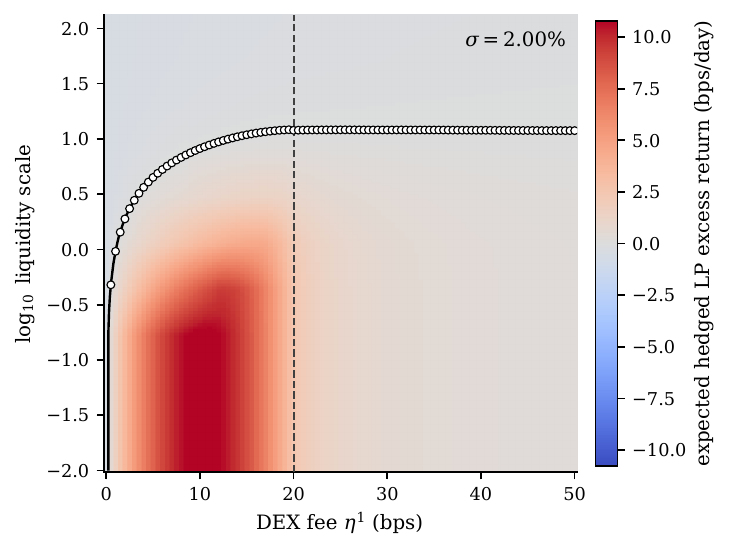}
    \end{minipage}\hfill
    \caption{(Left) Expected hedged LP excess return
$\mathcal R(L;\eta^1)-\bar r$ as a function of the liquidity scale $L$
and the DEX fee $\eta^1$, holding volatility fixed at $\sigma=2\%$ and
setting $\bar r=0$. (Right) Return surface as a heatmap,
with the solid black curve marking the zero-return locus
$\mathcal R(L;\eta^1)=0$. The vertical dashed line marks the CEX trading cost
$\eta^0$.}
    \label{fig:free-entry-return}
\end{figure}

\Cref{fig:free-entry-return} illustrates the competitive-entry
mechanism at a representative volatility level where the pool is generally profitable. The two panels depict the same off-equilibrium return function
$\mathcal R(L;\eta^1)-\bar r$. The heatmap emphasizes the zero-return level, while the three-dimensional representation shows how LP returns vary away from equilibrium. Regions above zero attract additional LP
capital, whereas regions below zero induce exit. The downward-sloping zero crossings shown in the figure indicate that the equilibrium is locally stable in the sense that the return is positive immediately below the equilibrium liquidity level and negative immediately above it.

The near-flat portion of the surface at low liquidity reflects the scale structure of the model. When the pool is shallow relative to fundamental demand, incoming orders are large enough to move the AMM price to the order routing boundary. In this region, routed volume scales approximately proportionally with $L$. The expected return on LP capital is therefore nearly invariant to the absolute liquidity scale.  Equivalently, this region corresponds to a large demand-to-liquidity ratio, for which the normalized return approaches a limiting value.

As liquidity increases, the pool eventually becomes deep relative to the fixed level of fundamental demand. Routed volume and fee revenue then cease to grow proportionally with the capital supplied, while initial TVL continues to increase linearly with $L$. Expected return per unit of LP capital consequently declines, producing the stable zero-return intersection. Thus, equilibrium arises in the transition from a regime in which returns are approximately scale-invariant to one in which additional liquidity dilutes the available trading revenue.

The fee changes both the height of the low-liquidity return plateau and the liquidity level at which returns cross zero. Increasing $\eta^1$
initially raises compensation per routed trade and expands the amount of liquidity that can be supported in equilibrium. At larger fees,
however, the reduction in fundamental participation increasingly offsets the additional revenue per trade. The positive returns therefore flatten over the upper part of the fee range. This pattern reflects the two mechanisms through which the fee affects equilibrium: directly through LP fee income and indirectly through the endogenous routing response of fundamental traders.

Throughout the remainder of this section, the AMM fee is treated as a fixed policy parameter. In \cref{sec:optimal.fees} below, we will study the fee that maximizes the return to installed LP capital and the fee that maximizes equilibrium TVL.

\subsection{Equilibrium Liquidity}
\label{sec:equilibrium.scale}

We next study how equilibrium liquidity responds to the AMM fee and the remaining market parameters. For each parameter configuration, we simulate the expected LP return $\mathcal R(L;\eta^1)$ over a grid of liquidity scales and identify its stable intersection with the benchmark return $\bar r$. Note that by \cref{sec:scale.invariance}, LP return depends only on the liquidity-to-demand ratio. Therefore, finding an equilibrium over $L$ is the same as finding the ratio that sets the pool return to $\bar{r}$.

At this stage, we treat $\eta^1$ as a policy parameter and remain
agnostic about the objective of the party choosing it. For every
candidate fee, the equilibrium condition determines a set of
liquidity levels $\eta^1 \mapsto L^*(\eta^1)$. This separates the question of how LP capital responds to a fee from the subsequent question of which equilibrium outcome a protocol, trader, or other market participant may prefer.

\begin{figure}[htb]
    \centering
    \begin{minipage}[c]{.54\textwidth}
        \centering
        \includegraphics[width=\linewidth]
        {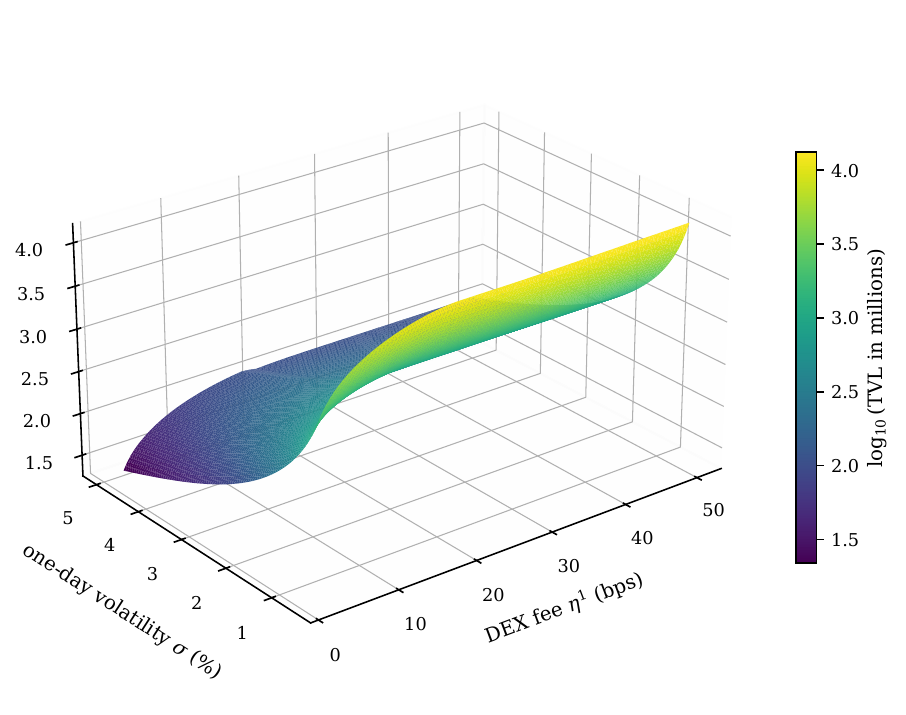}
    \end{minipage}
    \hfill
    \begin{minipage}[c]{.44\textwidth}
        \centering
        \includegraphics[width=\linewidth]
        {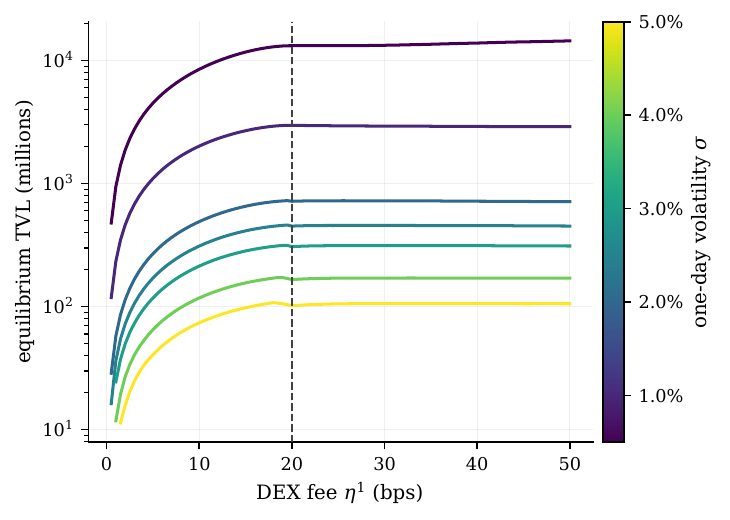}
    \end{minipage}
    \caption{
    (Left) Equilibrium TVL over the AMM-fee and volatility parameter space, with TVL plotted on a logarithmic scale. (Right) Fee cross-sections at representative volatility levels. The vertical dashed line marks the CEX trading cost $\eta^0=20$ basis points.}
    \label{fig:equilibrium.tvl}
\end{figure}

\Cref{fig:equilibrium.tvl} illustrates the two principal forces governing equilibrium liquidity. First, moving away from a near-zero AMM fee sharply increases the amount of capital that can be supported by the pool. When $\eta^1$ is small, fee income is insufficient to compensate LPs for inventory and adverse-selection losses, and the zero-return condition can therefore be satisfied only at a relatively shallow pool. Increasing the fee raises the revenue earned on routed trades and permits additional LP entry.

This effect is strongest at low fee levels. As the AMM fee approaches the CEX trading cost, the increase in equilibrium TVL begins to attenuate. A higher fee generates more revenue per routed trade but also makes the AMM less attractive to fundamental traders. The resulting decline in routed volume increasingly offsets the higher fee rate. Equilibrium TVL consequently becomes relatively flat over the upper part of the fee range and may decline slightly once the loss of fundamental participation dominates the additional revenue per trade. The dashed line in the right panel highlights that this flattening occurs near the point at which the AMM ceases to undercut the CEX on explicit trading costs.

Second, equilibrium TVL decreases sharply with volatility. Higher volatility causes the external price to move farther between trading opportunities and increases the losses borne by LPs when the AMM trades at a stale price. To restore the zero-return condition, less capital can remain in the pool. This effect is quantitatively large. We see that, over the parameter range displayed, an increase in one-day volatility shifts equilibrium TVL downward by several orders of magnitude. Volatility therefore affects not only the return on installed liquidity, as in the preceding analysis, but also the amount of liquidity that competitive LPs are willing to install.

\subsection{Equilibrium Comparative Statics}
\label{sec:equilibrium.comparative.statics}

Having characterized the amount of liquidity supplied in equilibrium, we next evaluate the market outcomes generated by that liquidity. For each fee and volatility configuration, we report the share of fundamental demand routed to the AMM and the all-in execution cost across both venues.

\begin{figure}[t]
    \centering
    \begin{minipage}[t]{0.49\textwidth}
        \centering
        \includegraphics[width=\linewidth]
        {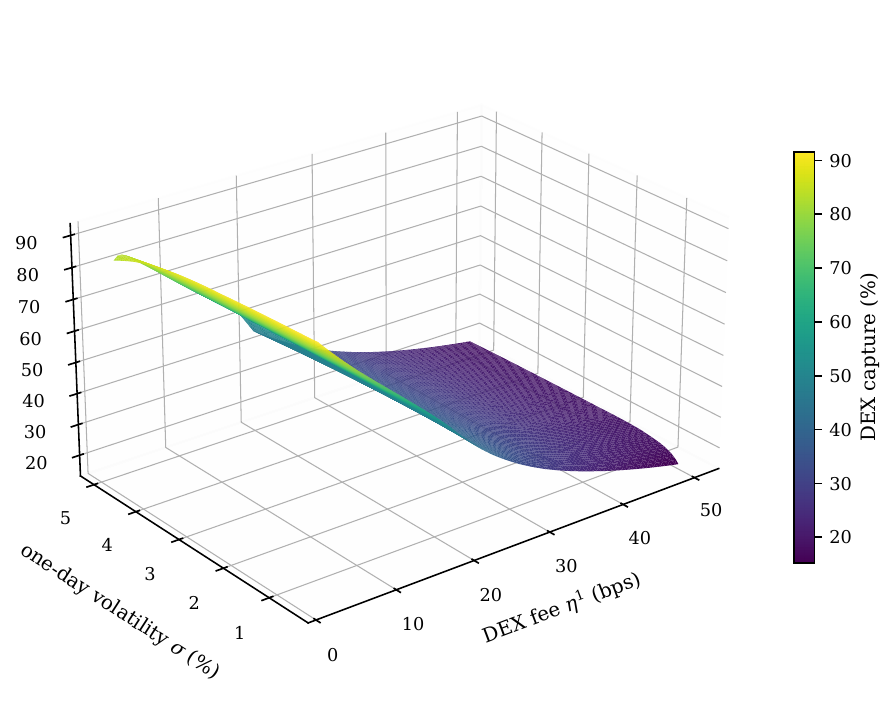}

    \end{minipage}
    \hfill
    \begin{minipage}[t]{0.49\textwidth}
        \centering
        \includegraphics[width=\linewidth]
        {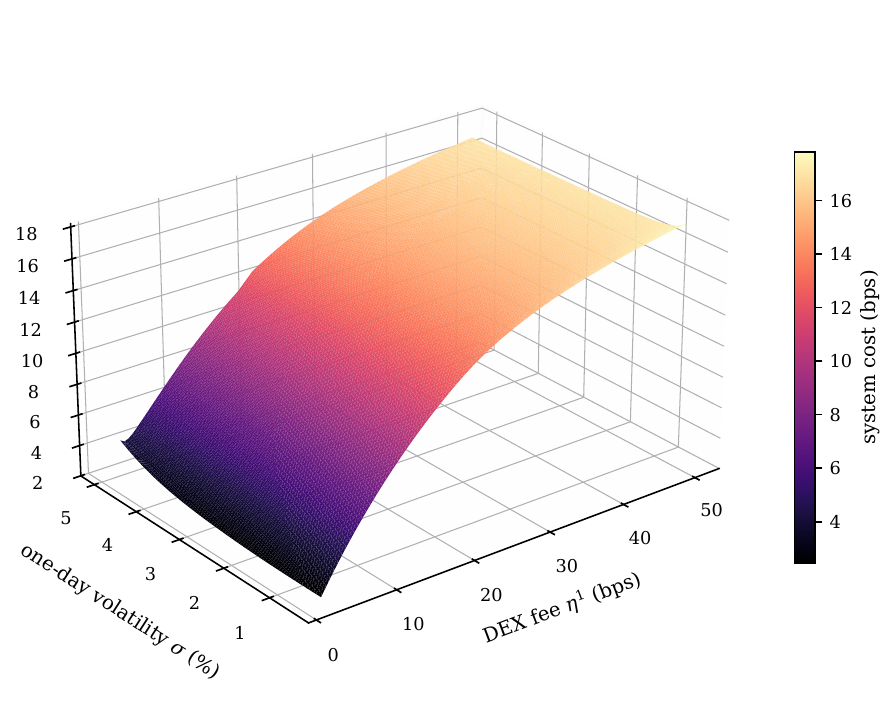}
    \end{minipage}

    \caption{
    (Left) Share of fundamental notional routed to the AMM at the unique stable finite equilibrium.
    (Right) Whole-system execution cost, including the all-in cost of trades executed on the AMM and the CEX trading cost incurred by residual orders routed to the CEX.  Execution costs are measured relative to the contemporaneous CEX midprice and normalized by total fundamental notional.}
    \label{fig:equilibrium.routing.execution}
\end{figure}

\Cref{fig:equilibrium.routing.execution} shows how the fee affects
traders both directly and through the endogenous response of equilibrium liquidity. The dominant pattern in the left panel is a decline in fundamental AMM participation as $\eta^1$ increases. A higher fee worsens the AMM's all-in execution price and causes a larger share of fundamental orders to be routed to the CEX. When $\eta^1$ approaches the CEX trading cost $\eta^0$, the AMM loses its systematic cost advantage. Fundamental demand capture nevertheless remains positive even when $\eta^1\geq\eta^0$, since routing depends on the complete execution price rather than on the fee alone. A sufficiently favorable discrepancy between the AMM price and the contemporaneous CEX price can still make the AMM the cheaper venue.

The effect of volatility on routing is more subtle. Higher volatility reduces equilibrium pool depth and therefore increases price impact, which tends to reduce AMM participation. At the same time, greater dispersion between the AMM and CEX prices creates more states in which the AMM quote is favorable to one side of the fundamental order flow. These opposing effects make the volatility dependence of average demand captured less pronounced than its fee dependence.

The right panel translates these routing decisions into execution costs for fundamental traders. Over the displayed parameter range, the execution cost of the whole system (considering all fundamental orders and both venues) generally increases with the AMM fee. The fee directly raises the cost of orders executed on the AMM and indirectly displaces additional flow to the CEX. At high AMM fees, the system increasingly resembles the counterfactual in which all fundamental orders are executed on the CEX. Execution costs can remain below the all-CEX benchmark, however, because traders continue to use the AMM selectively when its quoted price is sufficiently favorable.

The whole-system measure is preferable to an execution cost statistic conditioned only on AMM routing. At high fees, the small set of orders that continues to use the AMM is selected precisely because the AMM offers an unusually favorable quote. Conditional AMM execution quality may therefore appear strong even as the venue captures little fundamental flow. The system-wide measure accounts for both the quality of AMM executions and the orders displaced to the CEX.

Taken together we see that a fee supporting a large amount of equilibrium liquidity need not maximize fundamental participation or minimize trader execution costs. As a result, fee choice, which is the subject of the next section, depends intimately on the objective of the protocol designer.

\section{Fee Choice with Fixed and Equilibrium Liquidity}\label{sec:optimal.fees}

The AMM fee can be evaluated under two complementary assumptions about pool liquidity. We first retain the short-horizon convention of \cref{sec:profitability} and hold initial reserves fixed. Maximizing expected hedged PnL is then equivalent to maximizing expected LP return because initial TVL does not vary with the fee. We next allow pool depth to adjust through competitive entry and determine the fee that maximizes equilibrium TVL. This second criterion measures the amount of LP capital that the venue can support at the benchmark return. This is one natural metric for the pool's success that is not the LP return, which by definition equals $\bar r$ in equilibrium.

\subsection{Fee Choice with Fixed Liquidity}
\label{sec:fixed.fee.choice}

\begin{figure}[!htb]
    \centering
        \includegraphics[width=.49\linewidth]{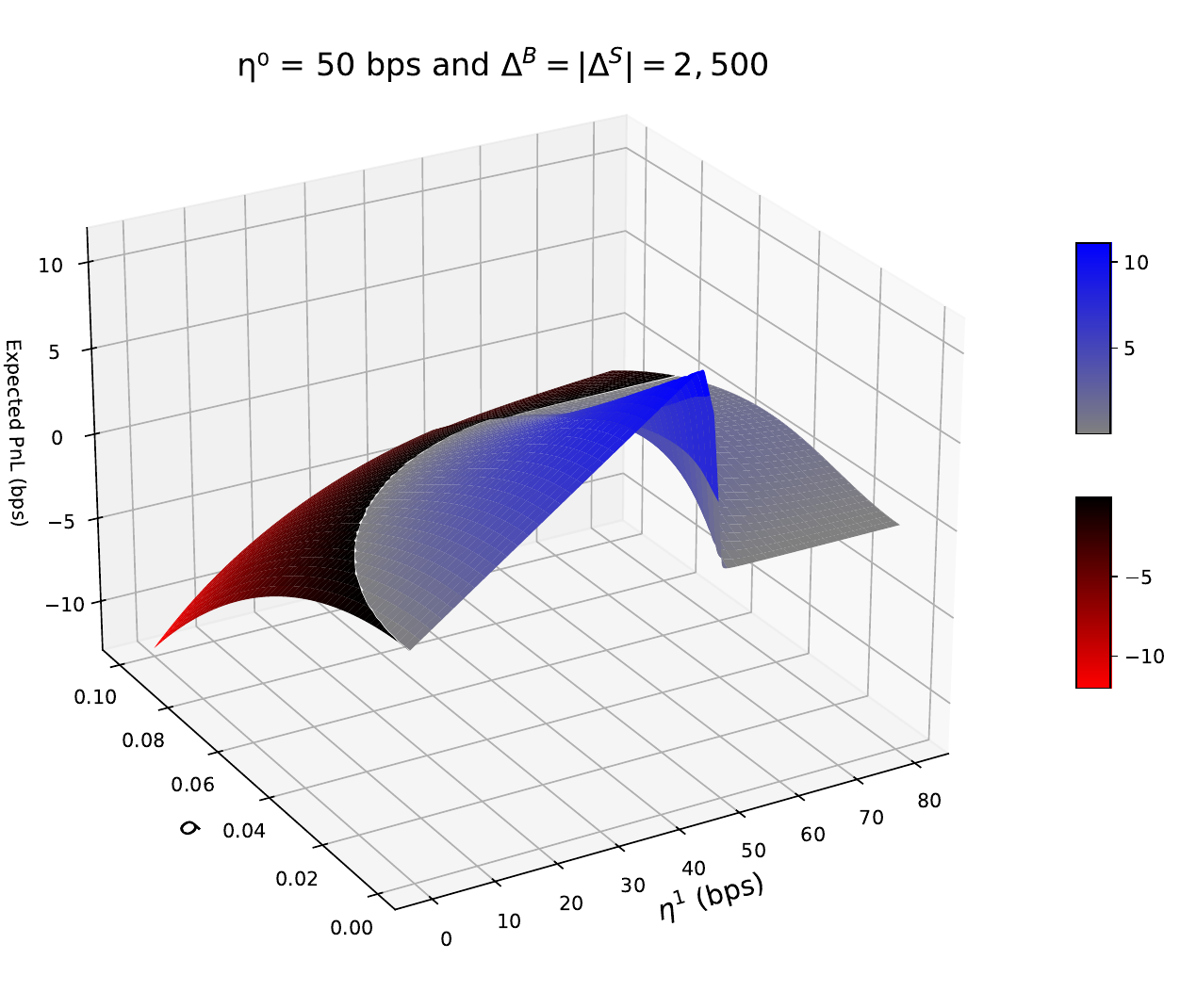} 
        \includegraphics[width=.49\linewidth]{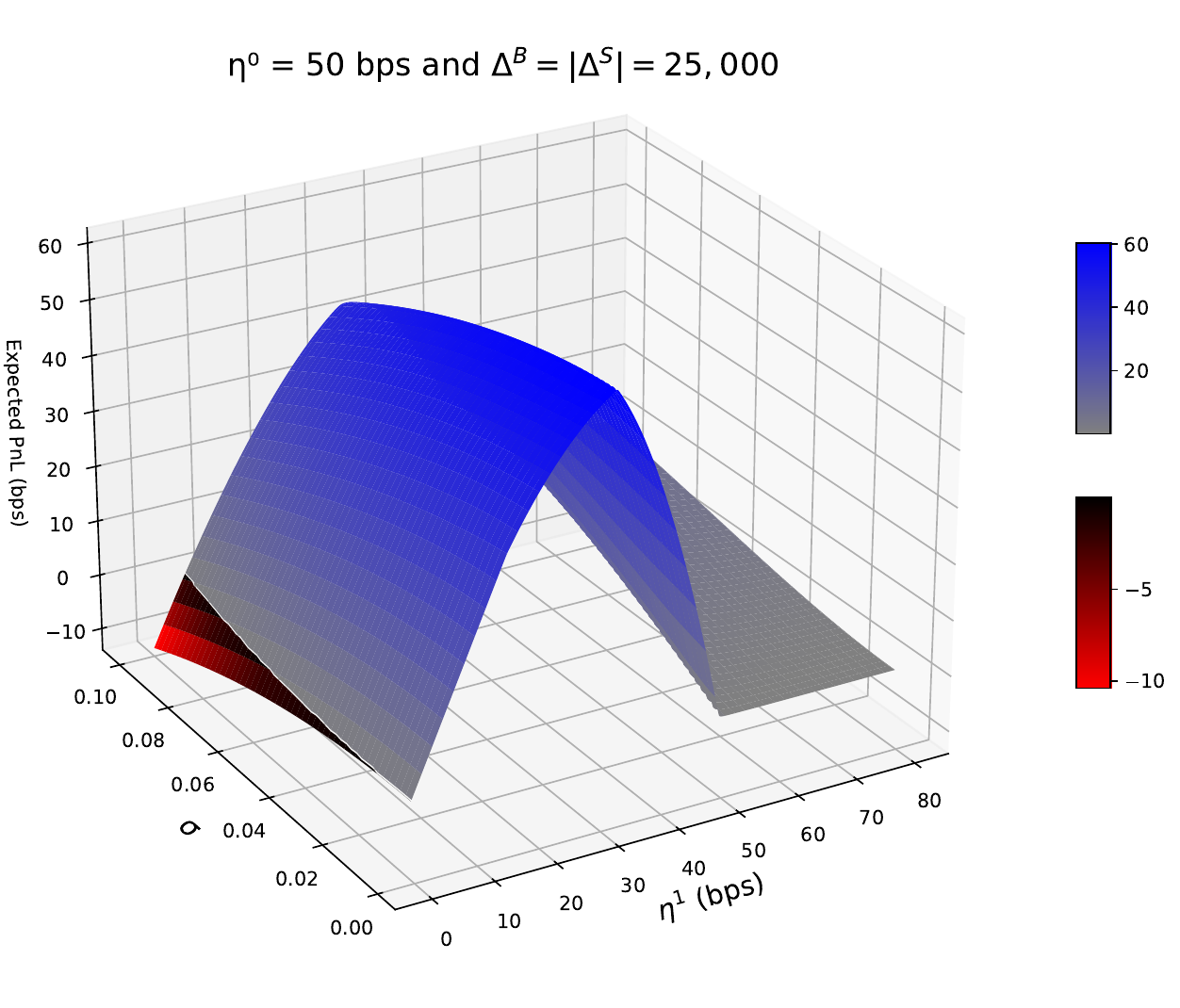}
        \includegraphics[width=.49\linewidth]{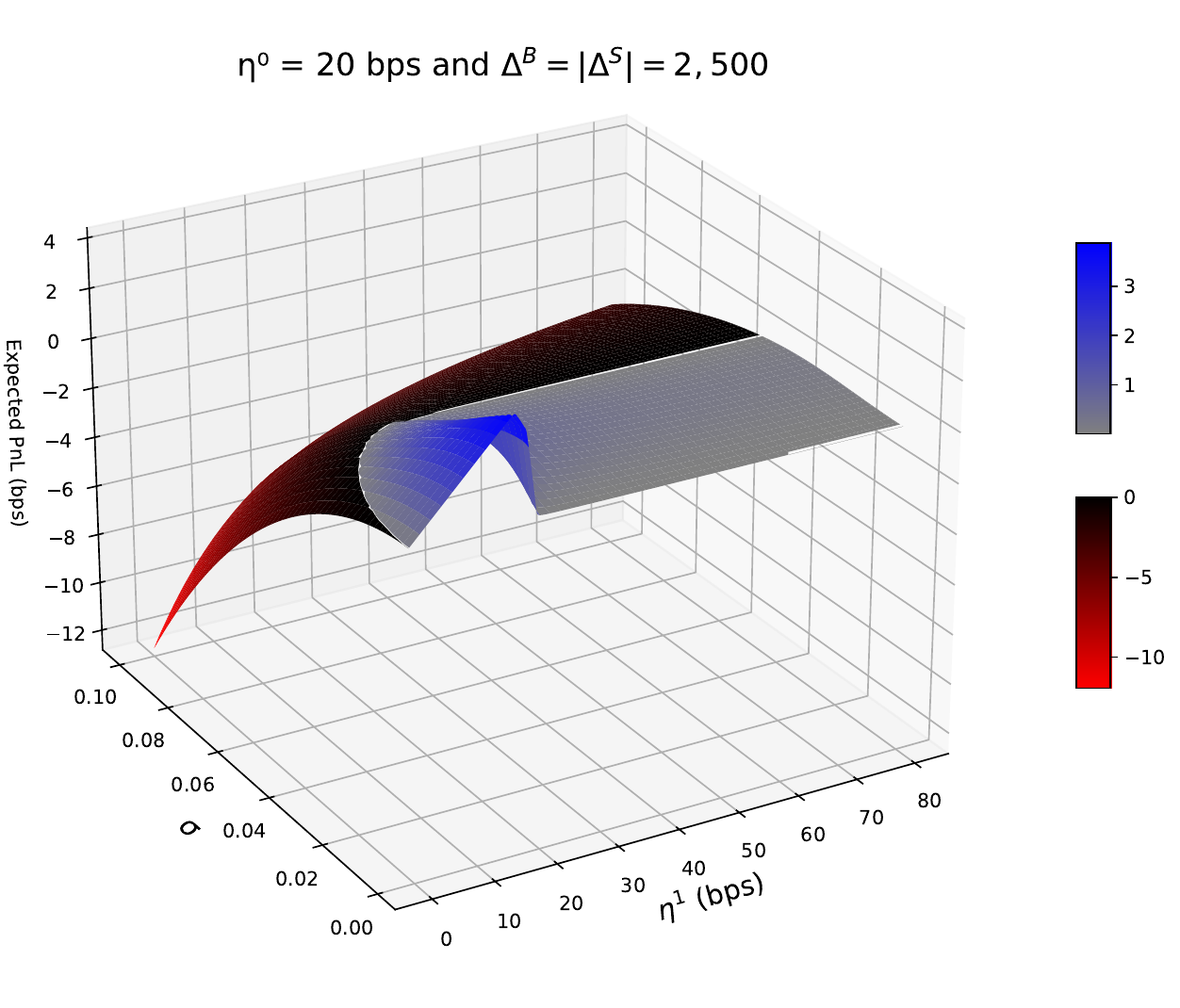} 
        \includegraphics[width=.49\linewidth]{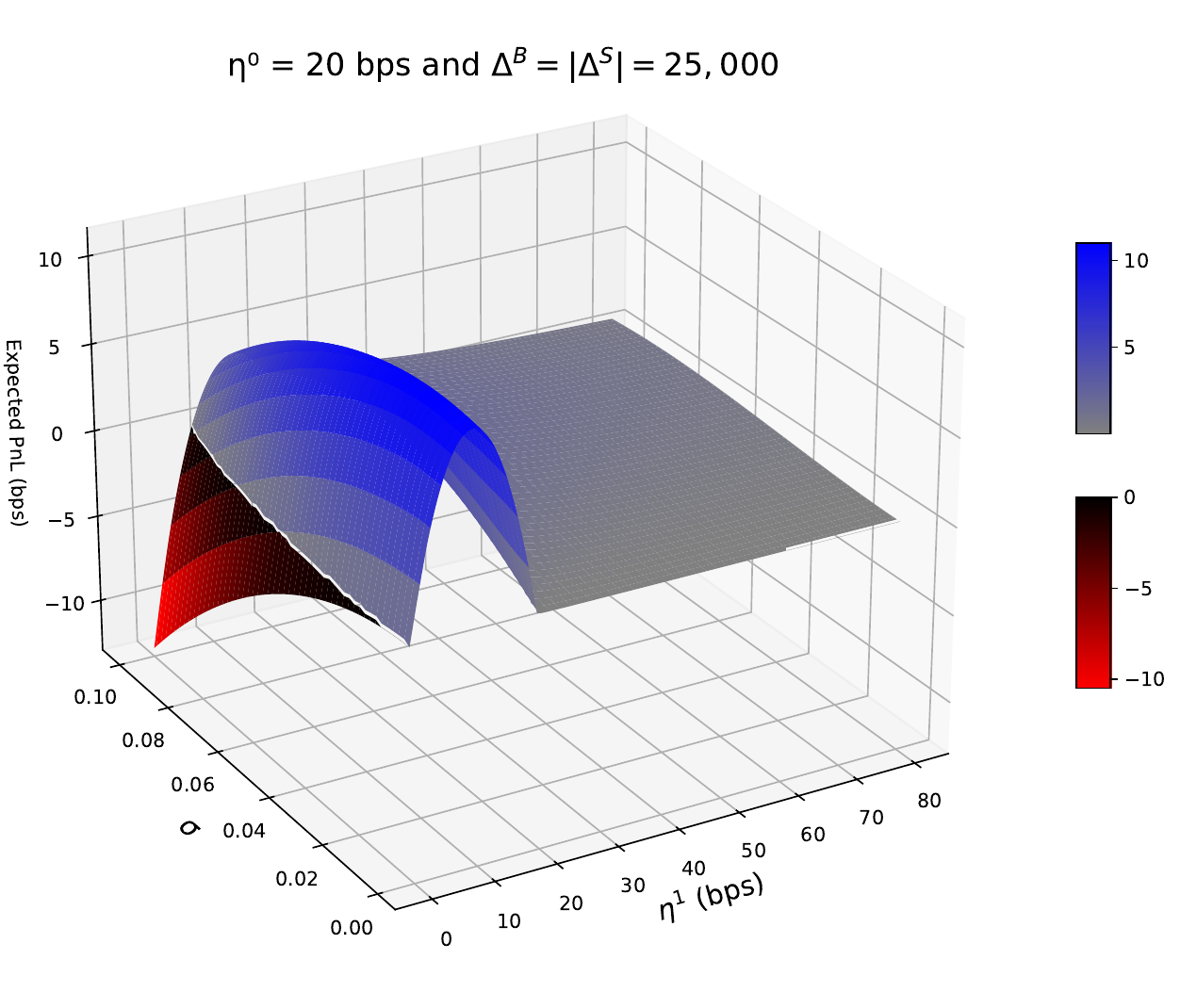}
            \caption{Expected PnL surface as a function of the AMM fee~$\eta^1$ and market volatility $\sigma$, shown under two different assumptions about fundamental demand and CEX fee $\eta^0$. The bottom row corresponds to the baseline fee of $\eta^0 = 20$ bps while the top row uses $\eta^0 = 50$ bps. The left column illustrates a demand of $\Delta^B = |\Delta^S| = 2,500$, below our baseline, while the right column increases demand to the (very high) level of $\Delta^B = |\Delta^S| = 25,000$. The expected PnL for our baseline of $\Delta^B = |\Delta^S| = 5,000$ is represented in the right panel of \cref{fig:optimal_fee_curves} below.
        }
        \label{fig:expected_pnl_surface}
\end{figure}

The discussion in Section \ref{sec:profitability} highlighted that the AMM fee $\eta^1$ governs a trade-off: higher fees increase per-trade revenues for the LP and defend against adverse selection; at the same time, they reduce the AMM’s competitiveness, lowering the volume of fundamental order flow it attracts and increasing price stickiness.  In this subsection, we distill these interactions and determine the AMM fee level that maximizes LP profitability for given market conditions when initial reserves are fixed.

\Cref{fig:expected_pnl_surface} plots the expected PnL surface as a function of $\eta^1$ and volatility $\sigma$, for varying levels of fundamental demand and different CEX fee benchmarks~$\eta^0$. A consistent pattern emerges: for a broad range of volatilities~$\sigma$, the optimal fee is strictly less than the CEX fee, with the PnL surface exhibiting an interior maximum below $\eta^0$. The maximum is particularly pronounced for moderate volatility. As the CEX fee increases, the peak of the surface shifts by a similar amount. In brief, the optimal fee for the AMM is to \emph{undercut} the CEX fee. 

At this point it may be important to recall that our CEX fee $\eta^0$ is a reduced-form representation of the effective execution cost faced by market participants, not merely the nominal fee posted by the exchange. Moreover, in practice, this cost (as well as the posted fee) varies substantially across traders due to tiered fee schedules, volume-based discounts and rebates, and optimal execution algorithms employed by sophisticated traders. Our parsimonious model absorbs these nuances into a single parameter $\eta^0$, representing an average or effective benchmark. Consequently, the aforementioned undercutting does not mean that the optimal AMM fee is lower, e.g., than a posted headline rate such as Binance’s taker fee, but rather that it should remain competitive with the distribution of effective CEX trading costs.\footnote{For very large demand, a proportional fee may not be a good proxy for price impact costs on the CEX. Additional price impact would likely increase the optimal AMM fee, as the AMM competes with the overall slippage on the CEX. We remark that adding nonlinear price impact on the CEX would substantially complicate the order routing discussed in \cref{sec:ecosystem}.} A calibration to real-world data is provided in \cref{sec:empirics}, and generalizations of the fee structure are also discussed in \cref{sec:extensions_fees}.

Returning to \cref{fig:expected_pnl_surface}, we observe that increased volatility depresses PnL across all fee levels, while increased demand raises it. At high levels of volatility, the interior maximum disappears and expected PnL turns \emph{negative} across all plausible fee values, indicating that the LP would prefer an infinite AMM fee to prevent trading. 

We also observe that greater demand smooths the relationship between PnL and optimal AMM fee, consistent with the notion that an abundance of available order flow reduces the sensitivity of performance to fee design. Of course, there is a limit to this effect. At the extremes of demand, the liquidity in the pool is insufficient to absorb any marginal flow. Likewise, at very high fee levels and low volatility, it is almost never optimal for traders to engage with the pool regardless of aggregate demand.

\begin{figure}[!htb]
    \centering
        \includegraphics[width=.48\linewidth]{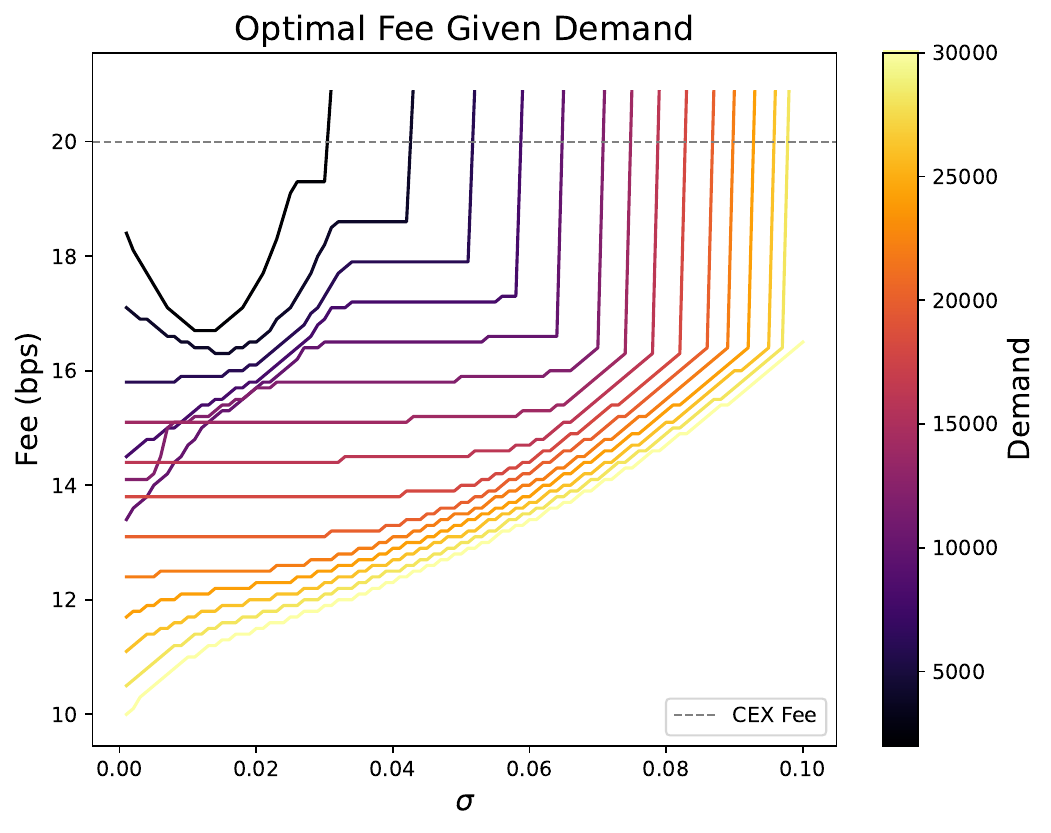}
        \includegraphics[width=.505\linewidth]{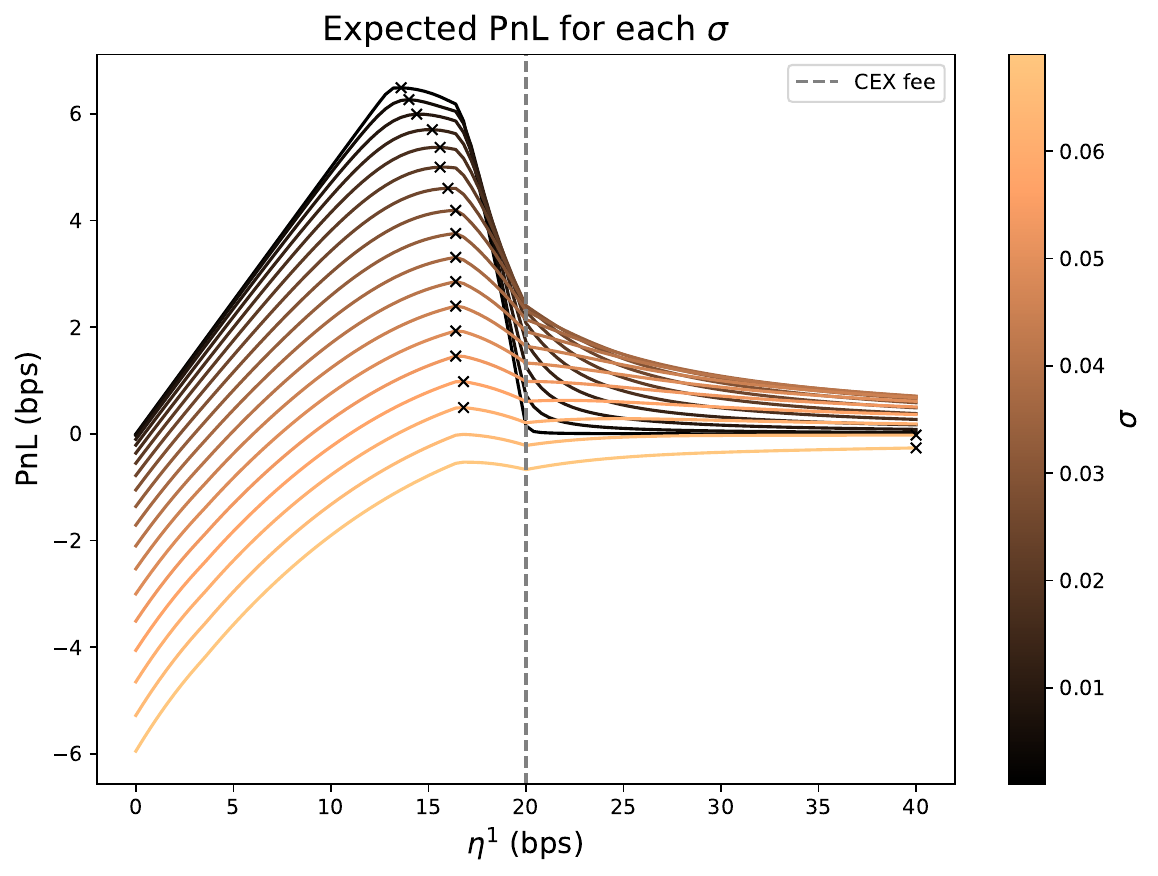} 
        \caption{(Left) Optimal AMM fee as a function of volatility, for several levels of total fundamental demand $\Delta^B+|\Delta^S|$. (Right) Expected PnL as a function of the AMM fee (for the baseline total demand of 10,000) with the numerically optimal fee marked by $\times$, for several levels of volatility. Where $\times$ is at 40 bps, the PnL is always negative and the true optimum is $\eta^1=\infty$.}
        \label{fig:optimal_fee_curves}
\end{figure}

\cref{fig:optimal_fee_curves} extracts from the PnL surface the functional relationship between the optimal fee and market volatility $\sigma$ for several levels of fundamental demand. Overall, we observe a fairly complex behavior in the left panel---but note that the figure shows a very wide parameter range. Next, we discuss this behavior, going from low to high demand, or from the top-left  to bottom-right curve. While we explain in detail how all the shapes---including extreme regimes---are implied by the model, we emphasize that the most important regime for practical purposes is likely closer to the benchmark regime discussed in \cref{sec:opt_fee_benchmark} below. %

\subsubsection{Low Demand} 

For low to moderate levels of demand (darkest curves in the left panel of \cref{fig:optimal_fee_curves}), the optimal fee exhibits a U-shape in~$\sigma$, which we explain first.

When there is no volatility (i.e., $\sigma=0$), the AMM can capture \emph{all} of the fundamental demand by undercutting the CEX by just enough to remain the preferred venue after accounting for price impact. While there is still a nonlinear relationship between LP profit and fee levels due to the influence of $\eta^1$ on demand, this is a good proxy for what happens when demand is low.  Indeed, if $\Delta^B=|\Delta^S|\approx 0$ and $\sigma=0$ then the buyers have next to no price impact and (since $S^1_t \approx S^0_t\equiv S^0_0=S^1_0$)
\[\mathrm{Fee \ Revenue} \approx \eta^1 (\Delta^B+|\Delta^S|)S^1_0 \mathds{1}_{\{\eta^1<\eta^0\}}.\]
This is clearly maximized by taking $\eta^1$ arbitrarily close (but not equal) to  $\eta^0$. 

For these lower demand levels, the gap between the CEX fee and the left endpoint of the curves in \cref{fig:optimal_fee_curves} (left panel) is approximately such that the AMM remains the preferred venue even after most of the fundamental buyers (or sellers) have pushed its price in one direction. As a result, the gap initially increases with fundamental demand. Since the external price is constant, there is no possibility of arbitrage in this environment. This story persists in very low-volatility environments. Indeed, since prices are relatively stable, the AMM can undercut the CEX by a modest amount and still attract the lion's share of the fundamental volume.

As volatility starts to increase, the fluctuations in the reference price become more of a concern as AMM quotes become stale more quickly. Nonetheless, the dominant theme for low volatility is volume capture. Remaining competitive in this environment requires a deeper fee discount, leading to a decline in the optimal fee (the decreasing part of the U-shape). At intermediate levels of volatility, the directional pressure from fundamental flow is no longer sufficient to keep prices aligned. In this regime, the optimal fee flattens out, reflecting a delicate balance between capturing order flow and earning adequate revenue per trade. 

Beyond a critical volatility threshold, adverse selection begins to dominate the LP’s decisions. The optimal fee starts to rise---gradually at first, and then more quickly---as the LP shifts focus from volume to per-trade PnL and aims to avoid severely unprofitable executions. Eventually, the optimal fee reaches a plateau below the CEX fee. This plateau represents the stability of the argmax also seen in the right panel of \cref{fig:optimal_fee_curves} (shown there for higher demand). It remains near this level until a tipping point is reached: when volatility becomes too severe, expected PnL is negative for all fee levels and essentially increasing in $\eta^1$. At this stage, the optimal fee jumps to $+\infty$, signaling that the LP would prefer to shut down AMM activity.

\subsubsection{Moderate to High Demand} \label{sec:mod.to.high.demand}
As fundamental demand increases, the dependence of the optimal fee on volatility $\sigma$ becomes monotonically increasing. We also observe that the optimal fee for $\sigma = 0$ decreases, toward approximately $\eta^0 / 2 = 10$ bps for very large demand.\footnote{For low volatility regimes, the monotonicity of the optimal fee in demand breaks down at an intermediate point. This is elucidated in Figure~\ref{fig:approx.infinite.demand} of Appendix~\ref{sec:inf.demand.opt}, which illustrates that a shift in the expected PnL curve’s maximum is responsible for this behavior.}

The limit of $\eta^0 / 2 = 10$ bps can be understood analytically as follows. In the regime of high demand, DEX liquidity remains finite, so the LP can only absorb a limited amount of volume per time step. With abundant demand, buyers push the price ratio to the buy boundary, and sellers push it to the sell boundary. Over time, the system oscillates between these two points. By assuming we initialize the system at one of these boundaries and randomizing the arrival of buyers and sellers, the AMM captures both sides of the market 50\% of the time, and one-sided trades otherwise. In expectation, this results in fee revenue equivalent to 75\% of a round-trip price movement from the buy boundary to the sell boundary and back. Since the CEX price remains constant for $\sigma=0$, both trade sizes and LP proceeds are deterministic and can be computed explicitly. 
By expanding for small $\eta^0, \eta^1$, we obtain the following leading-order approximation for the expected revenues (see Appendix \ref{sec:inf.demand.opt} for details)
\[
\frac{1}{N}\mathbb{E}\left[ \text{Fee Revenue} \right] \approx \frac{3}{4} X_0 \left( 2 \eta^1 \eta^0 - 2 (\eta^1)^2 \right).
\]
Maximizing this expression with respect to $\eta^1$ yields the approximate optimal fee
$
\eta^{1,*} \approx \eta^0/2
$
which is consistent with the behavior observed in \Cref{fig:optimal_fee_curves}.

At this fee level, the LP's profit boundaries and the buy-sell boundaries are nearly identical (see \Cref{tab:thresholds}). We conclude that, in this regime, the LP's objective shifts from maximizing order flow capture to optimizing fee revenue per unit of volume. 
This perspective also clarifies why the optimal fee becomes strictly increasing in volatility $\sigma$ as demand grows. In low-demand settings, higher volatility reduces the AMM's competitiveness by making prices stale, necessitating lower fees to maintain flow capture. However, in high-demand environments---where price ratios naturally hover near the profit boundaries due to high levels of fundamental trading---volatility introduces the risk of prices drifting into unprofitable regions. To mitigate this, the LP increases fees, providing a buffer that preserves favorable execution conditions.

That said, even with elevated demand, extreme volatility renders price swings so severe that continued AMM operation becomes unprofitable, motivating the LP to once more exit the market by setting prohibitively high fees, explaining the vertical part of the curves.

\subsubsection{Benchmark Performance and Robustness}\label{sec:opt_fee_benchmark} The right panel of \cref{fig:optimal_fee_curves} depicts the expected PnL as a function of the AMM fee across a range of $\sigma$ when demand is fixed at the baseline of $\Delta^B=|\Delta^S|=5{,}000$ (i.e., 10{,}000 units in total). The maxima corresponding to the left panel of the figure are indicated by $\times$ markers. This demand falls squarely between the two extremes discussed above. At this level, we no longer observe the decreasing part of the U-shape that was prevalent at the lowest total demands of 2{,}000-4{,}000 units. At the same time, we do not undercut the CEX as much as in the highest levels of 20{,}000-30{,}000 units. In this intermediate regime, we see that when volatility is low, PnL is sharply peaked, and too high of a fee can meaningfully impair performance. As volatility increases, the curvature around the PnL maximum decreases, indicating reduced sensitivity to fee choice near the optimum. Nonetheless, in the highest volatility regimes where PnL is always negative, undercutting the CEX is extremely undesirable and any fee that allows for a meaningful trading volume can result in significant losses.

Notably, for a broad range of intermediate volatility levels, a suitably chosen constant fee performs reasonably well. By contrast, for very high volatility (relative to demand) it is optimal for the LP to quote a prohibitively high fee, effectively shutting down trading. This behavior aligns with established practice in traditional financial markets: during market shocks, it is often prudent to widen spreads substantially---or even halt trading altogether---to mitigate risk.

\begin{figure}[!htb]
    \centering
        \includegraphics[width=.49\linewidth]{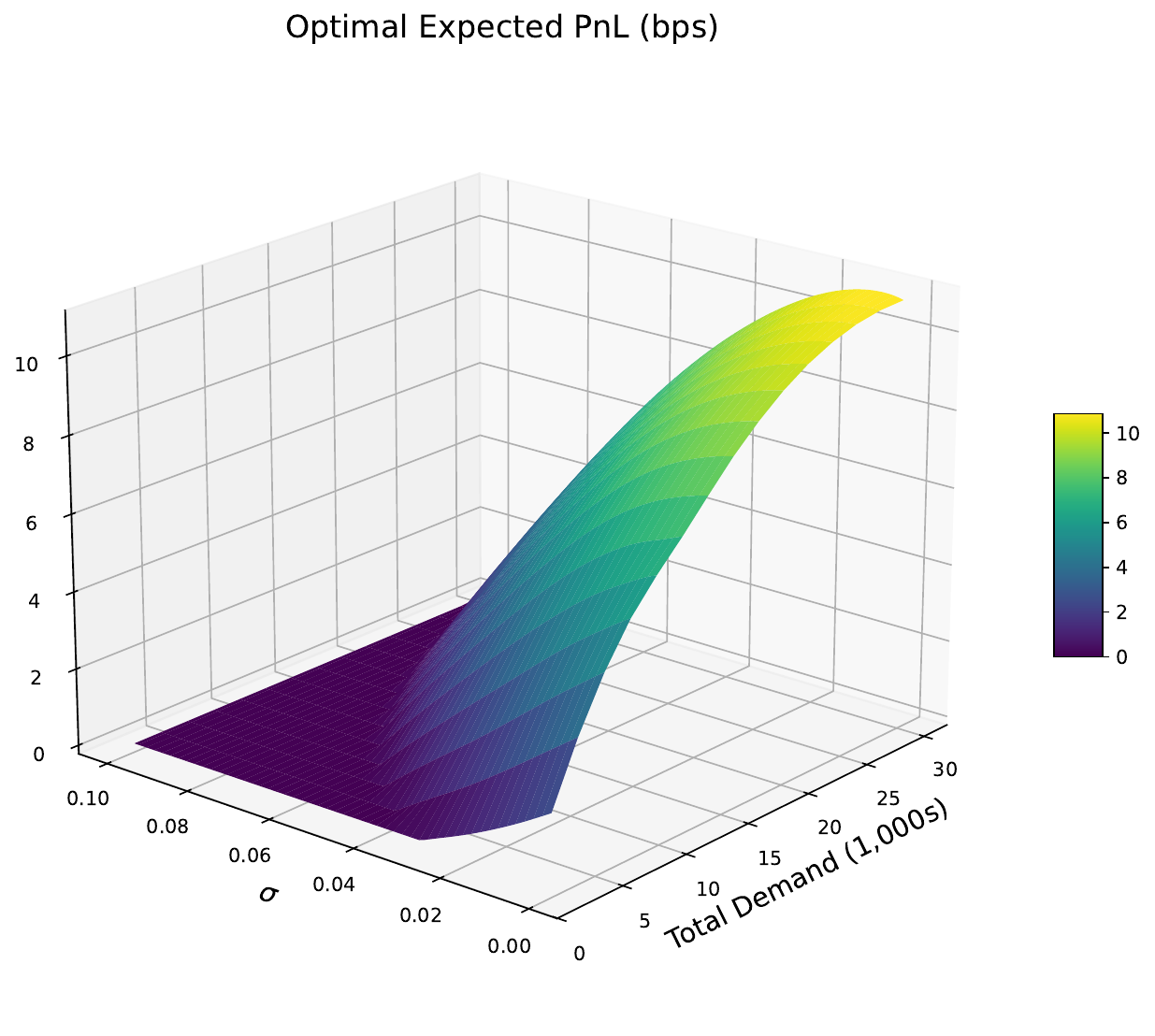}
        \includegraphics[width=.49\linewidth]{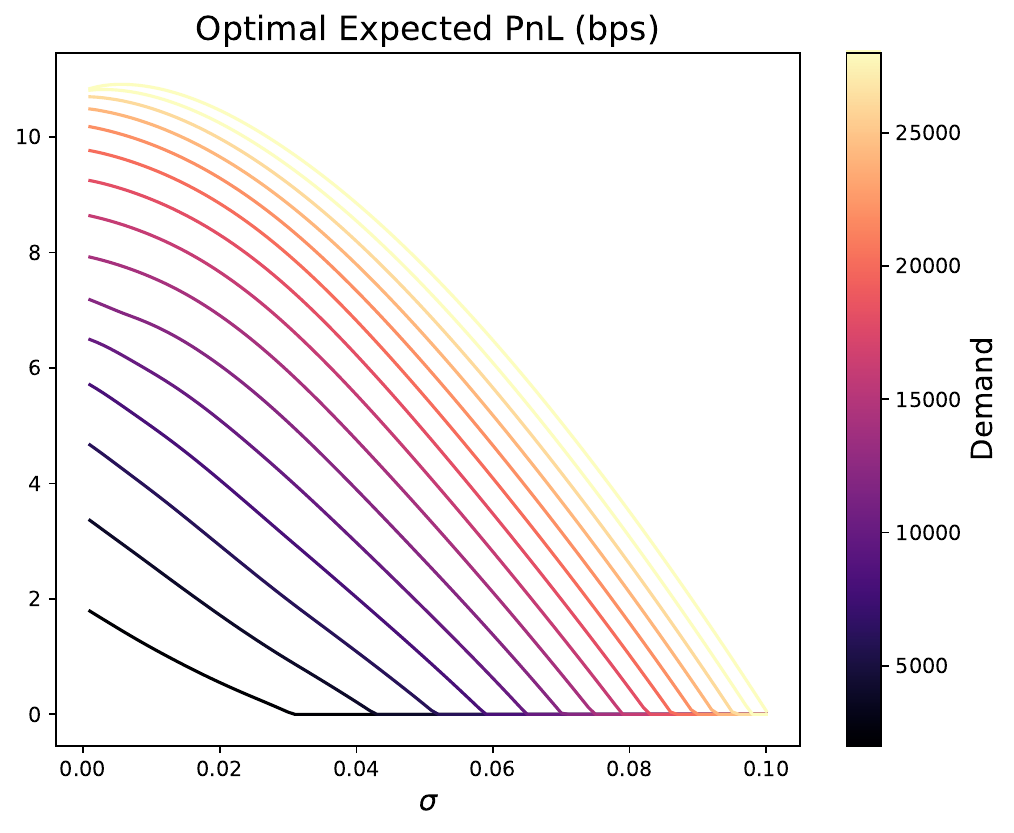}  
        \caption{(Left) Expected PnL surface under optimal fees from Figure~\ref{fig:optimal_fee_curves}, shown as a function of volatility $\sigma$ and total fundamental demand $\Delta^B + |\Delta^S|$. (Right) Cross-section of the same surface as a function of $\sigma$, with total demand levels indicated by the color gradient.}
        \label{fig:optimal_pnl}
\end{figure}

\Cref{fig:optimal_pnl} shows the expected PnL achieved under the optimal fee for each combination of fundamental demand and volatility. Consistent with our original findings in Figure \ref{fig:pnl_vs_demand_vol} for fixed fees, the optimal PnL increases with the volume of fundamental flow and decreases with volatility. As expected, the LP performs best in settings with sustained, bidirectional flow and relatively low price volatility. 

\begin{figure}[!htb]
    \centering
        \includegraphics[width=.49\linewidth]{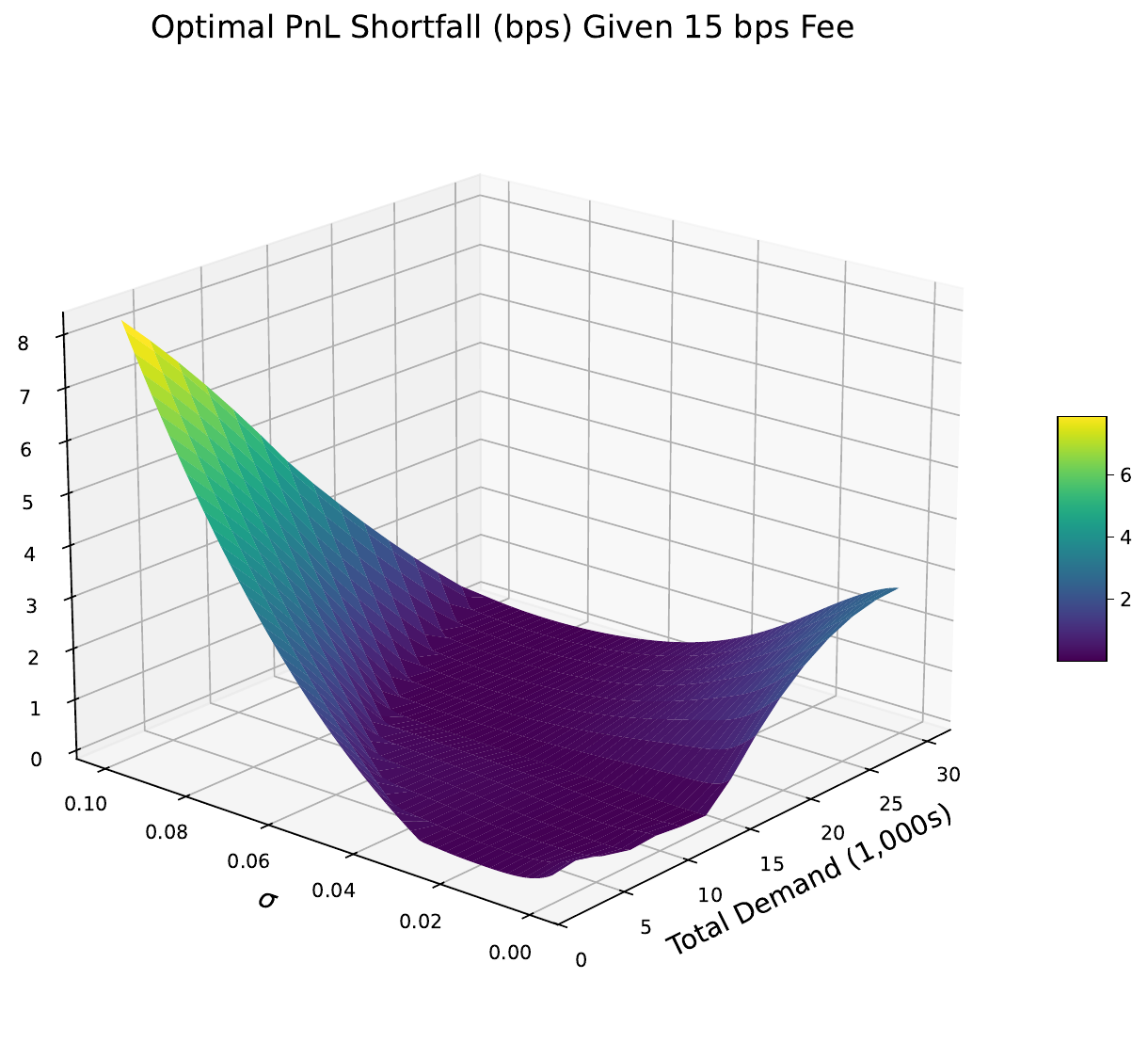}
        \caption{Shortfall in expected PnL relative to the optimum from Figure~\ref{fig:optimal_pnl} when applying a fixed 15 bps AMM fee, as in Figure~\ref{fig:pnl_vs_demand_vol}.}
        \label{fig:regret}
\end{figure}

While we have discussed the dynamics of the optimal AMM fee in detail, an important practical takeaway is that expected PnL is relatively robust around a well-chosen constant fee. As shown in the right panel of \Cref{fig:optimal_fee_curves}, setting the AMM fee to roughly 70–80\% of the CEX fee yields only modest regret across a broad range of volatility and demand levels.\footnote{This rule of thumb can change based on the reference CEX fee $\eta^0$. See Section \ref{sec:empirics} for an example where $\eta^0\approx 10$ bps.} Here, regret refers to the shortfall between the expected PnL obtained with a fixed fee and the maximum attainable PnL for each choice of market parameters (i.e., with the optimal fee for those parameters). We visualize this shortfall in Figure \ref{fig:regret} by applying a 15 bps AMM fee (approximately 75\% of the CEX fee, $\eta^0 = 20$ bps). We see that along the diagonal of the $(\sigma, \Delta^B + |\Delta^S|)$ plane, where volatility and demand increase together, the regret from using a constant fee remains minimal. This suggests that in dynamic environments, a fixed fee policy performs particularly well when demand and volatility are positively correlated. On the other hand, in extremely high-volatility environments where the LP would optimally choose to halt trading (i.e., set $\eta^1 = \infty$), the performance loss from maintaining a constant fee becomes substantial. Since cryptocurrency prices do experience such episodes, this regime cannot be neglected.

\subsection{Fee Choice with Equilibrium Liquidity}
\label{sec:equilibrium.fee.choice}

\begin{figure}[t]
    \centering
    \begin{minipage}[t]{0.5\textwidth}
        \centering
        \includegraphics[width=\linewidth]
        {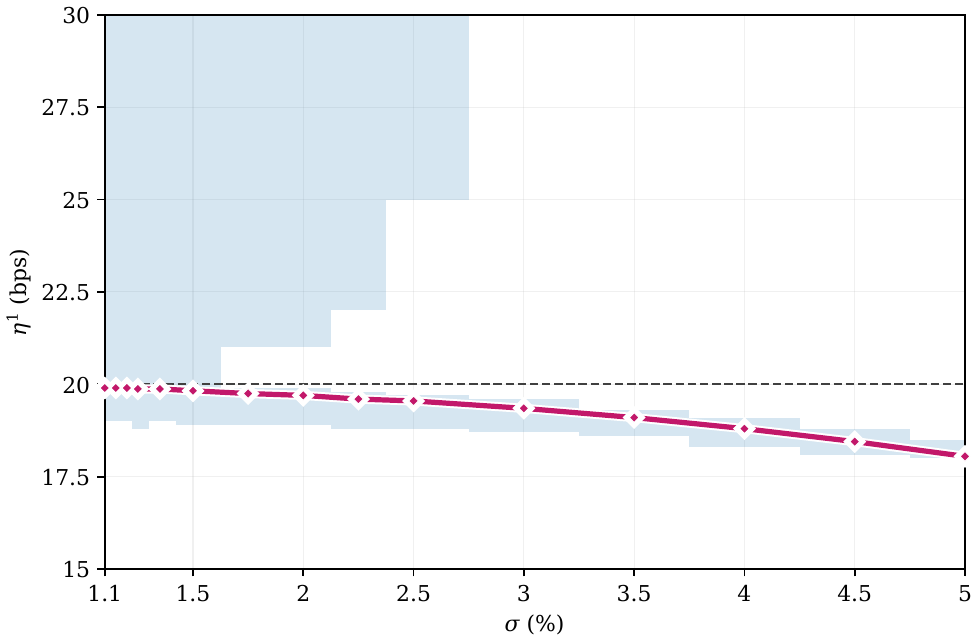}
        \vspace{-0.3em}
    \end{minipage}
    \hfill
    \begin{minipage}[t]{0.47\textwidth}
        \centering
        \includegraphics[width=\linewidth]
        {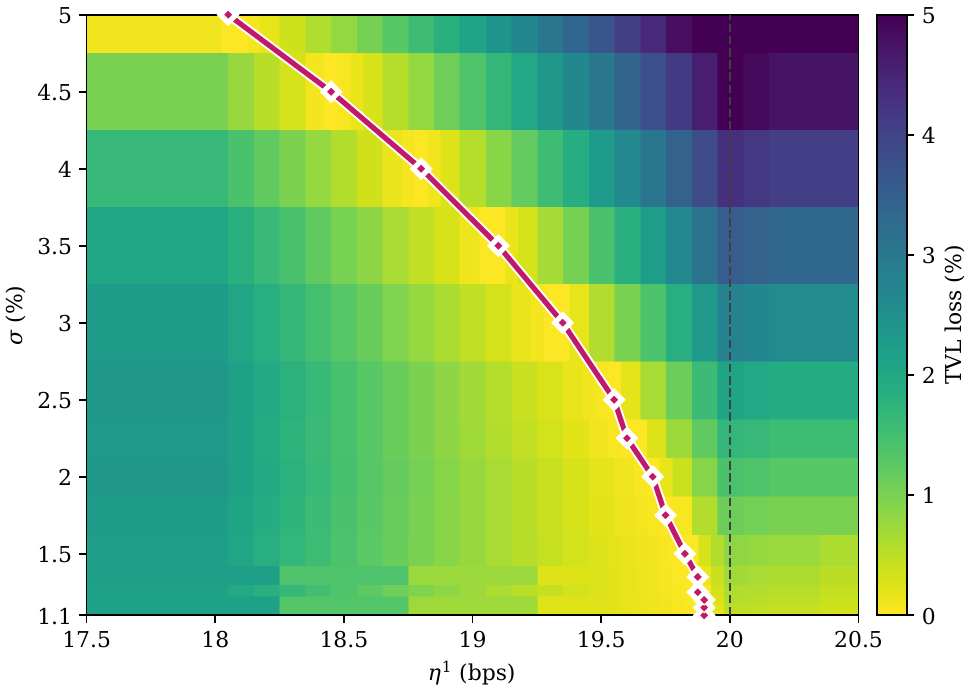}
        \vspace{-0.3em}
    \end{minipage}

    \caption{
    (Left) The DEX fee that maximizes equilibrium TVL over the simulated fee grid. The shaded region is the near-optimal set of fees that generate at least $99\%$ of maximum TVL. (Right) Percentage loss in equilibrium TVL relative to the fixed volatility maximum, with losses above $5\%$ capped for visual clarity. In both figures the dashed black line marks the CEX trading cost and the magenta line is the optimal DEX fee.}
    \label{fig:equilibrium.fee.choice}
\end{figure}

\Cref{fig:equilibrium.fee.choice} evaluates fee choice using equilibrium TVL as the objective. The fee maximization was performed over a fee grid ranging from 0 to 50 bps. Conditional on the existence of a unique equilibrium, the fee that maximizes equilibrium TVL lies immediately below the CEX trading cost throughout the volatility range considered. This undercutting preserves the DEX's explicit-cost advantage while extracting nearly the largest fee compatible with broad fundamental participation. The exact maximizing grid point varies modestly with volatility ($18$--$19.5$ basis points), but a fee just below, or equal to, the CEX cost remains close to optimal. At very low volatility the global fee optimum is not reliably identified on the simulated domain. The economic conclusion therefore appears to be that the TVL objective favors a small undercut of the competing venue.

The undercutting reflects the two opposing effects of the AMM fee. A higher fee increases the revenue earned on each routed trade and thereby permits the pool to support more LP capital. At the same time, it worsens traders' all-in execution prices and reduces the amount of fundamental flow routed to the AMM. A fee immediately below $\eta^0$ extracts as much revenue as possible without sacrificing the DEX's cost advantage over the CEX. The resulting combination of fee income and broad fundamental participation supports the greatest equilibrium pool depth.

The left panel also shows that the exact optimizer is not always robustly identified economically. At lower volatility, a relatively broad range of fees generates TVL within one percent of the maximum. In this region, adverse selection losses are modest and the decline in routed volume caused by a higher fee is largely offset by the greater revenue earned per trade. As volatility increases, the near-optimal region narrows and moves closer to a strict undercut of the CEX cost. 

The loss surface in the right panel reveals an additional asymmetry. Setting the AMM fee substantially below the CEX cost can produce a large reduction in equilibrium TVL because LP fee income becomes insufficient to compensate for inventory and adverse-selection losses. By contrast, moderately exceeding the TVL-maximizing fee often produces a much smaller loss in liquidity supplied since the reduction in routed volume is partially offset by the higher revenue collected on the trades that remain. 

That said, a small TVL loss does not imply that fees above the TVL-maximizing level are economically innocuous. As \cref{fig:equilibrium.routing.execution} shows, higher fees reduce the share of fundamental demand routed to the AMM and generally increase aggregate fundamental-trader execution costs across the two venues. A fee may therefore remain close to optimal for equilibrium pool depth while producing less attractive outcomes for traders.

\subsection{Comparison and Takeaways}\label{sec:opt_fee_takeaways}

The two optimization exercises answer different questions, but they are related insofar as they reflect how attractive the pool is to liquidity providers. With liquidity held fixed, maximizing expected hedged PnL is equivalent to maximizing the expected return on installed LP capital. Under competitive entry, expected return is instead pinned to the outside option. Maximizing equilibrium TVL then identifies the fee that supports the greatest pool depth at that return.

When liquidity is fixed, the maximizing fee depends materially on volatility and the demand-to-liquidity ratio. In the benchmark region, a constant fee around $70$--$80\%$ of the CEX trading cost produces relatively modest regret over a broad set of displayed market conditions. This conclusion does not extend to extreme volatility, where every active fee produces negative expected PnL and the fixed-liquidity objective favors a prohibitive fee that suppresses trading.

With competitive entry, pool depth can adjust over time in response to market conditions. While this adaptation may happen over longer timescales, the expectation is that market forces inevitably shape the economic decisions of liquidity suppliers. By the scale invariance in \cref{sec:scale.invariance}, a proportional increase in demand expands equilibrium liquidity and TVL by the same proportion. On the other hand, the preceding analysis shows that a higher volatility sharply reduces the amount of capital that can be supported. We find that the TVL-maximizing fee remains close to the CEX trading cost over the displayed range and typically undercuts it only slightly. 

We emphasize again that equilibrium depth is not a complete measure of welfare. A fee near the TVL maximum can support substantial LP capital while capturing less fundamental flow and imposing higher execution costs on traders than a lower fee. Therefore optimality in equilibrium is a delicate question that depends on the objective of the protocol (or, more broadly, the fee setter). That said, both optimization exercises reflect the trade-off among revenue per trade, adverse selection, and fundamental trader participation. What is more, over the benchmark parameter region, both favor undercutting the CEX, but the PnL criterion prescribes a larger undercut than the equilibrium-TVL criterion.

\section{Empirical Analysis}\label{sec:empirics}

This section evaluates the empirical relevance of our stylized model and examines how fee design affects AMM performance in realistic settings. Our goal is twofold: first, to test whether the model, once calibrated to market data, can replicate key features of observed price and volume dynamics; second, to assess the financial implications of fee optimization in simulations and when applied to historical data. To this end, we calibrate the model parameters using minute-level data for the ETH/USDC pair over January 2025, then simulate trading under two fee regimes: the standard Uniswap v2 fee and the PnL-optimal fee at the observed pool size. We subsequently apply the competitive-entry condition in the calibrated environment and determine the fee that maximizes equilibrium TVL. The results show that implementing the PnL-optimal fee leads to narrower trading bands and higher execution rates for both arbitrageurs and fundamental traders. From a financial perspective, the AMM's cumulative PnL improves significantly under the optimal fee, particularly when inventory risk is actively managed. Under competitive entry, the calibration implies substantially greater equilibrium reserves. These results confirm that fee adjustments can meaningfully enhance AMM profitability and pool depth, and that our modeling framework provides a practical tool to guide fee setting based on empirical data.

\subsection{Data}

Our dataset consists of minute-by-minute records of reserve balances for the ETH/USDC Uniswap v2 pool, extracted via Dune Analytics\footnote{\url{https://api.dune.com}} for the month of January 2025. Using these reserve quantities, we compute the implied DEX price at each timestamp via the constant product formula. The pool applies a fixed fee of 30 basis points and begins the month with reserves of \$23{,}000{,}000 and 6{,}903 ETH. We also obtain minute-level swap volume data for liquidity takers (LTs) in both directions (ETH to USDC and USDC to ETH) over this period.

In parallel, we collect minute-level historical price data for the ETH/USDC pair from Binance, covering the same month.\footnote{Binance price data: \url{https://data.binance.vision/?prefix=data/spot/monthly/klines/ETHUSDC/1m/}.} This centralized exchange data serves as a reference price to benchmark and calibrate our model against real market conditions.

\subsection{Calibration}

We calibrate our simulation model to the market data described above for the ETH/USDC pair over the month of January 2025.\footnote{We have repeated the same calibration procedure on earlier periods -- particularly in 2020 and 2021, when Uniswap v2 enjoyed greater liquidity -- as well as February and March 2025, and obtained qualitatively similar results. We focus here on the month of January 2025.} 
The central object of interest is the log price ratio process
\[
r_t := \log\left(R_t\right),
\]
computed at each minute from historical data. We construct an empirical histogram of this process and seek to reproduce its distribution through simulation.

\begin{figure}[!htbp]
    \centering
    \includegraphics[width=0.9\textwidth]{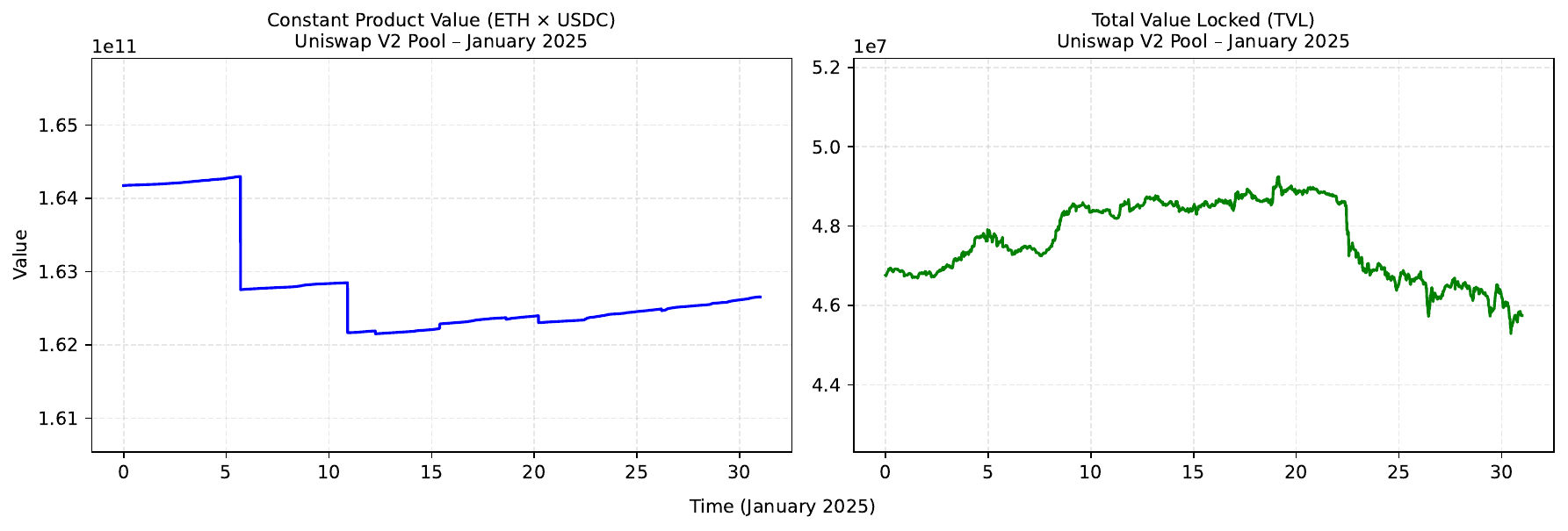}
    \caption{
Left: evolution of the constant product $X_tY_t$ over time in the Uniswap V2 ETH/USDC pool during January 2025. Right: corresponding total value locked (TVL) $X_t + S_t^1 Y_t = 2X_t$. 
}
    \label{fig:liq_jan25}
\end{figure}

For simplicity, we assume that $\Delta^B = |\Delta^S| =: \bar\Delta$, and that this rate is constant over time. Note that the AMM fee $\eta^1$ is publicly known. Specifically, we fix $\eta^1$ to the Uniswap v2 fee for the ETH/USDC pool, i.e.,
\[
\eta^1 = 30 \text{ bps}.
\]
Recall that for our model we assume that LPs are passive and precommit liquidity to the pool for the \emph{entire} trading horizon. We also treat fees as though they are disbursed directly to the LP. In v2 market data, liquidity is added and withdrawn, and fees contribute to pool liquidity. However, over the horizon we consider pool liquidity does not vary substantially (see Figure \ref{fig:liq_jan25}). Consequently, we fix the liquidity used for calibration to be $2.3\cdot 10^7\text{ USDC}$ and $6904 \text{ ETH}$, which roughly corresponds to the average values of the pool.

This allows us to reduce the calibration problem to three parameters:
\begin{itemize}
    \item the volatility $\sigma$ of the CEX price process $S^0$,
    \item the effective CEX transaction cost $\eta^0$,
    \item and the arrival rates $\bar \Delta$ of fundamental buyers and sellers, respectively.
\end{itemize}

\begin{figure}[!htbp]
    \centering
    \includegraphics[width=0.9\textwidth]{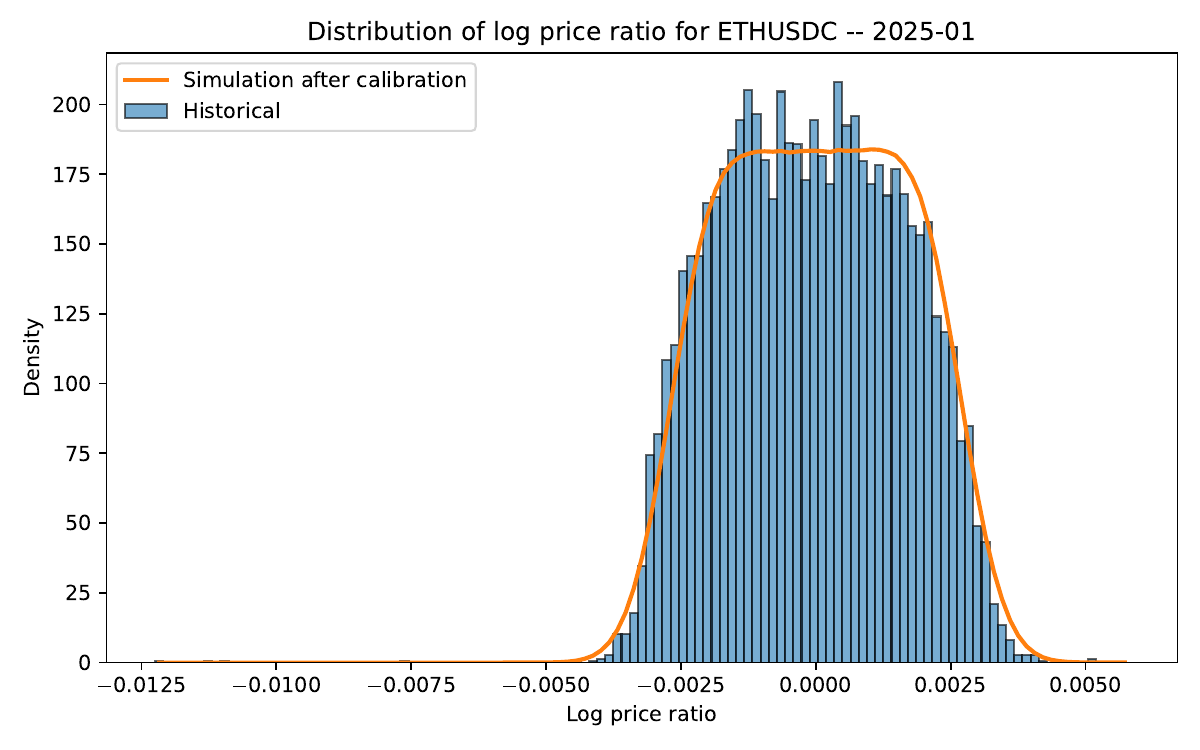}
    \caption{Comparison between the historical distribution of the log price ratio (bars) and the simulated distribution after calibration (solid line) for the ETH/USDC pair over January 2025. }
    \label{fig:calibration_hist_sim}
\end{figure}

Although historical volatility could be used for the value of $\sigma$, this leads to an over-estimation of the volatility due to clustering effects, and we rather rely here on a model-implied volatility that allows for a better fit to the price ratio. To estimate the parameters $(\sigma, \eta^0, \bar\Delta)$, we adopt an approach that targets the distribution of the log price ratio between the AMM and the CEX. This focus reflects our earlier finding that price ratio dynamics are central to determining LP profitability. 
Concretely, we minimize the $L^2$ distance $\left\| \hat{r}_t - r_t \right\|_2$, where $r_t$ is the empirical distribution of the log price ratio, computed from minute-level Binance (CEX) and Uniswap (AMM) prices. The simulated counterpart, $\hat{r}_t$, is generated from Monte Carlo simulations using the geometric Brownian motion dynamics \eqref{eqn:price.evolution} for the CEX and the stylized AMM trade process detailed in Section~\ref{sec:ecosystem}.
The minimization is carried out using a grid search implemented with the \texttt{Optuna} optimization library \cite{Optuna19}. This method produces consistent estimates for the volatility $\sigma$ and the effective transaction cost $\eta^0$, and accurately captures the empirical shape of the log-ratio distribution, as seen in Figure \ref{fig:calibration_hist_sim}. In particular, this demonstrates that despite its stylized nature, our model can produce parameter values that faithfully capture key features of real-world trading dynamics.
The resulting calibrated parameters are reported in Table~\ref{tab:calib}.

\begin{table}[h!]
    \centering
    \caption{Calibrated parameters with $\eta^1 = 30$ bps}
    \label{tab:calib}
    \begin{tabular}{lcc}
        \toprule
        Parameter & Value & Unit \\
        \midrule
        Buy/sell rate \(\bar\Delta \) & 19{,}068 & ETH/day \\
        CEX trading cost \(\eta^0\) & 6.58 & basis points (bps) \\
        Daily volatility \(\sigma\) & 2.60\% & $\text{day}^{-1/2}$ \\
        \bottomrule
    \end{tabular}
\end{table}

\medskip

Our model does not explicitly account for gas fees, which represent an additional cost for traders interacting with the AMM. As a result, the value obtained for the CEX trading cost $\eta^0$ in Table~\ref{tab:calib} is likely downward-biased. To partially correct for this, we repeat the calibration procedure while adjusting the AMM fee to $\eta^1 = 35$ bps, thereby approximating the additional gas cost incurred by DEX users. This yields the following estimates, presented in Table~\ref{tab:calib_35}, which provide a plausible range for $\eta^0$ that better reflects the all-in trading costs on both venues. Notably, the resulting value for $\eta^0$ is consistent with observed taker fees on Binance.

\begin{table}[h!]
    \centering
    \caption{Calibrated parameters with $\eta^1 = 35$ bps}
    \label{tab:calib_35}
    \begin{tabular}{lcc}
        \toprule
        Parameter & Value & Unit \\
        \midrule
        Buy/sell rate \(\bar\Delta \) & 10{,}338 & ETH/day \\
        CEX trading cost \(\eta^0\) & 11.78 & basis points (bps) \\
        Daily volatility \(\sigma\) & 2.60\% & $\text{day}^{-1/2}$ \\
        \bottomrule
    \end{tabular}
\end{table}

\medskip

The difference in the estimated arrival rate of fundamental traders $\bar\Delta$ across the two calibrations should not be overinterpreted. High levels of demand tend to produce very similar fits for the distribution of log price ratios when $\eta^1$ is much larger than $\eta^0$, making them statistically hard to distinguish. As such, our estimates are best understood as providing a plausible range of demand intensities consistent with the empirical price ratio distribution.

\begin{remark}
As a robustness check, we also performed the calibration by directly using the historical Binance price series for $S^0$, instead of simulating it as a geometric Brownian motion. In this setup, the volatility $\sigma$ is no longer a free parameter, and we calibrate only the effective CEX transaction cost $\eta^0$ and the net arrival rate $\bar\Delta$ of fundamental traders. We find that this alternative procedure yields similar ranges for both parameters (specifically, $\bar \Delta=11{,}684 \text{ ETH/day}$ and $\eta^0 = 9.87 \text{ bps}$), reinforcing the robustness of our baseline calibration and suggesting that the geometric Brownian motion assumption for $S^0$ does not materially affect the identification of key model features.
\end{remark}

\begin{remark}
While the proposed calibration successfully captures the microstructural dynamics of the AMM, we observe in simulation that the trading volume captured by the AMM is lower than its empirical counterpart (approximately 210 ETH/day versus 750 ETH/day in practice). One possible explanation is the presence of ``noise traders'' who interact with the AMM independently of relative prices or fees, for idiosyncratic or exogenous reasons. An alternative modeling route is thus to interpret the residual volume gap as arising from such non-strategic order flow. In this interpretation, matching the empirical trading volume would require adding roughly 540 ETH/day of noise-driven trades, which corresponds to less than 5\% of the fundamental volume estimated by our model. This addition has no meaningful impact on the qualitative results that follow, since optimal fee setting remains primarily driven by strategic, incentive-compatible order flow. For clarity of exposition, we omit noise traders from the main analysis, but develop the corresponding extension in Section~\ref{unconditional_flow}. We emphasize that it is expected that realized volume will differ since our model is conservative. Even if we ignore other motivations for trading with the AMM, our simulation logic assumes that traders have perfect knowledge of market prices and can always route optimally, which can fail to hold in practice.
\end{remark}

\subsection{Numerical Analysis}
\begin{figure}[!htbp]
    \centering
    \includegraphics[width=0.6\textwidth]{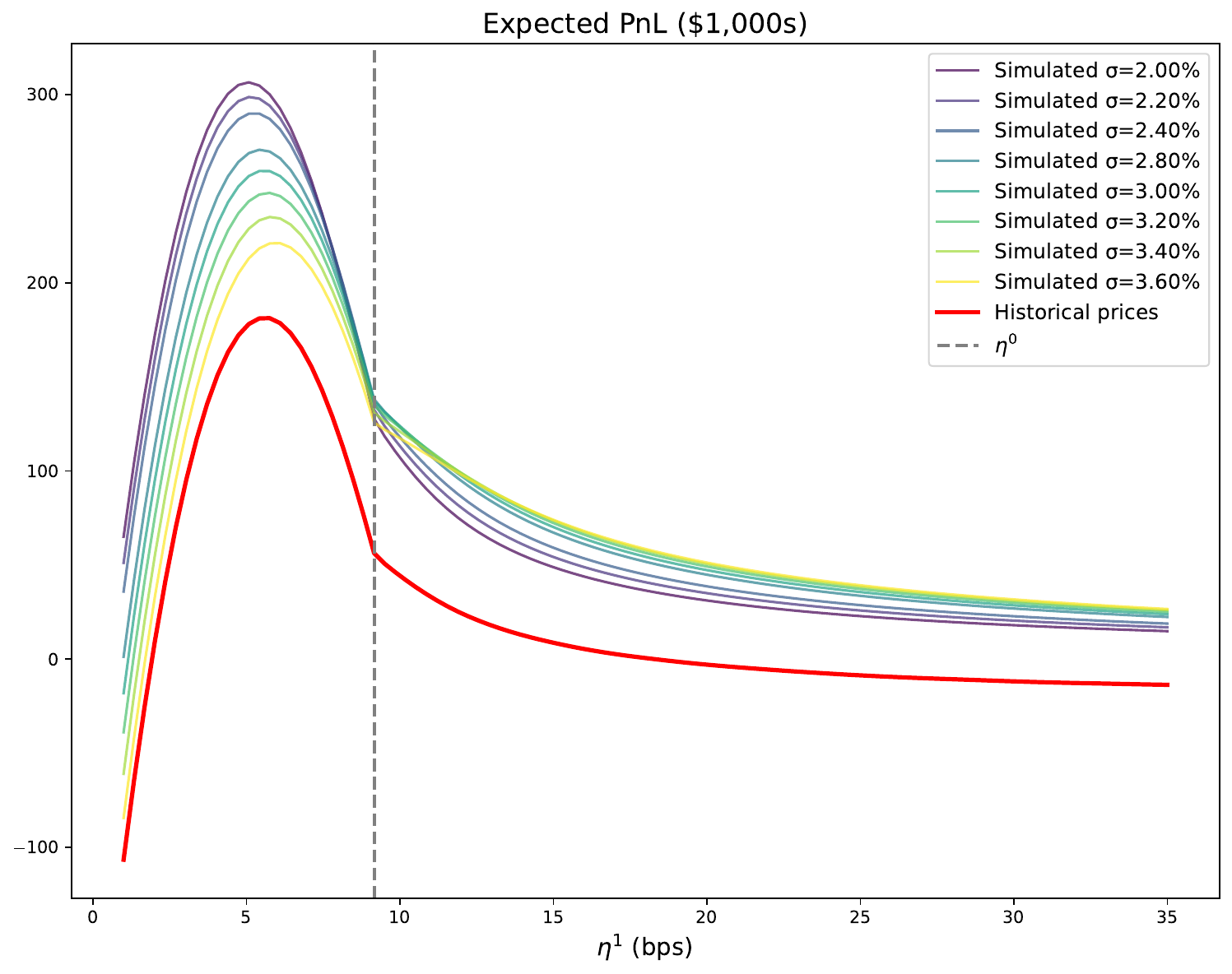}
    \caption{Expected PnL of liquidity providers as a function of the AMM fee $\eta^1$ using historical CEX prices (in red), compared to simulated prices under different volatility levels.}

    \label{fig:fee_hist_sim}
\end{figure}

We now perform a simulation based on the stylized model and the parameter values estimated in the previous subsection. For the CEX transaction cost, we set $\eta^0 = 9.18$ bps, which lies at the midpoint of the range identified through the two calibration procedures. Similarly, we set the fundamental demand rate to $\bar\Delta = 14{,}704$ ETH/day, representing an intermediate value between the two previously estimated demand levels.
The simulation is carried out on a minute-by-minute basis over an entire month. 
Figure~\ref{fig:fee_hist_sim} reports the expected PnL of liquidity providers as a function of the AMM fee~$\eta^1$, for various volatility levels in a neighborhood of the calibrated value. We also include a benchmark curve computed using historical CEX prices instead of simulated geometric Brownian motion paths, while keeping the same trade simulation procedure.

The results confirm that the optimal strategy consistently involves undercutting the CEX fee. Moreover, the maximizer of expected PnL remains stable across different volatility values, highlighting the robustness of our framework. Importantly, the fact that the optimal fee remains nearly unchanged when switching from simulated to historical prices suggests that our stylized price model suffices to capture the core tradeoffs involved in fee design. This provides strong support for using simulation-based approaches to optimize AMM parameters, even when abstracting away from complex real-world dynamics such as stochastic volatility and price jumps.

To assess the practical impact of fee design, we evaluate AMM performance on historical price data. Using the volatility estimated in calibration, we identified in the above analysis that the expected PnL is maximized around $\eta^1 = 5.42$ bps---slightly below the effective CEX fee $\eta^0$. As in the earlier analysis of Section \ref{sec:optimal.fees}, this value represents an optimal trade-off between competitiveness and profitability, encouraging volume without unduly sacrificing fee revenue.

To test the real-world relevance of this result, we run two simulations using historical minute-by-minute Binance data:

\begin{itemize}
    \item In the first, we set $\eta^1 = 30$ bps, matching the Uniswap v2 fee for ETH/USDC.
    \item In the second, we rerun the simulation with the optimal fee $\eta^1 = 5.42$ bps.
\end{itemize}

\begin{figure}[!htbp]
    \centering
    \includegraphics[width=0.9\textwidth]{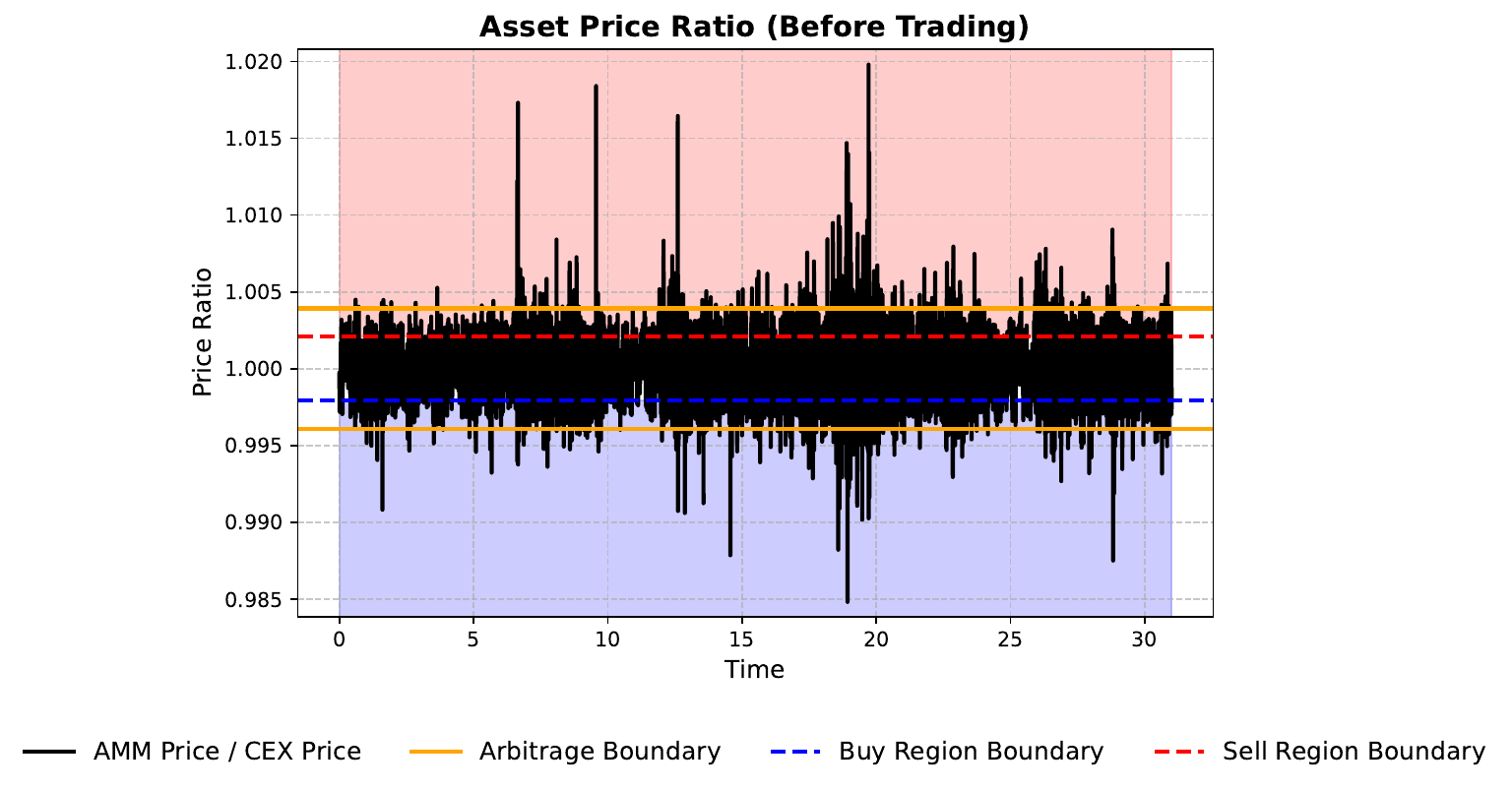}
    \caption{Asset price ratio between AMM and CEX prices before trading, based on historical Binance data, with arbitrage and trading region boundaries indicated.}
    \label{fig:price_ratio_hist}
\end{figure}

\begin{figure}[!htbp]
    \centering
    \includegraphics[width=0.9\textwidth]{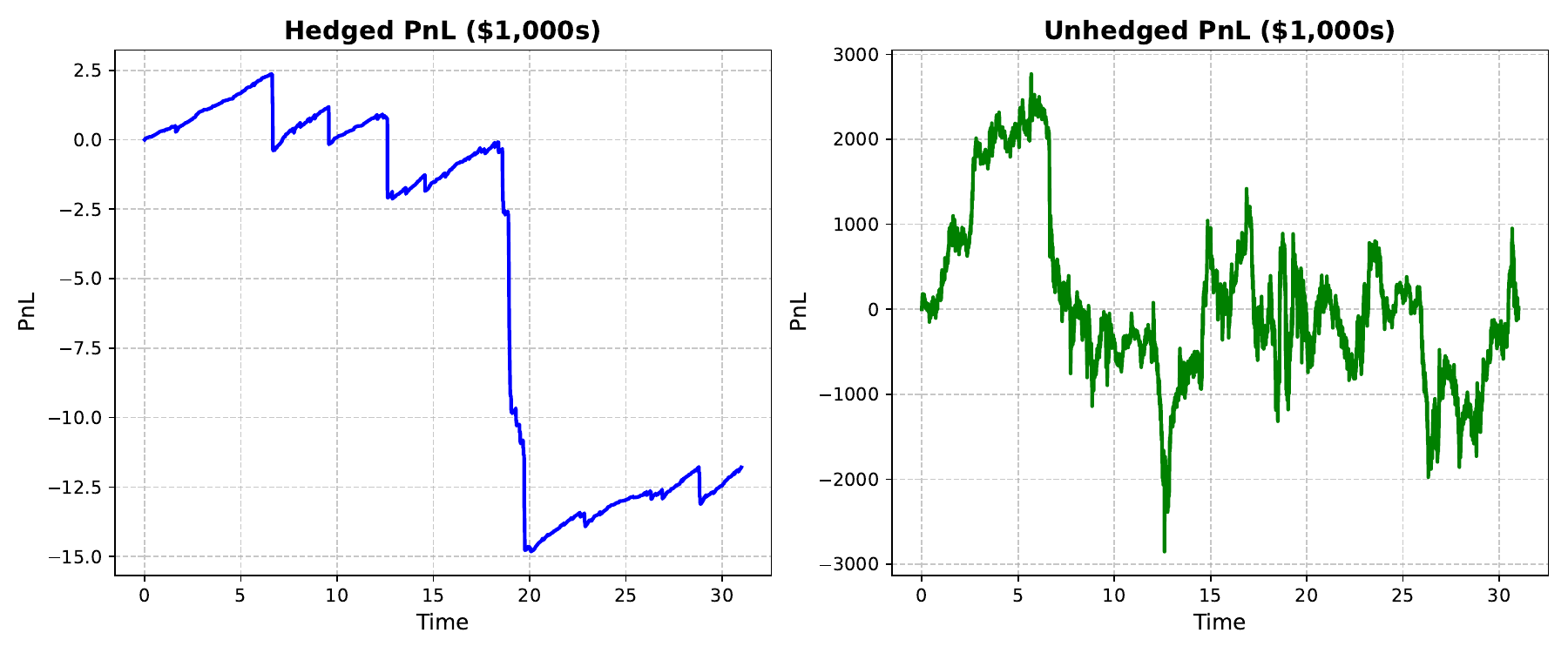}
    \caption{Cumulative profit and loss (PnL) for hedged (left) and unhedged (right) AMM strategies over the simulation period using historical data, expressed in thousands of USD, with fee $\eta^1 = 30$ bps.}
    \label{fig:pnl_hist}
\end{figure}

Figures~\ref{fig:price_ratio_hist} and~\ref{fig:pnl_hist} display the results under the standard fee $\eta^1 = 30$ bps. We observe wide arbitrage bands and sparse trading, reflecting the deterrent effect of high transaction costs. The resulting final PnL, both for the hedged and unhedged strategies, is negative due to limited flow and fewer opportunities to charge trading fees. Echoing the findings of Section~\ref{sec:optimal.fees}, PnL deteriorates markedly during the sharp increase in volatility observed between days 18 and~20 in Figure~\ref{fig:price_ratio_hist}.

\begin{figure}[!htbp]
    \centering
    \includegraphics[width=0.9\textwidth]{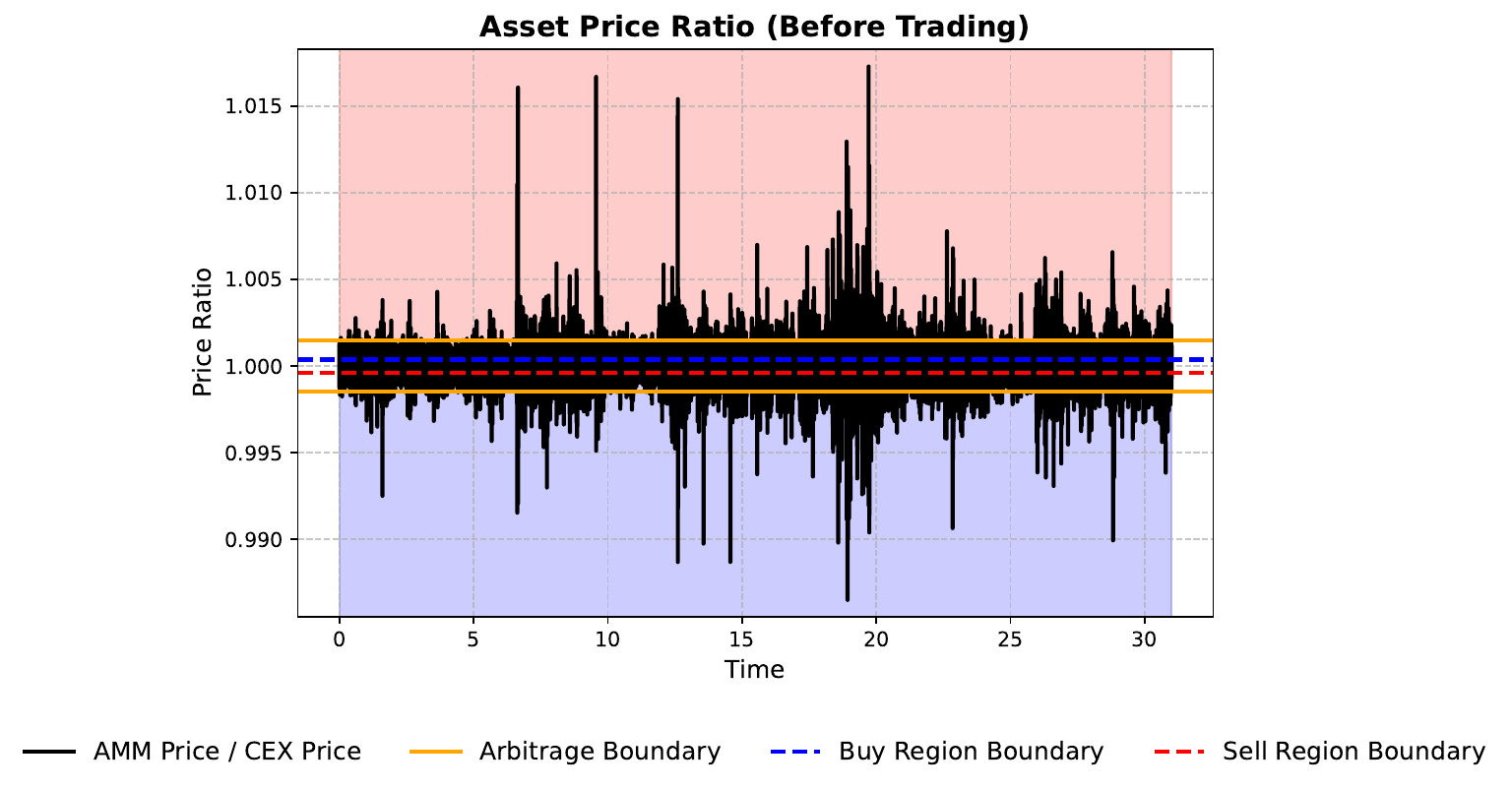}
    \caption{Asset price ratio between AMM and CEX prices before trading with optimal DEX fee $\eta^1 = 5.42$ bps, based on historical Binance data.}
    \label{fig:price_ratio_hist_opt}
\end{figure}

\begin{figure}[!htbp]
    \centering
    \includegraphics[width=0.9\textwidth]{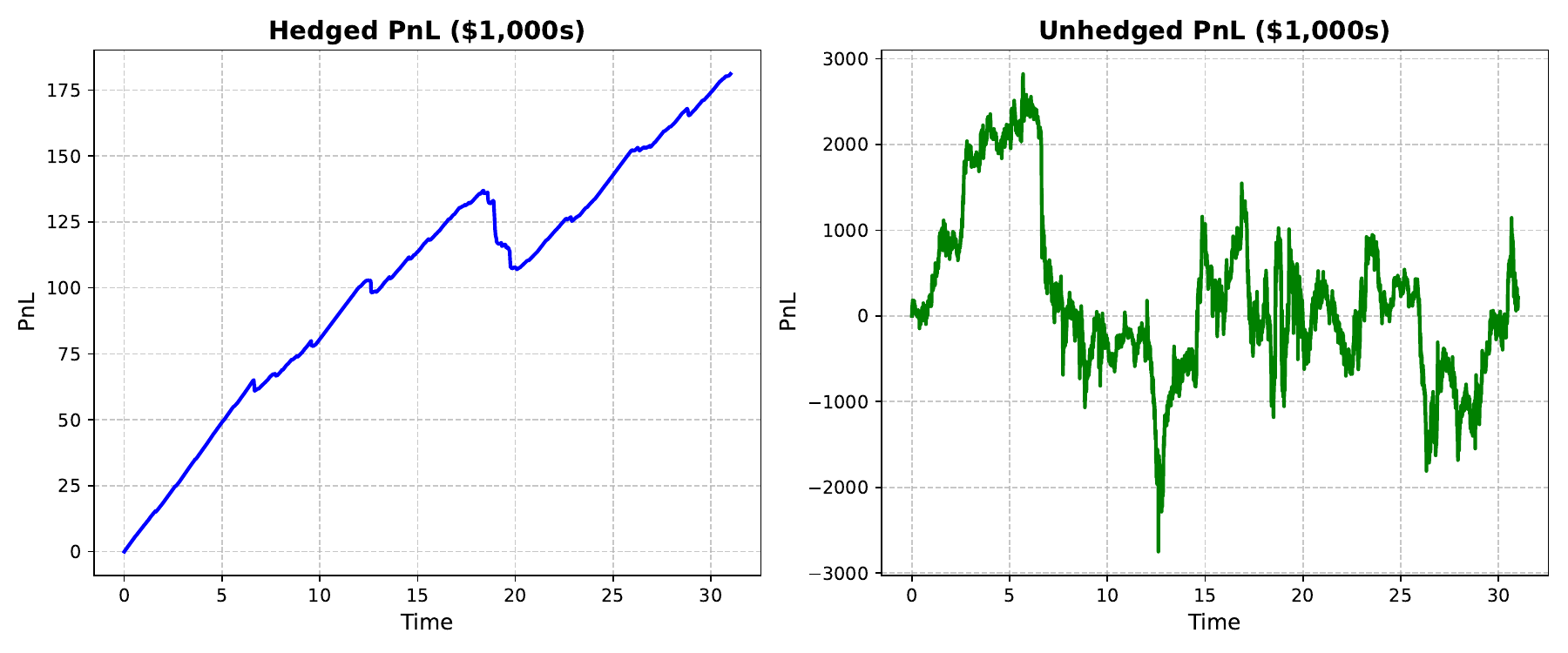}
    \caption{Cumulative profit and loss (PnL) for hedged (left) and unhedged (right) AMM strategies over the simulation period with optimal fee $\eta^1 = 5.42$ bps using historical data.}
    \label{fig:pnl_hist_opt}
\end{figure}

Next, we rerun the same experiment with the optimal fee $\eta^1 = 5.42$ bps. The resulting dynamics are shown in Figures~\ref{fig:price_ratio_hist_opt} and~\ref{fig:pnl_hist_opt}. With lower fees, we observe a noticeable narrowing of both arbitrage and trading regions. This leads to more frequent arbitrage interventions and significantly increased volume from fundamental traders. Nonetheless, PnL still declines during the volatility spike on days 18--20 which marks a period when the LP would have benefited from levying a much larger fee.\footnote{This coincides with the days leading up to the U.S.\ presidential inauguration (January 20, 2025).}

Overall, the hedged PnL improves substantially and becomes positive under the optimal fee, while the unhedged PnL remains largely unchanged---being mostly driven by the underlying price path. This highlights that fee optimization is most effective when inventory risk is actively managed. By lowering its fees, the AMM invites more arbitrage activity, but more importantly it unlocks a much larger share of profitable flow from fundamental traders. The result is a robust increase in PnL.

These findings reinforce a key insight from our earlier analysis: excessive protection against arbitrage can come at the expense of participation from profitable, non-arbitrage trading flows. Striking the right balance is therefore essential. Rather than minimizing arbitrage outright, effective fee design should aim to maximize total value captured by liquidity providers, even if that requires tolerating some arbitrage as part of a broader, healthy market ecosystem. Our empirical results suggest that modestly undercutting the effective CEX fee during stable conditions, combined with significantly increasing fees during periods of heightened volatility, provides a practical path to improving LP performance. This principle aligns with the design choices seen in recent protocols such as Uniswap v3, which introduced multiple fee tiers to reflect differing asset profiles, and Uniswap v4, which further enables dynamic fee customization via hooks---underscoring the growing relevance
of adaptive fee mechanisms in AMM design.

The experiments above are conducted at fixed initial reserves. Hence, the fee $\eta^1=5.42$ bps should be interpreted as maximizing LP profitability conditional on the observed pool size, rather than as an equilibrium fee when liquidity provision is endogenous. A positive hedged PnL may attract additional liquidity to the pool, at least over longer time horizons. This ultimately dilutes the return earned per unit of liquidity and pushes hedged PnL towards the opportunity cost of capital (assumed to be approximately zero here over short horizons).

To explore this equilibrium adjustment numerically, we first fix the fee at the value $\eta^1=5.42$ bps obtained from the fixed-reserve optimization, and search for a reserve level at which the final hedged PnL over the historical sample is zero. This is reached when the initial reserves are increased by a factor of approximately $4.95$. Thus, within this numerical experiment, the fee that maximizes LP PnL at the observed liquidity level would be compatible with a substantially larger pool once liquidity provision is allowed to adjust.

We may also extend this exercise by allowing the fee itself to vary. For each candidate fee $\eta^1$, we compute the associated equilibrium reserve level and search for the fee that maximizes the equilibrium TVL.  Under our calibrated environment, the liquidity-maximizing fee is found to be $\eta^{1}=6.75$ bps,
and the corresponding equilibrium reserve level is approximately $5.69$ times larger than the initial reserve level. Thus, once liquidity provision is allowed to adjust endogenously, the fee that supports the largest equilibrium TVL remains below the calibrated CEX fee, but differs from the one that maximizes LP PnL when reserves are held fixed.

It should be emphasized that these results should be interpreted as an equilibrium benchmark conditional on our calibrated environment. In particular, the historical CEX price path, the fundamental trading intensities, and the effective CEX transaction cost are kept fixed as liquidity and fees vary, and the equilibrium condition is imposed on the final hedged PnL over a finite historical sample. In this sense we interpret the average performance over the historical window as a rough approximation to short-run expectations of market behavior. In practice, the realized equilibrium environment may differ and consequently, this does not provide a complete model of how the market will respond. Nonetheless, it does suggest that there is a substantial gap between the current liquidity supplied and the amount that could be supported by lowering the pool fees.

\section{Extensions to the Model Design}\label{sec:extensions}

In this section we briefly address several possible extensions of our baseline model. We fix the parameters specified at the beginning of \cref{sec:profitability} as a reference point and assess the impact of each extension in isolation. %
The figures below hold initial liquidity fixed and therefore isolate the effect of each extension on the return to installed LP capital. As we will see, our fixed-liquidity findings are largely robust to perturbations in the model design. The corresponding competitive-entry calculations, together with sensitivity to the outside return, are reported in \cref{app:equilibrium.extensions}. Trading frequency, drift, and arbitrageur costs preserve the broad baseline fee pattern, whereas outside returns, hedging costs, fee-insensitive noise flow, and strong mean reversion have more pronounced effects on equilibrium depth.

\subsection{Trading Frequency}

As the number of trading periods $N$ increases---holding total daily fundamental demand constant---the market effectively distributes the same aggregate flow across a greater number of smaller trades. As a result, higher trading frequencies allow the AMM to better absorb fundamental order flow, as each individual trade imposes only a small price impact. This allows the AMM to remain competitive for a larger fraction of the total volume and in turn, enhances LP profitability by increasing the volume captured at each fee level.

\begin{figure}[!htbp]
    \centering
    \includegraphics[width=0.49\textwidth]{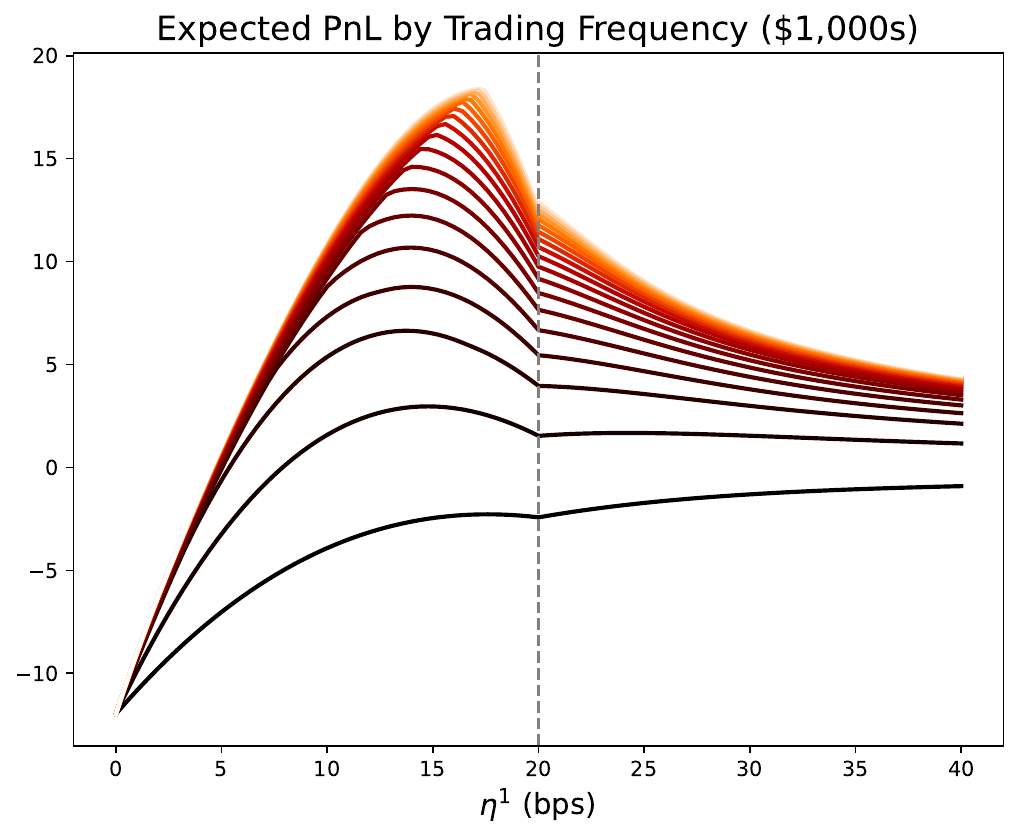}
    \includegraphics[width=0.49\textwidth]{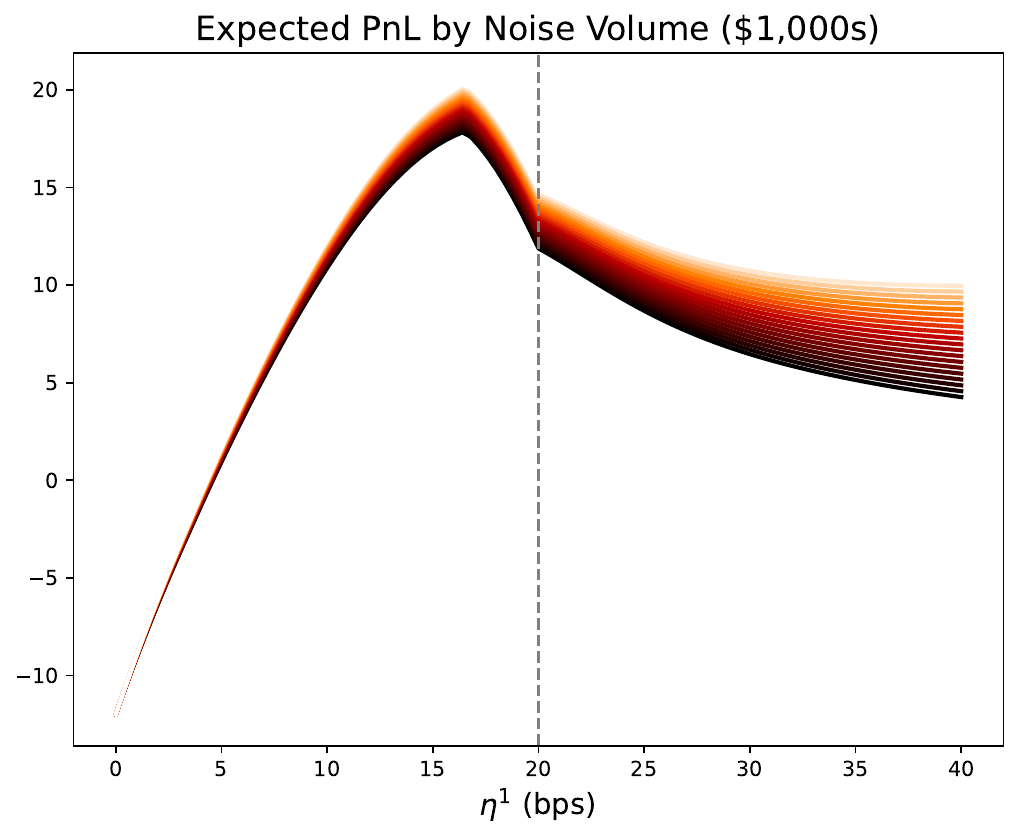}
    \caption{(Left) Expected PnL of the LP as the trading frequency $N$ increases from $100$ to $2000$. (Right) Expected PnL as the noise volume increases from $0$ to $500$ units of $Y$ per day. In both panels the increase in the parameter (i.e., $N$, noise volume) is represented through a color gradient from dark to light.}
    \label{fig:expected.pnl.trading.freq}
\end{figure}

This overall performance improvement driven by the trading frequency is illustrated in the left panel of Figure \ref{fig:expected.pnl.trading.freq}. At sufficiently high frequencies, we heuristically observe convergence toward a continuous-time limit. A rigorous analytic treatment of this limiting regime using stochastic process theory will be reported in future work.

\subsection{Noise Traders} \label{unconditional_flow}

As noted in the Introduction, there likely exist market participants who trade on the DEX even in situations where informed fundamental traders choose not to. We refer to these participants as \emph{noise traders}, though their behavior can stem from a range of plausible exogenous motives. For example, some traders may prefer the DEX due to its reduced compliance burden, seeking to avoid know-your-customer (KYC) requirements on centralized exchanges for privacy or convenience reasons. Moreover, it is likely that many traders lack perfect information about cross-venue pricing, causing them to frequently route orders in a manner that appears suboptimal ex-post.

There are various ways to model such participants. For simplicity, we consider noise trader flow to be unconditional and insensitive to any reasonable price discrepancies across venues. Specifically, we assume that in each period, a fixed volume of noise trades arrives on the DEX, divided evenly between buy and sell orders. These trades occur after fundamental traders transact, and their symmetric nature ensures they exert no direct influence on the price ratio distribution. Their sole effect is to increase the fee revenue earned by the LP.

Naturally, the assumption of unconditional noise flow is valid only over a restricted fee range. As $\eta^1 \to \infty$, execution costs on the DEX would preclude all participation, including that of noise traders. Absent such a constraint, rational traders would abstain from interacting with the DEX, while noise-driven volume would yield unbounded profits to the LP. Accordingly, we limit noise flow to fee levels that are economically plausible.

The right panel of Figure~\ref{fig:expected.pnl.trading.freq} illustrates the impact of introducing noise trader flow up to $5\%$ of the baseline total unsigned fundamental demand (i.e., $500$ units of $Y$ per day). Even when this flow is completely insensitive to $\eta^1$, the qualitative conclusions from Section~\ref{sec:optimal.fees} remain intact for all reasonable fee levels. In particular, while noise flow raises LP profitability across the board, the optimal AMM fee continues to lie strictly below the fee charged on the CEX.

\subsection{Alternative Fees}\label{sec:extensions_fees}

We can extend the baseline framework to incorporate more realistic fee structures by introducing two additional parameters: an arbitrageur-specific fee $\eta^A$ and a hedging fee $\eta^H$ applied to the LP’s CEX trades. These fees capture a more nuanced distinction between market participants. For one, arbitrageurs are often sophisticated actors who may face lower effective fees than other traders. Likewise, LPs typically incur strictly positive costs when rebalancing inventory that \emph{differ} from those faced by fundamental traders. While perhaps non-zero, the hedging fee $\eta^H$, which represents the LP’s effective execution cost on the CEX, still ought to be lower than the standard fee $\eta^0$ due to internal netting, tiered fee schedules, or other preferential arrangements.

\begin{figure}[!htbp]
    \centering
    \includegraphics[width=0.49\textwidth]{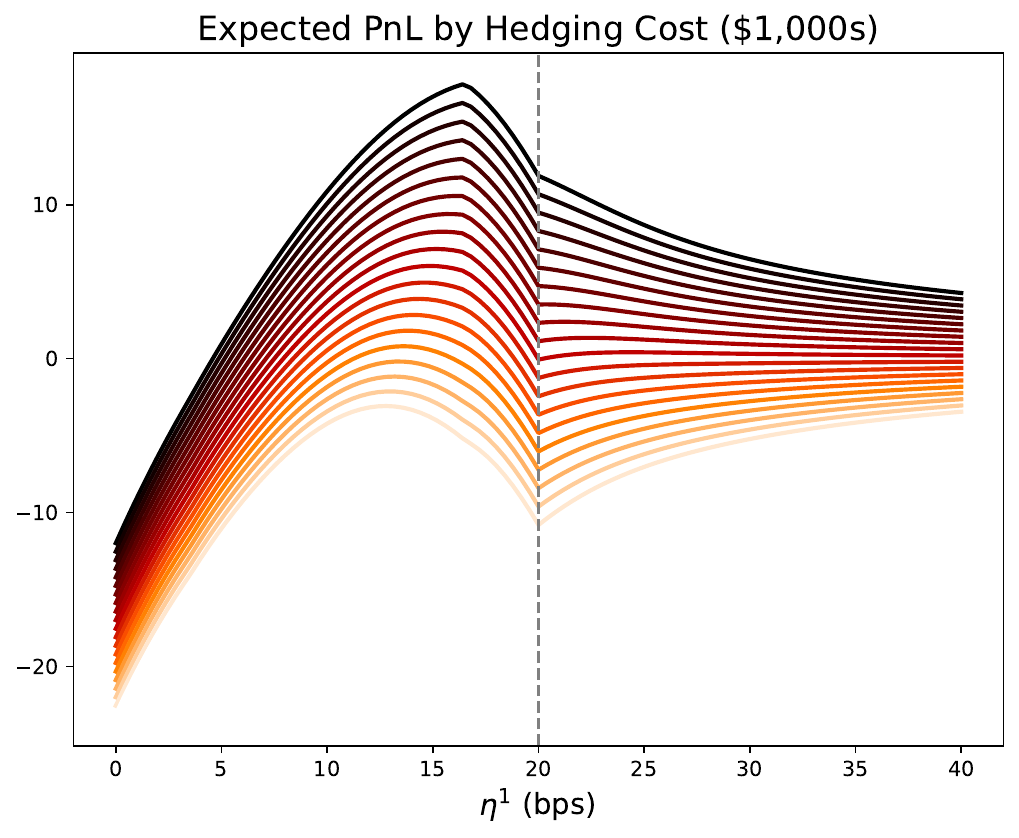}
    \includegraphics[width=0.49\textwidth]{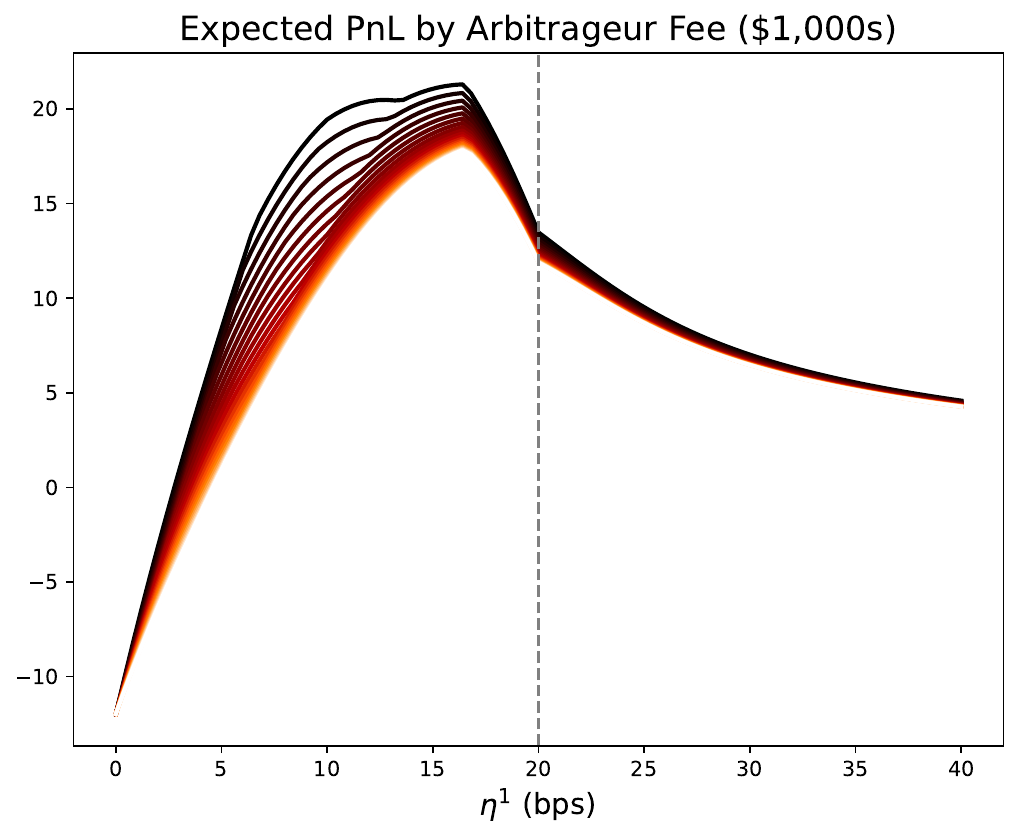}
    \caption{(Left) Expected PnL as the hedging cost $\eta^H$ increases from $0$ to $\eta^0$. (Right) Expected PnL as the arbitrageur fee $\eta^A$ increases from~$0$ to~$\eta^0$. All of the curves are illustrated on a color gradient from dark to light as $\eta^H$ and $\eta^A$ increase.}
    \label{fig:expected.pnl.trading.fees}
\end{figure}

In view of the tiers for the nominal fees that CEXs levy on their customers' trades based on monthly trading volume, one could also consider a distribution of fundamental trader fees instead of a single value. However, modeling this introduces ambiguity around how to specify trader heterogeneity and so, we do not pursue this latter extension here in detail. In effect, varying fundamental trader fees simply alters the decision thresholds for routing volume to the DEX, and is partly covered by our discussion. Specifically, fundamental traders with a high CEX fee would route almost all their flow to the DEX, making them similar to noise traders (see Section~\ref{unconditional_flow}), whereas fundamental traders with a low CEX fee would route almost all their flow to the CEX and hence become inconsequential for our analysis. In summary, having a spectrum of fundamental trader fees is similar to reducing fundamental demand and adding noise traders.

The introduction of a hedging fee $\eta^H$ modifies the analysis of Section \ref{sec:ecosystem} by increasing the LP's effective cost of inventory adjustments on the CEX. Accounting for this fee, the marginal change in the LP's hedged PnL for a small buy trade ($\Delta > 0$) becomes
\[
\text{Change in Hedged PnL} \approx \Delta \cdot S_t^0 \left[ (1 + \eta^1) R_t - (1 + \eta^H) \right],
\]
and for a small sell trade ($\Delta < 0$),
\[
\text{Change in Hedged PnL} \approx \Delta \cdot S_t^0 \left[ (1 - \eta^1) R_t - (1 - \eta^H) \right].
\]
The updated profitability thresholds are then given by
\[
R_t \geq \frac{1 + \eta^H}{1 + \eta^1} \quad \text{(buy trades)}, \qquad R_t \leq \frac{1 - \eta^H}{1 - \eta^1} \quad \text{(sell trades)}
\]
with net effect being that $\eta^H$ compresses the LP's profit region. As illustrated in Figure \ref{fig:expected.pnl.trading.fees}, since this addition does not change the ratio distribution or volume traded, the additional cost can only decrease the profitability of the LP.

On the other hand, introducing a lower arbitrageur fee $\eta^A<\eta^0$ pushes the arbitrage boundaries \emph{inwards} towards the profit region. In the extreme case $\eta^A = 0$ and $\eta^H = 0$, the arbitrage and profit boundaries coincide. Thus, in this regime arbitrageurs reflect the price ratio into the region where trading is profitable for the LP. As demonstrated in Figure \ref{fig:expected.pnl.trading.fees}, this general intuition persists for strictly positive values of $\eta^A$, and LP profitability improves as $\eta^A$ decreases. We collect these observations about the changes to the trading regions of Table \ref{tab:thresholds} in the updated Table \ref{tab:gen.thresholds}.

\begin{table}[ht]
\centering
\caption{Generalized Trading and Profitability Thresholds}
\label{tab:gen.thresholds}
\begin{tabular}{@{}lcc@{}}
\toprule
 & \textbf{Trade Condition} & \textbf{Profit Condition} \\ 
\midrule
\textbf{Fundamental Buyers} & 
$R_t \le \dfrac{1 + \eta^0}{1 + \eta^1}$ & 
$R_t \ge \dfrac{1+\eta^H}{1 + \eta^1}$ \\ 

\textbf{Fundamental Sellers} & 
$R_t \ge \dfrac{1 - \eta^0}{1 - \eta^1}$ & 
$R_t \le \dfrac{1-\eta^H}{1 - \eta^1}$ \\ 

\textbf{Arbitrage Buyers} & $R_t\leq \dfrac{1-\eta^A}{1+\eta^1}$ & N/A\\

\textbf{Arbitrage Sellers} & $R_t\geq \dfrac{1+\eta^A}{1-\eta^1}$ &  N/A\\
\bottomrule
\end{tabular}
\end{table}

\subsection{Alternative Price Dynamics}

\begin{figure}[!htbp]
    \centering
    \includegraphics[width=0.49\textwidth]{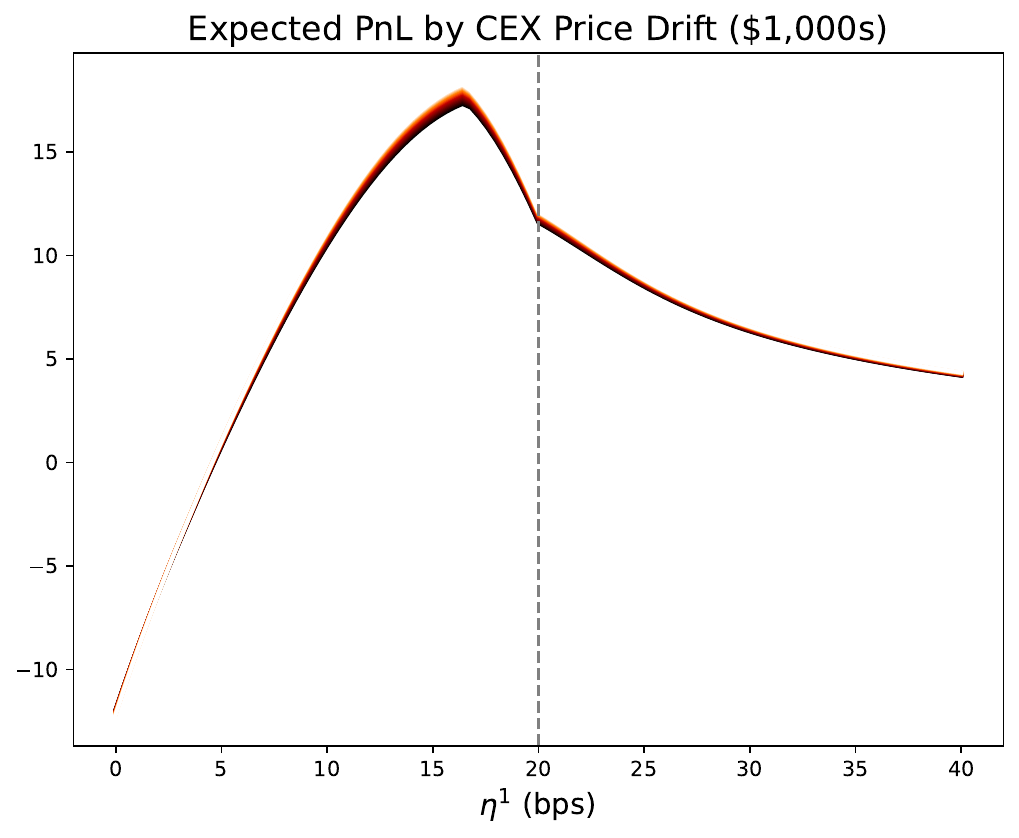}
    \includegraphics[width=0.49\textwidth]{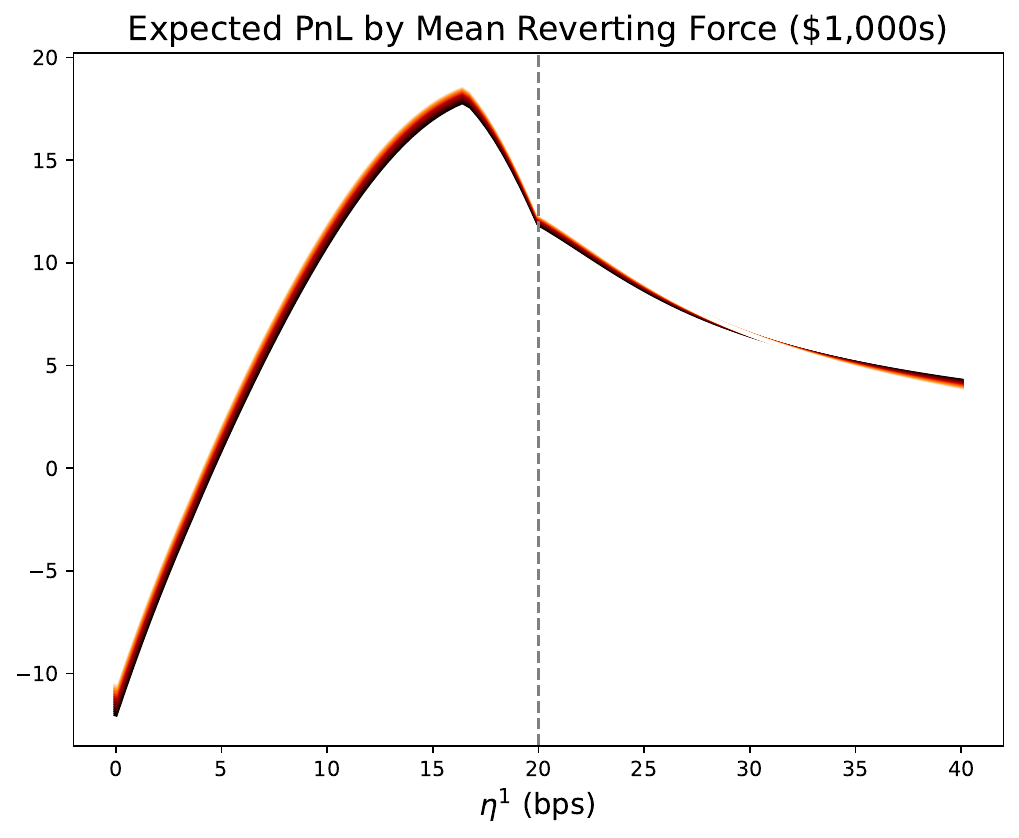}
    \caption{(Left) Expected PnL for drift $\mu\in[-0.04,0.04]$. (Right) Expected PnL for mean reverting force $\kappa\in[0,100]$. In both cases the PnL curves are shown on a color gradient from dark to light as $\mu$ and $\kappa$ increase. }
    \label{fig:expected.pnl.diff.dynamics}
\end{figure}

Building on the empirical analysis of Section \ref{sec:empirics}, we can investigate the influence of adding additional characteristics to the price series dynamics. Figure \ref{fig:expected.pnl.diff.dynamics} illustrates the sensitivity of the LP's PnL to prices that exhibit a trend (left) and mean reverting dynamics (right).

We implement the trend by augmenting the reference price dynamics in \eqref{eqn:price.evolution} to include a drift parameter $\mu\in\mathbb{R}$,
\begin{equation}\label{eqn:price.evolution.drift}
    \log(S_t^0/S_0^0) = \big(\mu-\tfrac{1}{2} \sigma^2\big) t + \sigma W_t, \ \ \ t\geq0.
\end{equation}
Similarly, the mean reverting dynamics are implemented by modeling $S^0_t$ as the exponential of an Ornstein-Uhlenbeck (OU) process,
\begin{equation}\label{eqn:price.evolution.mr} \log(S_t^0/S_0^0) = \int_0^t \kappa (\theta - \log(S_s^0)) \, ds + \sigma W_t,  \ \ \ t\geq0.
\end{equation}
Here, $\kappa>0$ is the mean reverting force and $\theta \in \mathbb{R}$ is the reference level. For Figure \ref{fig:expected.pnl.diff.dynamics} we fix $\theta = \log(S_0^0) = \log(3{,}000)$.

We see only minor variations of the expected PnL surface for reasonable choices of the new parameters $\mu$ and $\kappa$. Near the optimal fee, the expected (hedged) PnL appears to improve modestly as the trend goes from $\mu=-4\%$ to $\mu=4\%$. 
We also see a slight improvement to profitability for fees $\eta^1<\eta^0$ when introducing a mean reverting force.\footnote{We note that when the price process is not a martingale, the hedging strategy exhibits a non-zero expected PnL. In Figure~\ref{fig:expected.pnl.diff.dynamics}, we observe a cancellation between two effects: the impact of price dynamics on the unhedged PnL and their effect on the performance of the hedging portfolio. For example, when the price of asset $Y$ increases, both the pool and the hedging portfolio experience gains since they are long $Y$, and these gains partially offset each other. While an LP with perfect knowledge of the price dynamics might opt not to hedge, we do not assume such clairvoyance. To be consistent with our main analysis we focus on the baseline case in which the LP hedges by default.} 
Consistent with the analysis of Section \ref{sec:empirics}, we conclude that the general form of the PnL surface and the overall findings of Section \ref{sec:optimal.fees} for fixed-liquidity are generally robust to changes in the price dynamics. Importantly, the existence of a maximum in $\eta^1$ below the (effective) CEX fee persists in this setting.

\section{Conclusion}\label{sec:conclusions}

Using a parsimonious yet flexible reduced-form model of CEX--DEX interaction, we study passive liquidity provision to a CPMM. At the center of our analysis is the expected PnL of a representative liquidity provider, seen as a function of market conditions including volatility, demand, liquidity, and fees. We highlight the mechanisms underlying these dependencies through extensive comparative statics. Our simulation can be calibrated to market data, reproducing realistic price ratio distributions and yielding quantitative, actionable insights. We investigate in particular the AMM fee that maximizes expected LP PnL and its relation to the CEX trading cost. We further use the same PnL function to endogenize liquidity through competitive LP entry and study the fee that maximizes TVL. 

Our results closely link the AMM fee level to the trading cost on the CEX and suggest that fees should be competitive with those costs during business-as-usual periods, in particular, potentially lower than the 30 bps fee that was implemented in Uniswap~v2 and is still often considered a default for ETH/USDC and comparable pairs. This aligns with the recent popularity of 5 bps Uniswap~v3 and~v4 pools.\footnote{For ETH/USDC, 5 bps far dominates 30 bps on Uniswap~v4 as of June 30, 2025, with \$130.6M vs.\ \$5.3M total value locked. Uniswap~v3 has \$94M total value locked on 5~bps vs.\ \$51M on 30 bps. Source: \url{https://app.uniswap.org/explore/pools/ethereum}} In our model, the PnL-optimal fee remains remarkably stable throughout normal market conditions. However, our results also suggest that passive liquidity providers need protection during periods of high volatility, such as a high fee that discourages adverse selection on the AMM. Complementary to theoretical studies in toy models, our insights are directly implementable in practice. Specifically, Uniswap~v4 hooks could be used to implement a threshold-type dynamic AMM fee which takes two possible values, one that competes with CEX trading costs and a higher rate designed to limit adverse selection in periods of extreme volatility.

\section*{Acknowledgements}

\ifthenelse{\boolean{bl_anonymize_for_subm}}
{
(Removed for anonymization.)
}
{
The authors thank Patrick Chang, Fay{\c{c}}al Drissi, Ruizhe Jia, Roger Lee, Xin Wan, and numerous conference participants at the Fields Institute and the University of Chicago for helpful discussions. The authors also acknowledge support from a Columbia University CFDT Research Grant and the use of computing resources from Columbia University's Shared Research Computing Facility project, which is supported by NIH Research Facility Improvement Grant 1G20RR030893-01, and associated funds from the New York State Empire State Development, Division of Science Technology and Innovation (NYSTAR) Contract C090171, both awarded April 15, 2010.  S.~Campbell is supported in part by an NSERC Fellowship (PDF‑599675-2025), and 
M.~Nutz is supported in part by NSF Grants DMS-2606731, DMS‑2407074, and DMS‑2106056. J. Milionis's research is supported in part by NSF awards CNS-2212745, CCF-2332922, CCF-2212233, DMS-2134059, and CCF-1763970, by an Onassis Foundation Scholarship, and an A.G.\ Leventis educational grant.
}

\section*{Disclosures}

The first and fourth authors are partly supported by a research grant funded by DeFi companies. The third author is currently working with blockchain companies not directly involved in AMMs, advising companies on marketplace and incentive design, and has worked with companies involved in creating automated market making protocols, including some referenced in this work. The second author declares that they do not have any competing interests. Notwithstanding, the ideas and opinions expressed herein are those of the authors, rather than of any of these companies or their affiliates, and should not be construed as or relied upon in any manner as investment, legal, tax, or other advice.

\begingroup
\sloppy
\printbibliography
\endgroup

\crefalias{section}{appendix}
\appendix

\section{Pseudocode for Simulation}\label{se:simulation-algorithm}

\begin{algorithm}[H]
\caption{Pseudocode for Simulating Market Dynamics}
\KwIn{Initial reserves $(X_0, Y_0)$, CEX prices $S^0$, fees $\eta^0$, $\eta^1$, demand sizes $\Delta^B$, $\Delta^S$, time horizon $T$, time steps $N$, market instances $M$}
\KwOut{Reserve paths $(X_t, Y_t)$, DEX prices $S^1_t$, fee revenues, hedging values $H_t$}
\vspace{0.5em}

\textbf{Initialize market instances and trading flags:} \\
\For{each price path $j = 1,\dots,M$}{
    Set reserves: $(X_0^j, Y_0^j)$ \;
    Set hedge: $H_0^j \leftarrow X_0^j + Y_0^j S_0^{0,j}$ \;
    \For{each time step $i = 1,\dots,N$}{
        Randomly assign $\texttt{buyer\_first}_{i,j} \in \{\texttt{True}, \texttt{False}\}$ \;
        Set $\texttt{seller\_first}_{i,j} = \neg \texttt{buyer\_first}_{i,j}$ \;
    }
}

\textbf{Simulation loop:}\\

\For{each time step $i = 1,\dots,N$}{
    \For{each price path $j = 1,\dots,M$}{
        Update hedge portfolio value: $H_i^j \leftarrow H_{i-1}^j + Y_{i-1}^j \cdot (S_i^{0,j} - S_{i-1}^{0,j})$ \;

        \textbf{Arbitrage step:}
        \\
        \If{arbitrage opportunity exists}{
            Compute arbitrage volume $\Delta^A_{i,j}$ \;
            Execute arbitrage: update $(X_i^j, Y_i^j, S_i^{1,j})$ and record fees \;
        }

        \textbf{Fundamental trading step:}
        \\
        \If{$\texttt{buyer\_first}_{i,j}$}{
            Compute buy volume on DEX: $\Delta^{B,DEX}_{i,j}$ \;
            Execute buy trade: update $(X_i^j, Y_i^j, S_i^{1,j})$ and record fees \;
            Compute sell volume on DEX: $\Delta^{S,DEX}_{i,j}$ \;
            Execute sell trade: update $(X_i^j, Y_i^j, S_i^{1,j})$ and record fees \;
        }
        \ElseIf{$\texttt{seller\_first}_{i,j}$}{
            Compute sell volume on DEX: $\Delta^{S,DEX}_{i,j}$ \;
            Execute sell trade: update $(X_i^j, Y_i^j, S_i^{1,j})$ and record fees \;
            Compute buy volume on DEX: $\Delta^{B,DEX}_{i,j}$ \;
            Execute buy trade: update $(X_i^j, Y_i^j, S_i^{1,j})$ and record fees \;
        }
    }
}
\end{algorithm}

\newpage 

\section{Summary Statistics for Ratio Histograms}\label{se:summary.stats}

\begin{table}[h]
\caption{Summary statistics for varying $\eta^1$ in Figure \ref{fig:histograms}}
\label{tab:stats_by_eta1}
\begin{tabular}{lrrrrr}
\toprule
$\eta^1$ (bps) & 5 & 10 & 15 & 20 & 25 \\
\midrule
$\mathbb{E}$[profit] (\$) & 946.51 & 10881.14 & 17031.02 & 11877.55 & 8682.79 \\
$\mathbb{E}$[fees] (\$) & 12937.55 & 22872.52 & 29022.89 & 23869.56 & 20672.64 \\
$\mathbb{E}$[volume] & 8626.36 & 7625.19 & 6450.42 & 3979.41 & 2757.35 \\
$\mathbb{P}$(profit) & 24.74\% & 51.44\% & 72.89\% & 85.93\% & 91.17\% \\
$\mathbb{P}$(buy–sell)  & 66.79\% & 51.44\% & 29.15\% & 0.00\% & 0.00\% \\
$\mathbb{P}$(arb) & 9.35\% & 3.72\% & 1.59\% & 0.77\% & 0.48\% \\
\bottomrule
\end{tabular}
\end{table}

\begin{table}[h]
\caption{Summary statistics for varying $\eta^0$ in Figure \ref{fig:histograms}}
\label{tab:stats_by_eta0}
\begin{tabular}{lrrrrr}
\toprule
$\eta^0$ (bps) & 0 & 5 & 20 & 35 & 60 \\
\midrule
$\mathbb{E}$[profit] (\$) & -3482.90 & -1576.77 & 17031.02 & 26137.29 & 29612.52 \\
$\mathbb{E}$[fees] (\$) & 8507.31 & 10417.23 & 29022.89 & 38116.95 & 41543.90 \\
$\mathbb{E}$[volume] & 1891.54 & 2316.01 & 6450.42 & 8471.36 & 9233.20 \\
$\mathbb{P}$(profit)  & 64.77\% & 71.42\% & 72.89\% & 55.46\% & 29.95\% \\
$\mathbb{P}$(buy–sell) & 0.00\% & 0.00\% & 29.15\% & 70.03\% & 84.97\% \\
$\mathbb{P}$(arb) & 35.19\% & 14.73\% & 1.59\% & 0.64\% & 0.33\% \\
\bottomrule
\end{tabular}
\end{table}

\begin{table}[h]
\caption{Summary statistics for varying $\sigma$ in Figure \ref{fig:histograms}}
\label{tab:stats_by_sigma}
\begin{tabular}{lrrrrr}
\toprule
$\sigma$ & 2.50\% & 3.25\% & 4.00\% & 4.75\% & 5.50\% \\
\midrule
$\mathbb{E}$[profit] (\$)  & 27382.31 & 22319.82 & 17031.02 & 11676.99 & 6223.08 \\
$\mathbb{E}$[fees] (\$) & 32068.21 & 30237.50 & 29022.89 & 28585.86 & 28891.86 \\
$\mathbb{E}$[volume] & 7126.94 & 6720.19 & 6450.42 & 6353.62 & 6422.21 \\
$\mathbb{P}$(profit)  & 94.00\% & 83.94\% & 72.89\% & 63.39\% & 55.97\% \\
$\mathbb{P}$(buy–sell) & 47.45\% & 36.71\% & 29.15\% & 23.97\% & 20.43\% \\
$\mathbb{P}$(arb) & 0.01\% & 0.29\% & 1.59\% & 4.17\% & 7.59\% \\
\bottomrule
\end{tabular}
\end{table}

\begin{table}[h]
\caption{Summary statistics for varying demand in Figure \ref{fig:histograms}}
\label{tab:stats_by_demand}
\begin{tabular}{lrrrrr}
\toprule
$\Delta^B=|\Delta^S|$ & 1000 & 3000 & 5000 & 7000 & 9000 \\
\midrule
$\mathbb{E}$[profit] (\$) & -4133.82 & 5570.90 & 17031.02 & 28459.36 & 36430.06 \\
$\mathbb{E}$[fees] (\$) & 7836.39 & 17556.25 & 29022.89 & 40453.67 & 48426.54 \\
$\mathbb{E}$[volume] & 1742.01 & 3902.03 & 6450.42 & 8991.11 & 10764.58 \\
$\mathbb{P}$(profit)  & 46.80\% & 63.50\% & 72.89\% & 76.30\% & 78.27\% \\
$\mathbb{P}$(buy–sell) & 16.51\% & 24.31\% & 29.15\% & 30.82\% & 31.90\% \\
$\mathbb{P}$(arb) & 11.62\% & 4.27\% & 1.59\% & 0.77\% & 0.44\% \\
\bottomrule
\end{tabular}
\end{table}

\newpage

\section{Model Properties}\label{sec:model.properties}

\subsection{Competitive Arbitrage}\label{app:competitive.arbitrage}

The arbitrage rule~\eqref{eq:arbTradeSize} admits an interpretation as a competitive equilibrium: when the current AMM and CEX prices permit a profitable marginal round trip, free entry among arbitrageurs induces arbitrage trades until the AMM price reaches the no-arbitrage boundary. Moreover, the subsequent fundamental trades do not create another contemporaneous arbitrage opportunity.

\begin{proposition}\label{prop:competitive.arbitrage}
Suppose arbitrage trades are arbitrarily divisible and immediately offset on the CEX. Then, under competitive arbitrage, the unique aggregate AMM trade
is given by~\eqref{eq:arbTradeSize}. After this trade, the fundamental trades
\eqref{eq:buyTradeSize}--\eqref{eq:sellTradeSize} preserve the no-arbitrage region.
\end{proposition}

\begin{proof}
Set $s=S_t^0$, $p=S_t^1$, and $y=Y_t$. At AMM marginal price $q$, the profits per unit
from selling $Y$ to the AMM and buying it on the CEX, or buying $Y$ from the AMM and
selling it on the CEX, are
$
(1-\eta^1)q-(1+\eta^0)s,
$
and 
$
(1-\eta^0)s-(1+\eta^1)q
$, 
respectively. Thus no profitable marginal round trip exists precisely when
\[
q\in[\underline p,\overline p],
\qquad
\underline p:=s\frac{1-\eta^0}{1+\eta^1},
\qquad
\overline p:=s\frac{1+\eta^0}{1-\eta^1}.
\]

If $p>\overline p$, selling $Y$ to the AMM decreases its marginal price continuously
according to~\eqref{eq:newMarginalPrice}. Marginal profit is strictly positive while the
price exceeds $\overline p$ and strictly negative below it. Competition therefore forces
the terminal price to $\overline p$. Similarly, if $p<\underline p$, buying $Y$ from the
AMM raises its price to $\underline p$, while no arbitrage occurs for
$p\in[\underline p,\overline p]$. Solving~\eqref{eq:newMarginalPrice} in these three cases
gives
\[
\Delta_t^A
=
\left[
y\left(1-\sqrt{\frac{p}{\overline p}}\right)
\right]\wedge0
+
\left[
y\left(1-\sqrt{\frac{p}{\underline p}}\right)
\right]\vee0,
\]
which is~\eqref{eq:arbTradeSize}. It remains to consider fundamental trading. The buy and sell routing boundaries from
\eqref{eq:buyTradeSize}--\eqref{eq:sellTradeSize} are
\[
p_B:=s\frac{1+\eta^0}{1+\eta^1},
\qquad
p_S:=s\frac{1-\eta^0}{1-\eta^1},
\]
satisfying 
$\underline p\le p_S$ and $p_B\le\overline p$. Starting from the no-arbitrage interval, a fundamental buyer either does not trade or
raises the AMM price only until $p_B$; hence it cannot move the price above
$\overline p$. Likewise, a fundamental seller either does not trade or lowers the price
only until $p_S$, and therefore cannot move it below $\underline p$. Thus each
fundamental trade, and hence their sequential composition, preserves
$[\underline p,\overline p]$.
\end{proof}

\subsection{Proof of Scale Invariance}
\label{app:scale.invariance}

\begin{proof}[Proof of Proposition~\ref{prop:scale.invariance}]

To prove the claimed scaling properties we proceed by induction over the ordered sequence of DEX transactions. Suppose that immediately before a DEX transaction, $X_t(L)=LX_t(1)$ and $Y_t(L)=LY_t(1)$. This holds initially by construction. Since the AMM marginal price is the ratio of the reserves,
\[
    S_t^1(L)
    =
    \frac{X_t(L)}{Y_t(L)}
    =
    \frac{X_t(1)}{Y_t(1)}
    =
    S_t^1(1).
\]
As the CEX price is common to the two models, the AMM--CEX price ratio is also unchanged.

The arbitrage quantity in \eqref{eq:arbTradeSize} is the current risky-asset reserve multiplied by a function of the price ratio and the fee parameters. It therefore scales by $L$: $\Delta_t^A(L)=L\Delta_t^A(1)$. Similarly, in \eqref{eq:buyTradeSize} and \eqref{eq:sellTradeSize}, both the desired fundamental volume and the quantity available before reaching the relevant price boundary scale by $L$. Hence,
\[
    \Delta_t^{B,\mathrm{DEX}}(L)
    =
    L\Delta_t^{B,\mathrm{DEX}}(1),
    \qquad
    \Delta_t^{S,\mathrm{DEX}}(L)
    =
    L\Delta_t^{S,\mathrm{DEX}}(1).
\]

Let $\Delta(L)=L\Delta(1)$ denote the size of any one of these transactions. Its average execution price given by \eqref{eq:executionPrice} is unchanged:
\[
    \frac{X_t(L)}
         {Y_t(L)-\Delta(L)}
    =
    \frac{LX_t(1)}
         {LY_t(1)-L\Delta(1)}
    =
    \frac{X_t(1)}
         {Y_t(1)-\Delta(1)}.
\]
It follows from the reserve update \eqref{eq:reservesUpdate} that $X_{t+\delta t}(L)=LX_{t+\delta t}(1)$ and $Y_{t+\delta t}(L)=LY_{t+\delta t}(1)$. Induction over the ordered sequence of DEX transactions therefore shows that reserves and trade quantities scale by $L$ throughout the simulation, while the DEX price and the AMM--CEX price ratio remain the same.

We now assess the consequences for the LP return. The fee collected on each transaction is $\eta^1P_t[\Delta]|\Delta|$.
Because the average execution price is unchanged and the trade quantity scales by $L$, cumulative fee revenue scales by $L$. The initial and terminal pool values marked at the CEX price are linear in the reserves, so unhedged LP PnL also scales by $L$. Finally,
\[
\sum_{i=1}^N
    Y_{(i-1)\delta t}(L)
    \left(
        S_{i\delta t}^0-S_{(i-1)\delta t}^0
    \right) =
    L\sum_{i=1}^N
    Y_{(i-1)\delta t}(1)
    \left(
        S_{i\delta t}^0-S_{(i-1)\delta t}^0
    \right),
\]
so hedged LP PnL scales by $L$ as well. Since $\mathrm{TVL}_0(L)=L\,\mathrm{TVL}_0(1)$,
hedged and unhedged LP returns relative to initial TVL are invariant to the scaling.
\end{proof}

\subsection{Infinite Demand and $\sigma=0$ Approximation}\label{sec:inf.demand.opt}

\begin{figure}[!h]
    \centering
    \includegraphics[width=0.49\textwidth]{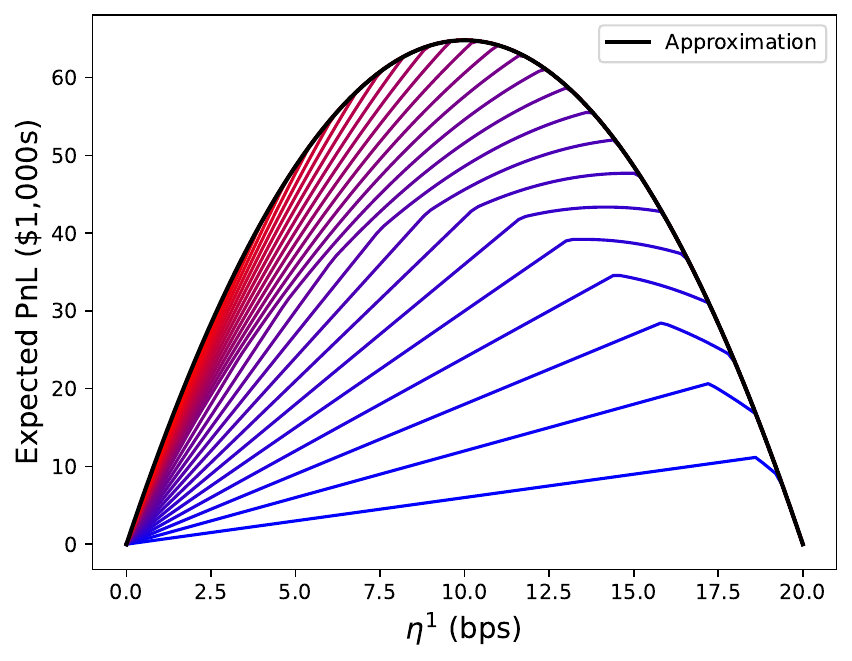}
    \caption{Illustration of the infinite demand approximation in \eqref{eqn:infinite.demand.approx} (black curve) for parameters $\sigma=0$ and $\eta^0 = 20$bps. The expected PnL curves for various demand levels $\Delta^B=|\Delta^S|\in \{1000,2000,\dots,25000\}$ are provided on an increasing gradient from blue to red.}
    \label{fig:approx.infinite.demand}
\end{figure}

Since demand is infinite and purchases/sales occur only at the sell/buy boundaries, respectively, we have the following volumes (when non-zero)
\[
\Delta_t^{B,DEX} = Y_t \left(1 - \sqrt{\frac{(1 - \eta^0)(1 + \eta^1)}{(1 - \eta^1)(1 + \eta^0)}} \right), \ \ \ 
\Delta_t^{S,DEX} = Y_t \left(1 - \sqrt{\frac{(1 + \eta^0)(1 - \eta^1)}{(1 + \eta^1)(1 - \eta^0)}} \right).
\]
The corresponding fee revenue collected by the LP is
\[
\text{Buy Revenue} = \eta^1 P_t[\Delta^{B,DEX}_t] \Delta^{B,DEX}_t, \qquad 
\text{Sell Revenue} = \eta^1 P_t[\Delta^{S,DEX}_t] \left|\Delta^{S,DEX}_t\right|,
\]
and the expected fee revenue per period (when initializing the price ratio at either the buy or sell boundary) is thus given by
\[
\mathbb{E}\left[ \text{Fee Revenue per period} \right] = \mathbb{E}\left[ \text{Buy Revenue} + \text{Sell Revenue} \right].
\]
Substituting the trade expressions and simplifying (using that we pick up approximately 75\% of a round trip and that $X_t\approx X_0$) yields
\begin{equation}\label{eqn:infinite.demand.approx}
\frac{1}{N}\mathbb{E}\left[ \text{Fee Revenue} \right] \approx \frac{3}{4} X_0 \eta^1 \left[ \sqrt{\frac{(1 + \eta^0)(1 - \eta^1)}{(1 + \eta^1)(1 - \eta^0)}} - \sqrt{\frac{(1 - \eta^0)(1 + \eta^1)}{(1 - \eta^1)(1 + \eta^0)}} \right].
\end{equation}
Expanding \eqref{eqn:infinite.demand.approx} to second order in $\eta^0$ and $\eta^1$ gives the expression reported in Section \ref{sec:mod.to.high.demand}. An illustration of the approximation accuracy of \eqref{eqn:infinite.demand.approx} for various demands is given in Figure \ref{fig:approx.infinite.demand}. 

As an aside, we recall the breakdown in monotonicity of the optimal fee with respect to demand observed in Figure~\ref{fig:optimal_fee_curves} of Section~\ref{sec:optimal.fees}. Figure~\ref{fig:approx.infinite.demand} exposes the precise mechanism by which monotonicity fails: the curve’s maximum translates abruptly from left to right at an intermediate demand level.

\section{Competitive Entry under the Model Extensions}
\label{app:equilibrium.extensions}

\begin{figure}
    \centering
        \includegraphics[width=.43\linewidth]{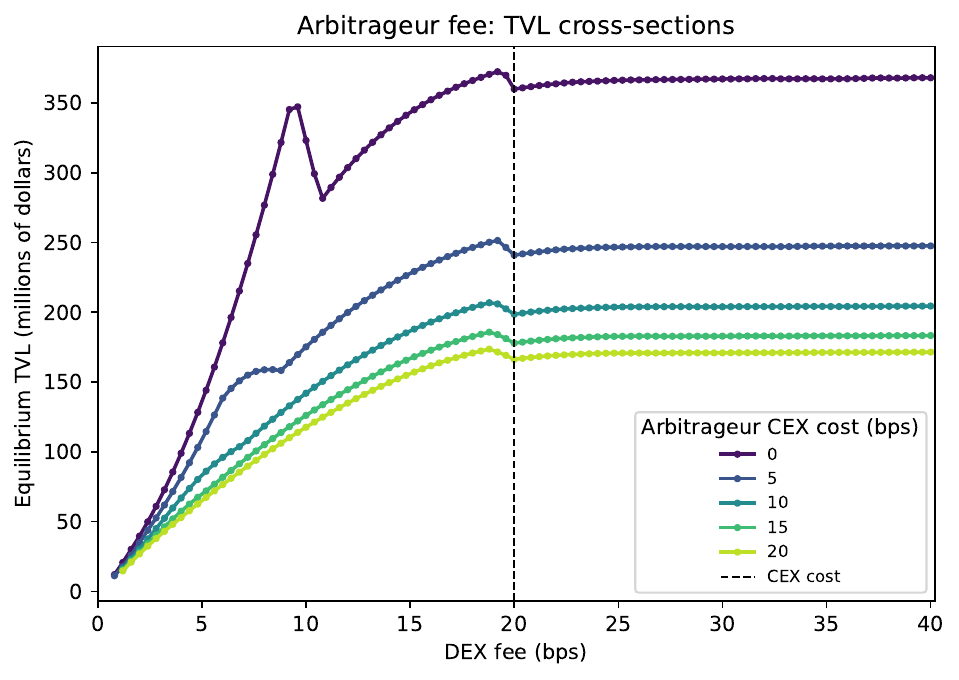}
        \includegraphics[width=.43\linewidth]{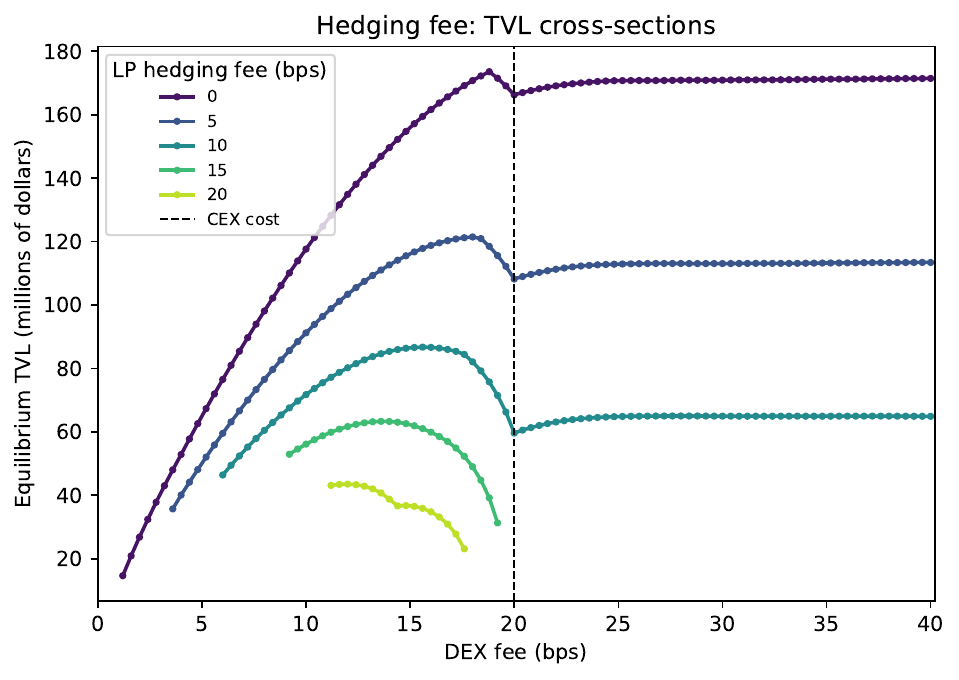}
        \includegraphics[width=.43\linewidth]{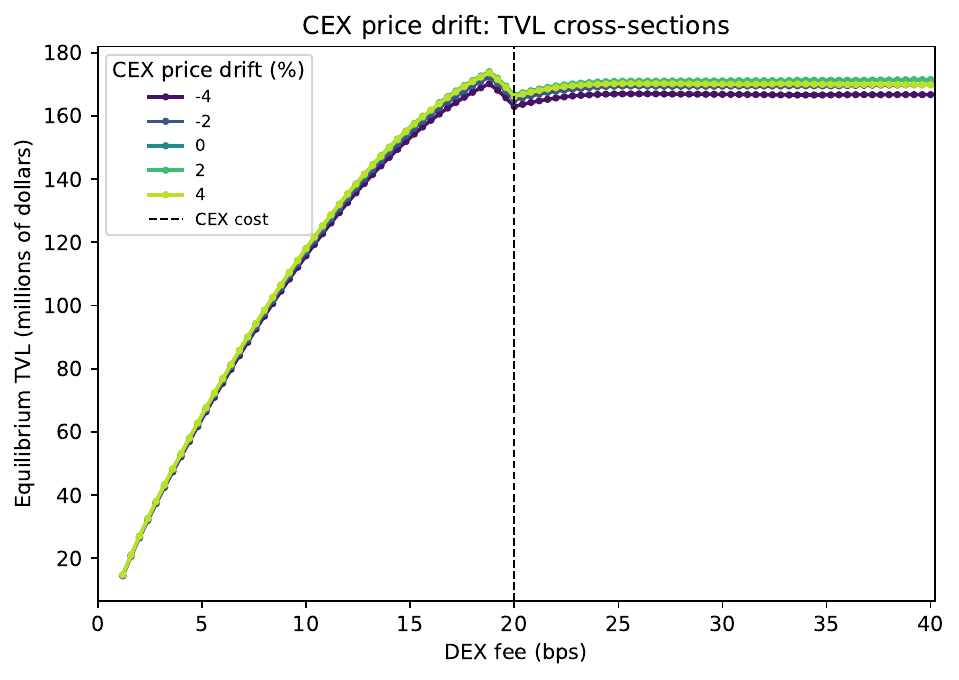}
        \includegraphics[width=.43\linewidth]{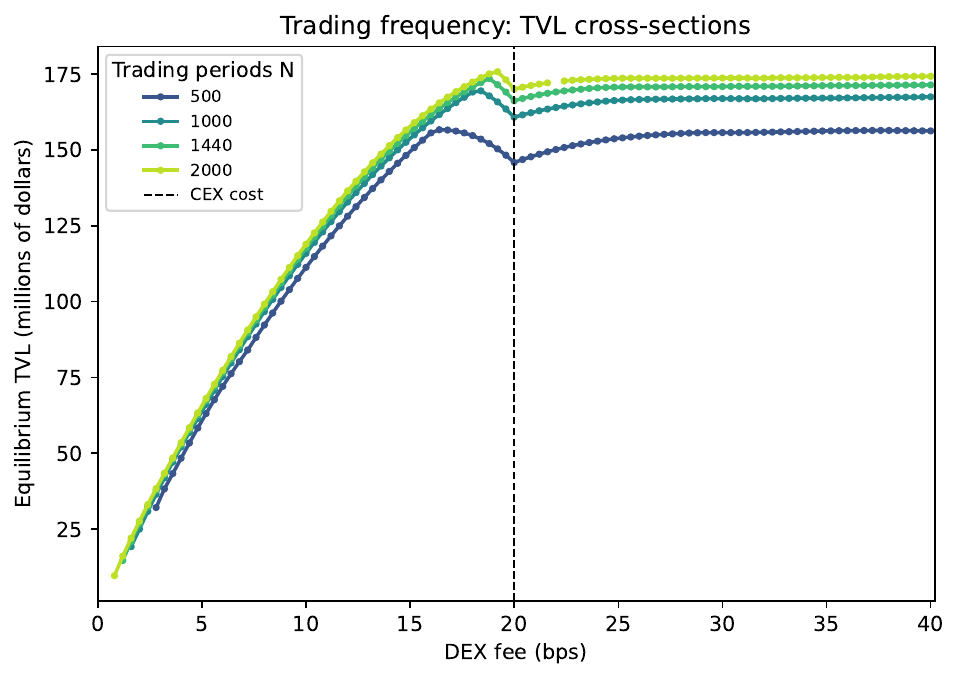}
        \includegraphics[width=.43\linewidth]{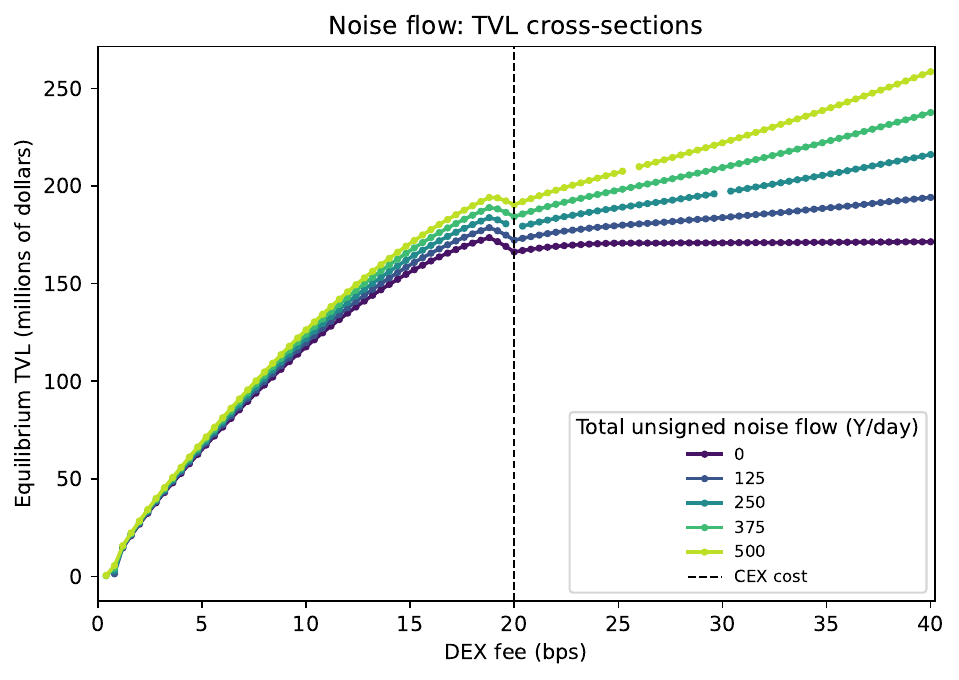}
        \includegraphics[width=.43\linewidth]{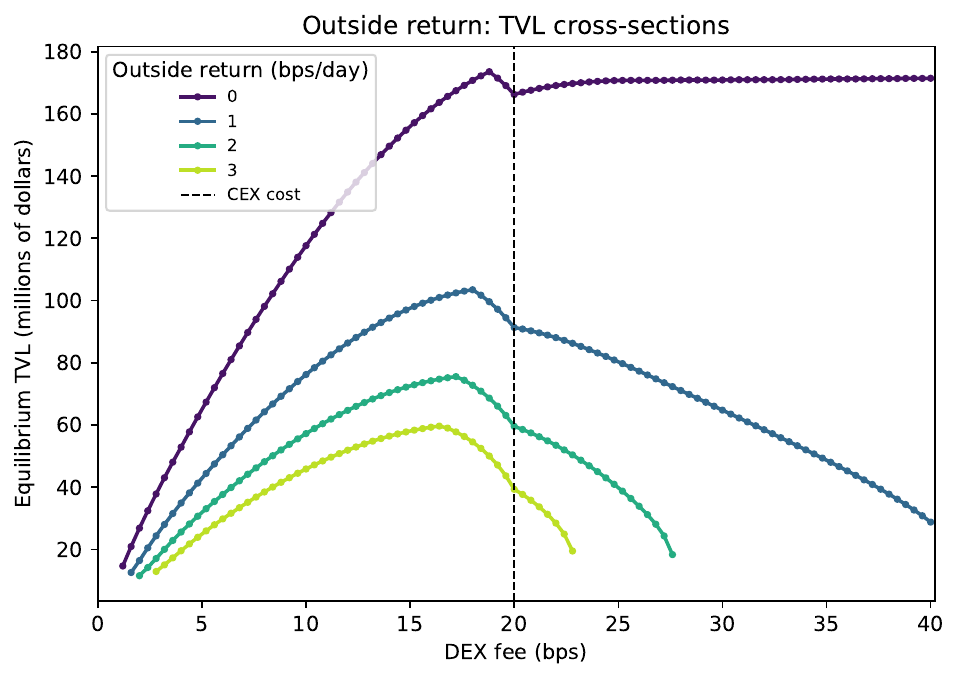}
        \caption{Equilibrium TVL under the model extensions. Reading from left to right and top to bottom, the panels vary the arbitrageur fee, LP hedging fee, CEX price drift, 
        trading frequency, 
        noise flow, and outside return. Curve values are shown only when there is a unique stable equilibrium.}
        \label{fig:equilibrium.extensions}
\end{figure}

In this appendix we report the competitive-entry counterparts to the extension exercises in \cref{sec:extensions}. In each panel of \cref{fig:equilibrium.extensions}, we vary one parameter at a time and otherwise retain the benchmark specification. We also vary the outside return $\bar r$, which plays no role in the fixed-liquidity PnL calculations but enters directly into the competitive-entry condition.

As in the equilibrium analysis of \cref{sec:equilibrium}, curves are shown only at fee values for which the numerical procedure identifies a unique, stable, finite equilibrium on the liquidity search grid. Missing portions of a curve therefore do not correspond to zero TVL.

The trading-frequency and drift panels are broadly consistent with the fixed-liquidity results. We see that changing either parameter has only a modest effect on equilibrium TVL, while the TVL-maximizing fee remains close to, but slightly below, the CEX trading cost. Changes in the arbitrageur cost have a larger effect on the level of equilibrium TVL. As $\eta^A$ decreases, the tighter arbitrage bounds keep the price ratio closer to the LP profit region and allow the pool to support more liquidity. Nonetheless, the TVL-maximizing fee remains close to $\eta^0$.

We also note two kinks in the arbitrageur-fee panel. At $\eta^1=\eta^0$, the AMM loses its strict fee advantage over the CEX and the routing of fundamental trades changes. The common kink there coincides with this change in routing regime, and we can therefore interpret it as a feature of the routing rule. A second, more pronounced kink for $\eta^A=0$ appears near $\eta^1=\eta^0/2$. It is at approximately this value that the post-arbitrage price boundaries cross the routing thresholds for opposite-side fundamental flow. For AMM fees below this level, both fundamental buyers and sellers initially route their marginal volume to the AMM, regardless of which side trades first. The loss of this two-sided flow appears to create the additional local maximum in TVL for the curves corresponding to small $\eta^A$.

By contrast, the outside return enters directly into the competitive-entry condition. We see that equilibrium TVL decreases as $\bar r$ increases. In particular, increasing the one-day outside return from $0$ to $3$ basis points materially lowers equilibrium TVL. Importantly, it also moves the TVL-maximizing fee farther below $\eta^0$ and narrows the fee range over which the numerical procedure identifies a stable, finite equilibrium. Since a positive outside return is plausible, this finding reinforces the broader conclusion that the AMM fee should undercut the CEX trading cost for TVL maximization.

At fixed liquidity, the hedging fee changes neither order routing nor the AMM price path. While hedging reduces the variance of LP PnL, the associated fee can only lower the LP's \emph{expected} hedged return. Once pool depth adjusts through competitive entry, however, this reduction in return leads to a shallower pool and can thereby affect routing. Moreover, as $\eta^H$ approaches $\eta^0$, the set of price ratios for which fundamental flow is both routed to the AMM and profitable for the LP narrows. Setting the AMM fee farther below the CEX trading cost expands the range of price ratios over which marginal flow from both fundamental buyers and sellers is routed to the pool. Consistent with this mechanism, at higher hedging costs the TVL-maximizing fee moves farther below $\eta^0$, and the corresponding curves turn downward before $\eta^1$ reaches $\eta^0$. For the largest values of $\eta^H$, our numerical procedure also identifies stable, finite equilibria only over an intermediate fee range.

The interpretation of the noise-flow panel requires some care since noise demand is assumed to be fee-insensitive. An increase in $\eta^1$ therefore raises fee revenue from this flow without reducing its volume, allowing a larger amount of liquidity to enter. We consequently see equilibrium TVL continue to increase after the AMM fee exceeds the CEX trading cost for every positive level of noise flow shown. Since the corresponding curves are still increasing at the upper edge of the displayed fee grid, the largest displayed value does not identify a TVL-maximizing fee. Indeed, the fee-insensitive specification is not suitable for identifying a global optimum because it ignores the natural demand response that ought to appear at implausibly large fees. Nonetheless, for moderate values of noise flow, the figures are broadly similar near and below the CEX fee.

Unlike the fixed-liquidity analysis in \cref{sec:extensions}, we do not illustrate the case with mean-reverting prices. Under the Ornstein--Uhlenbeck dynamics specified there, the CEX price is not a martingale and the hedging strategy can have nonzero expected PnL. While the competitive-entry condition remains well-defined with a zero return outside option, at such an equilibrium, expected unhedged LP PnL equals the expected
gain on the hedging portfolio, \emph{which need not be zero}. The resulting equilibrium outcomes can therefore differ
substantially from the baseline since mean reversion affects the performance of the hedge. Mean reversion can also create trading opportunities on the CEX with positive expected returns, which suggests that assuming a zero outside return may be economically implausible and misspecify the true opportunity cost. Similar considerations apply when we add a drift to the price, but at the small scale at which it is introduced in \cref{fig:equilibrium.extensions}, the effects are much more limited. Notably, the conclusion that the TVL-maximizing fee lies below $\eta^0$ need not persist under strong short-run mean reversion.

\clearpage
\section{Supplementary Empirical Evidence on Modeling Assumptions}
\label{app:modeling-assumptions}

This appendix reports two supplementary empirical exercises meant to support key modeling choices in the paper. First, we assess whether the approximation of nearly constant liquidity is reasonable over the short horizons relevant for our empirical calibration. Second, we examine whether a robust reduced-form relation between aggregate LP provision and exogenous market volatility is clearly visible in the data for the large Uniswap v3 ETH/USDC pools that are most relevant for our discussion.

\subsection{Short-Horizon Stability of Pool Liquidity}

We begin by assessing whether the constant-liquidity approximation used in the empirical calibration is reasonable over short horizons. To this end, we consider the two main Uniswap v3 ETH/USDC pools on Ethereum, corresponding to the 5 bps and 30 bps fee tiers, over the period from January to June 2025. Using daily data, we compute the distributions of (unsigned) relative TVL changes over one-day and seven-day horizons.

\begin{figure}[!htbp]
    \centering
    \includegraphics[width=\textwidth]{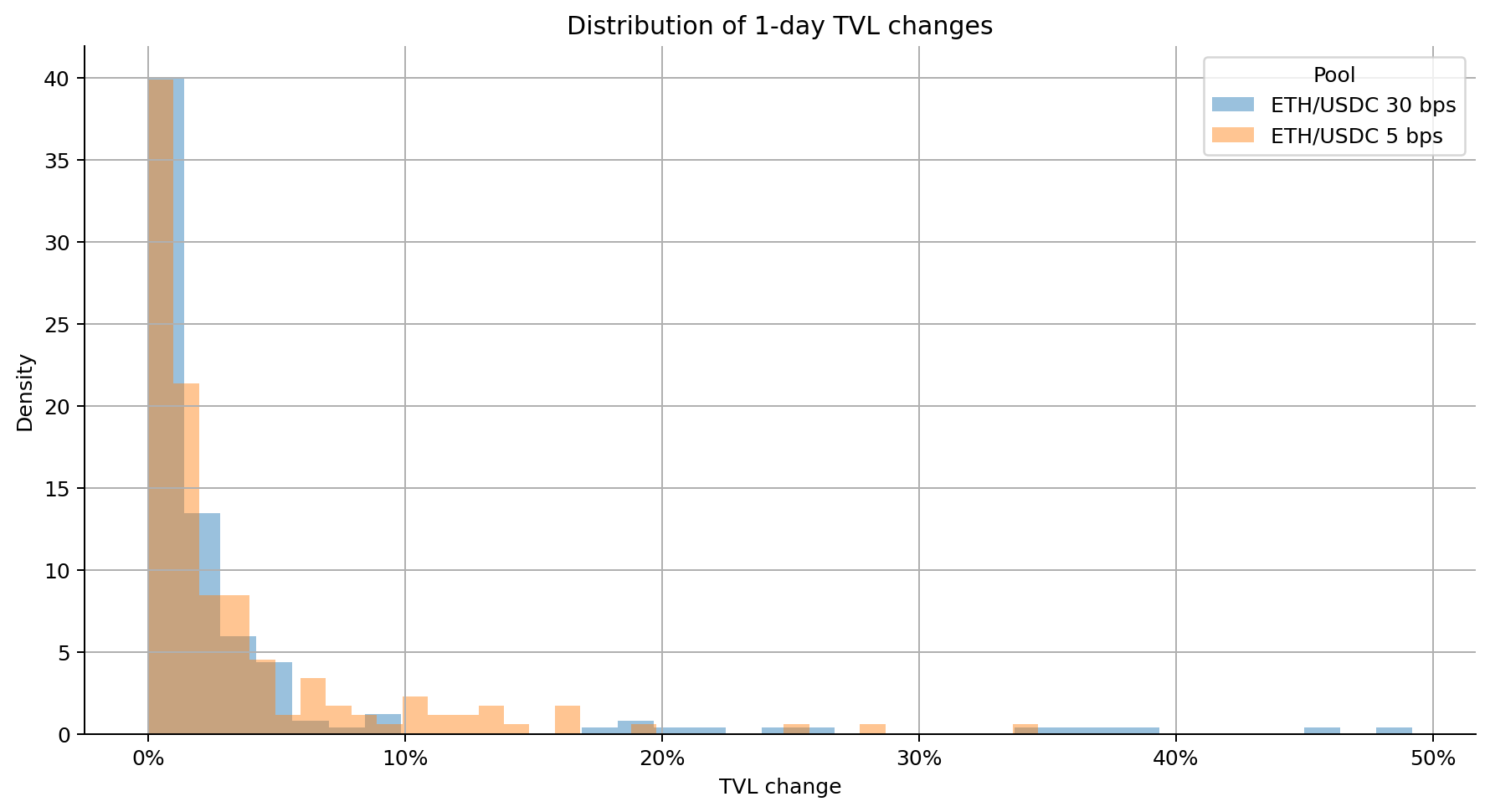}
    \caption{Distribution of TVL changes over one day for the two main Uniswap v3 ETH/USDC pools on Ethereum over the first six months of 2025. The distribution is tightly concentrated near zero, indicating that TVL remains very stable at the daily horizon.}
    \label{fig:tvl_moves_1d}
\end{figure}

\begin{figure}[!htbp]
    \centering
    \includegraphics[width=\textwidth]{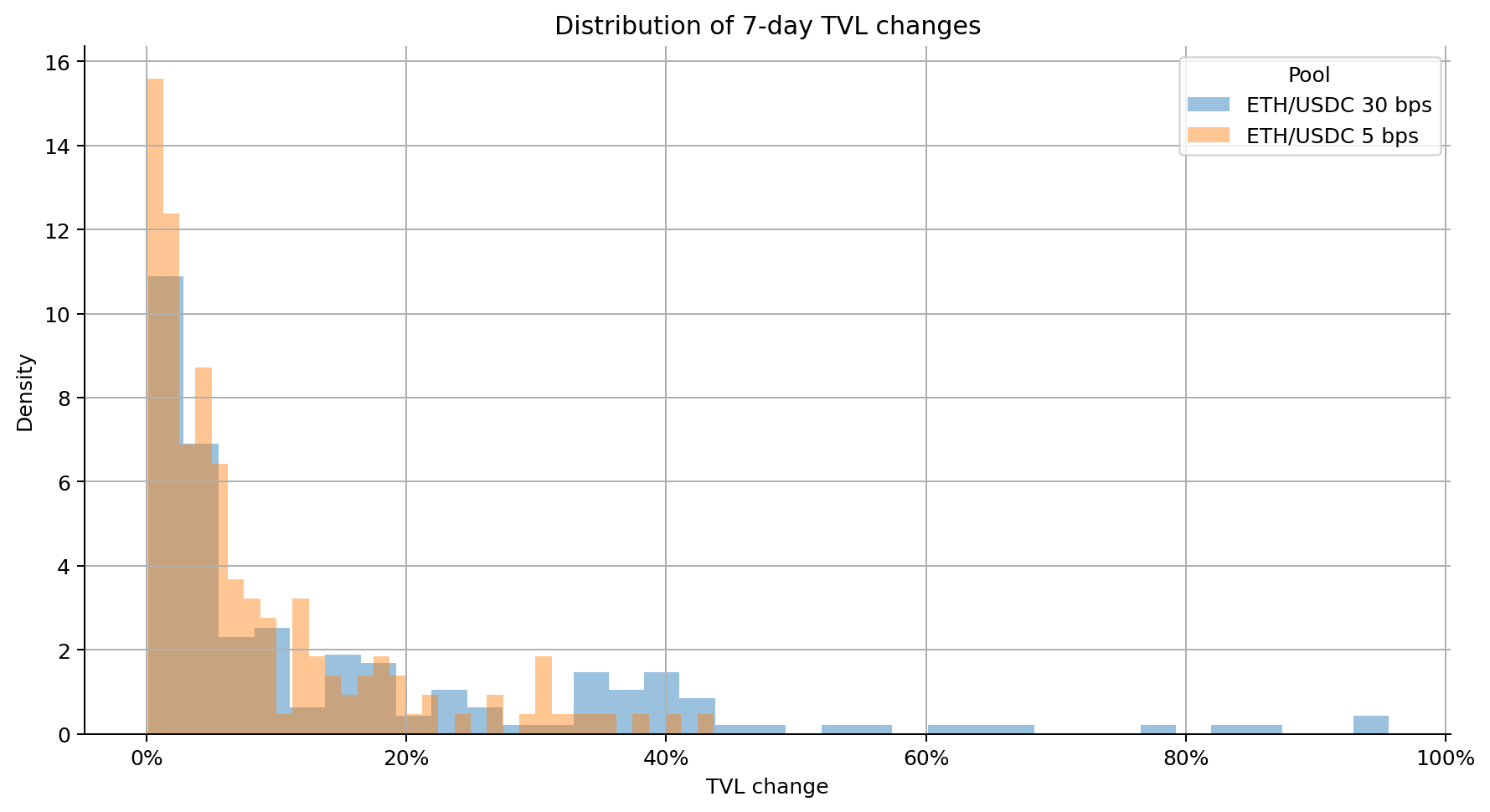}
    \caption{Distribution of TVL changes over seven days for the two main Uniswap v3 ETH/USDC pools on Ethereum over the first six months of 2025. Although naturally more dispersed than at the daily horizon, the distribution remains concentrated near zero, indicating that TVL remains fairly stable over one week.}
    \label{fig:tvl_moves_7d}
\end{figure}

Figures~\ref{fig:tvl_moves_1d} and~\ref{fig:tvl_moves_7d} show that these changes are tightly concentrated around zero, especially at the daily horizon, and remain moderate even over one week. This suggests that, for the purpose of our reduced-form calibration over a one-month window, treating pool liquidity as approximately constant is a reasonable approximation of the short-run behavior of these large pools. 

\vspace{2cm}
\subsection{LP Provision and Exogenous Volatility}

We next report a supplementary empirical exercise on the relation between aggregate LP provision and market volatility. The purpose of this exercise is deliberately modest. Rather than estimating a structural model of LP behavior, we simply ask whether a robust reduced-form relation is clearly visible in the data for the major Uniswap v3 ETH/USDC pools that are most relevant for our discussion.

\begin{figure}[!htbp]
    \centering
    \includegraphics[page=1,width=0.92\textwidth]{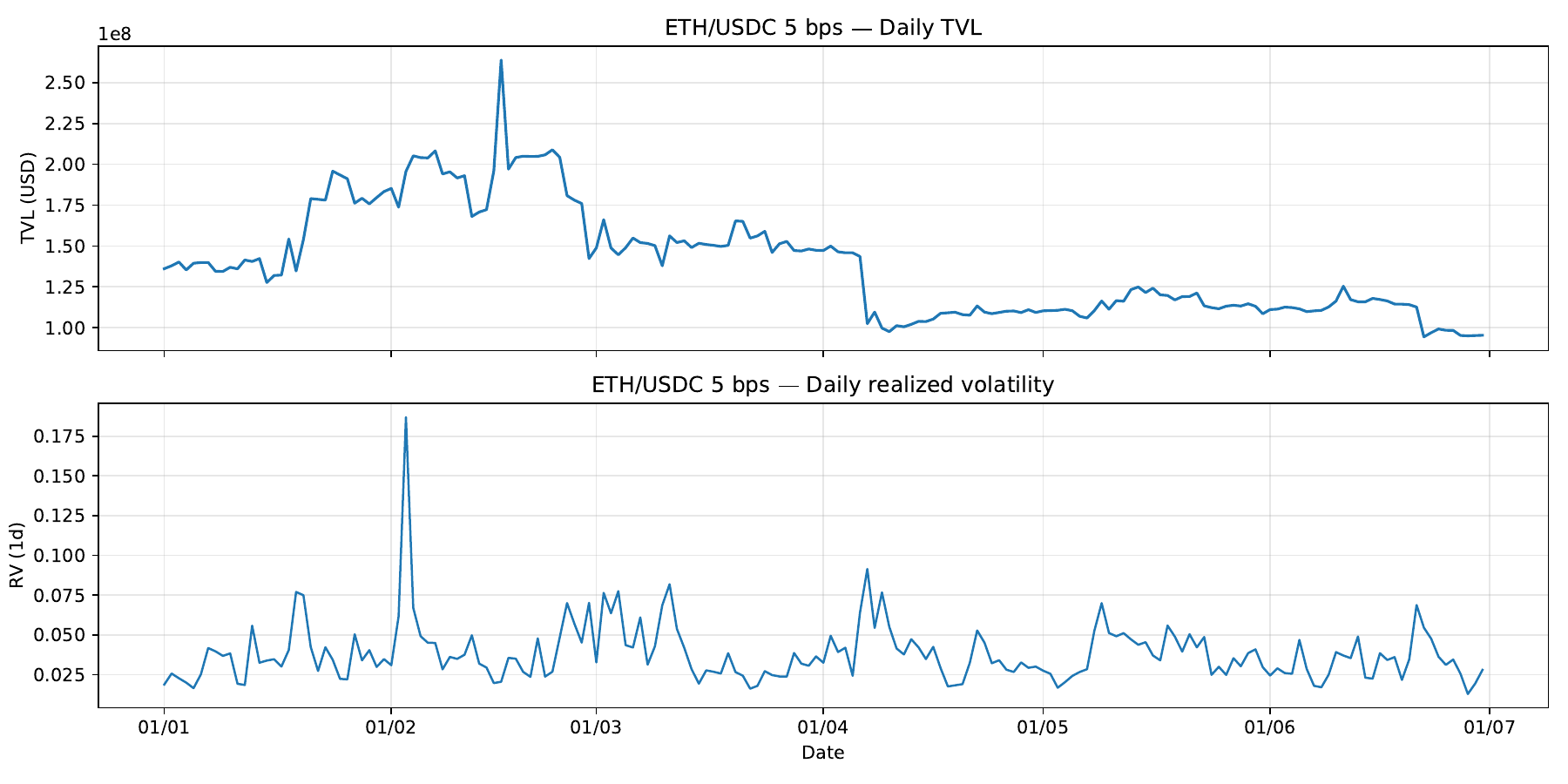}
    \caption{ETH/USDC 5 bps pool: daily pool size and daily realized volatility. The figure does not reveal systematic liquidity withdrawals during volatility spikes.}
    \label{fig:lpvol-5bps-dailyvol}
\end{figure}

\begin{figure}[!htbp]
    \centering
    \includegraphics[page=1,width=0.92\textwidth]{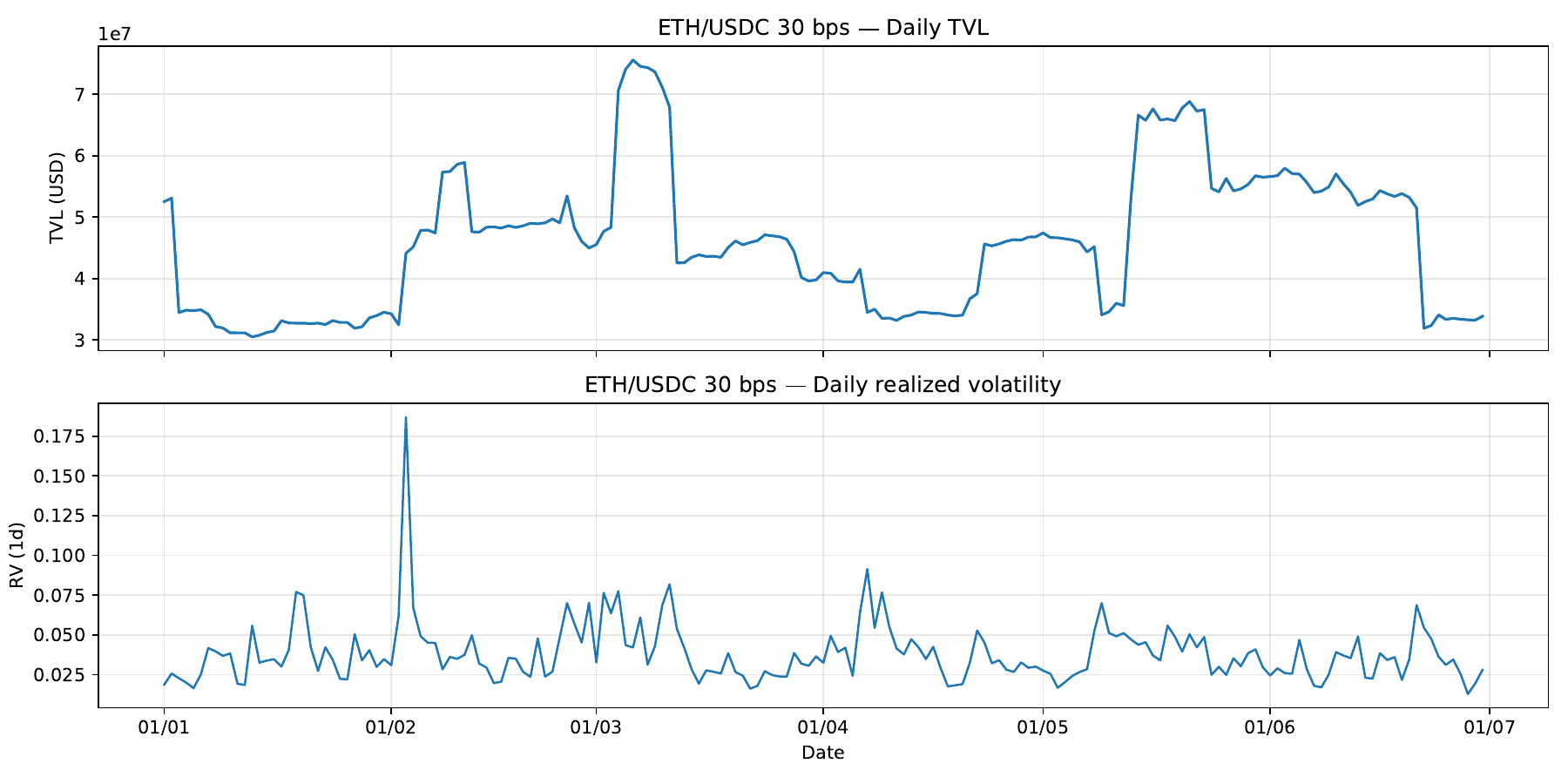}
    \caption{ETH/USDC 30 bps pool: daily pool size and daily realized volatility. As in the 5 bps pool, no clear co-movement between pool size and volatility is visually apparent.}
    \label{fig:lpvol-30bps-dailyvol}
\end{figure}

\begin{figure}[!htbp]
    \centering
    \includegraphics[page=1,width=0.92\textwidth]{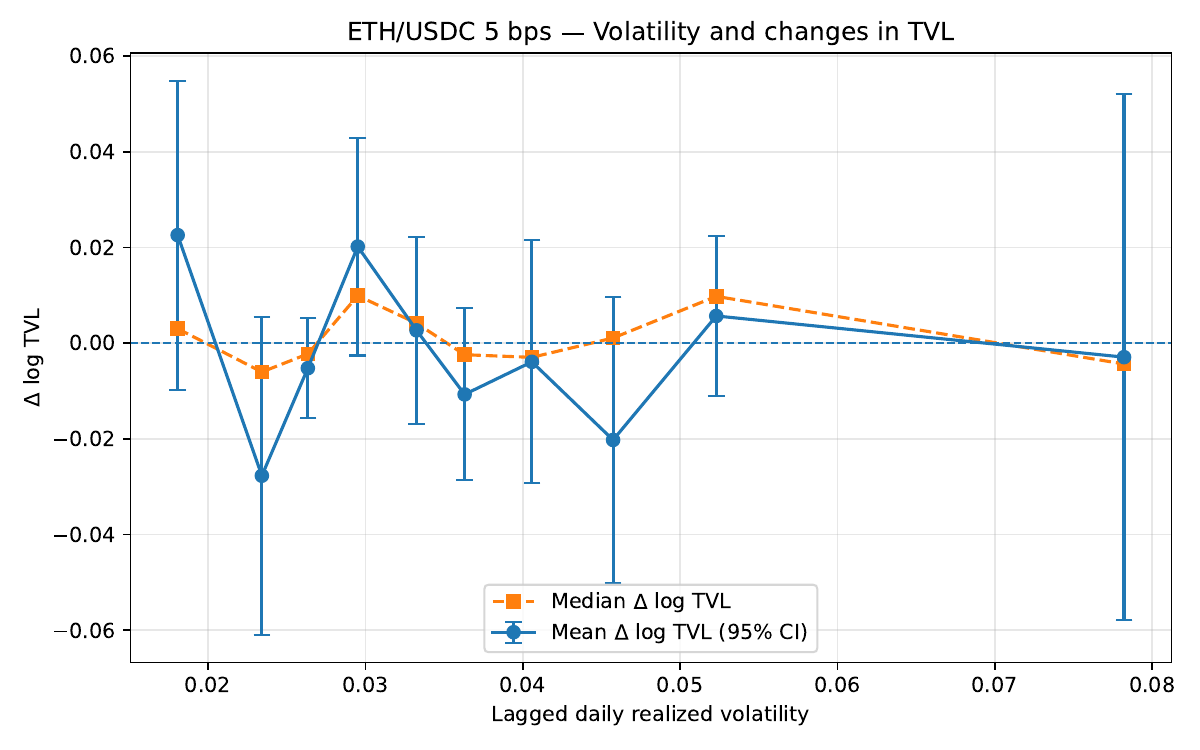}
    \caption{ETH/USDC 5 bps pool: binned changes in pool size against lagged daily realized volatility. Average and median changes remain close to zero across bins, with no clear monotone pattern.}
    \label{fig:lpvol-5bps-bins}
\end{figure}

\begin{figure}[!htbp]
    \centering
    \includegraphics[page=1,width=0.92\textwidth]{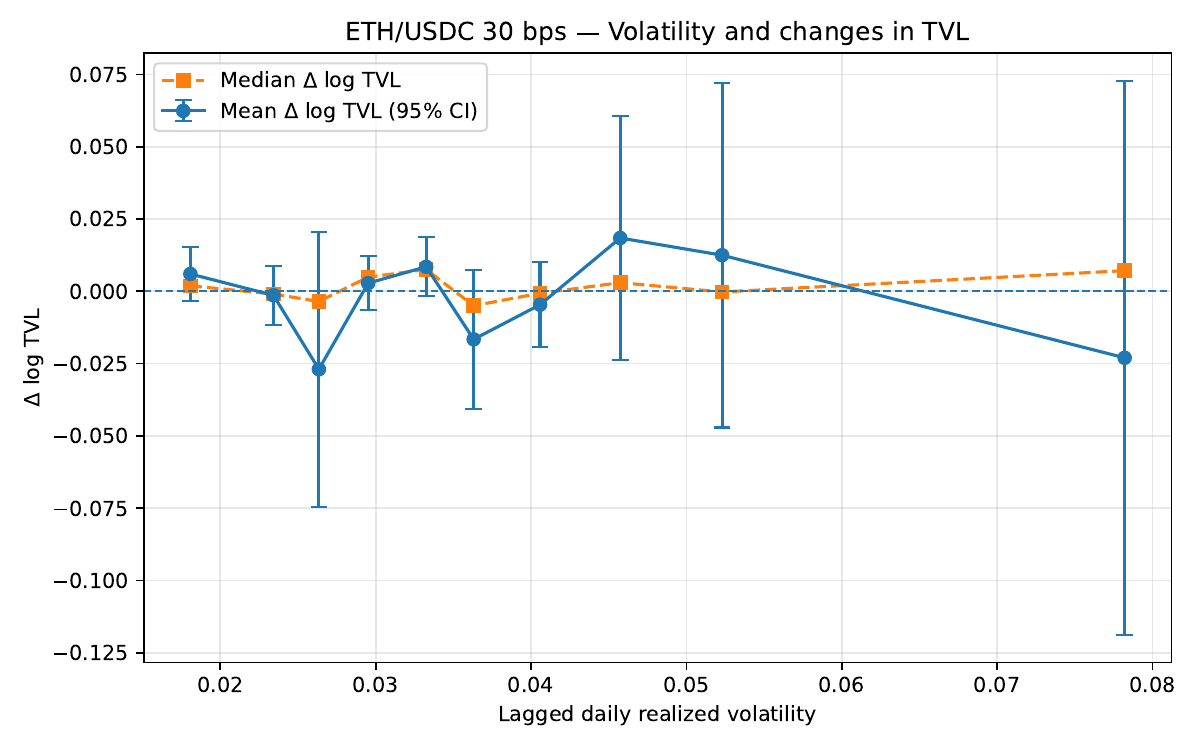}
    \caption{ETH/USDC 30 bps pool: binned changes in pool size against lagged daily realized volatility. The estimated relation is essentially flat, with wide confidence intervals and no systematic trend.}
    \label{fig:lpvol-30bps-bins}
\end{figure}

We again consider the two main Uniswap v3 ETH/USDC pools on Ethereum, corresponding to the 5 bps and 30 bps fee tiers, over the period from January to June 2025. For each pool, we construct a daily measure of aggregate pool size and compare it with daily realized ETH/USDC volatility computed from minute-level Binance prices. Figures~\ref{fig:lpvol-5bps-dailyvol} and~\ref{fig:lpvol-30bps-dailyvol} plot daily pool size together with daily realized volatility for the 5 bps and 30 bps pools, respectively. The figures do not reveal systematic liquidity withdrawals during volatility spikes, nor any visually stable co-movement between pool size and realized volatility. Figures~\ref{fig:lpvol-5bps-bins} and~\ref{fig:lpvol-30bps-bins} then report binned plots of subsequent changes in pool size against lagged daily realized volatility. In both fee tiers, the estimated mean and median changes remain close to zero across volatility bins, with no clear monotone pattern.

The closest dedicated empirical paper of which we are aware is \cite{chen2026liquidity}. Their paper studies a large cross-section of Uniswap v2 pools and reports statistically significant relations between net liquidity provision and volatility, with signs that differ across stable and unstable pools. Our exercise differs from theirs in an important respect, namely in the construction of the volatility variable. In their empirical analysis, volatility is measured from the pool price itself. In an AMM, however, thinner liquidity mechanically amplifies price impact and slippage. As a result, volatility measured from the pool price may partly reflect the current liquidity state of the pool, rather than a purely exogenous driver of LP decisions. By contrast, our objective here is precisely to assess whether exogenous market volatility helps explain LP provision. For that reason, our empirical exercise relies on an external reference measure of ETH/USDC volatility rather than on the pool price alone. In that sense, our null result should be interpreted as evidence that, once one mitigates this mechanical feedback, the direct reduced-form link between volatility and liquidity provision becomes much weaker in the large ETH/USDC pools that we analyze.

Taken together, these diagnostics do not indicate a robust or economically meaningful reduced-form relation between aggregate LP provision and exogenous market volatility in these large ETH/USDC pools. This exercise is only meant as a supplementary robustness check, not as a thorough empirical study. Nevertheless, it suggests that  our modeling choice is not contradicted by the data we examine.

We have repeated the same exercise on several additional Uniswap v3 pools, namely ETH/USDT, BTC/USDC, BTC/ETH, LINK/ETH, and DAI/ETH, and obtained the same qualitative conclusion: liquidity provision does not appear to be responsive to changes in volatility over short time horizons.

\clearpage
\end{document}